\documentclass[11pt,letterpaper]{article}

\usepackage[margin=1in]{geometry}
\usepackage[hyphens]{url}
\usepackage{graphicx}
\usepackage{natbib}
\usepackage{booktabs}
\usepackage{longtable}
\usepackage{array}
\usepackage{tabularx}
\newcolumntype{Y}{>{\raggedright\arraybackslash}X}
\usepackage{caption}
\usepackage{xcolor}
\usepackage{tikz}
\usetikzlibrary{positioning,arrows.meta,shapes.geometric,fit,backgrounds,calc}
\usepackage[section]{placeins} 
\usepackage{hyperref}
\hypersetup{colorlinks=true,linkcolor=black,citecolor=blue!50!black,urlcolor=blue!50!black}
\makeatletter
\renewcommand{\fps@table}{htbp}
\renewcommand{\fps@figure}{htbp}
\makeatother
\newcommand{\substance}{\textsc{Substance}}
\newcommand{\motion}{\textsc{Motion}}
\newcommand{\control}{\textsc{Control}}
\newcommand{\regime}[2]{\noindent{\small\textit{Regime:} \textsc{#1}. #2}\par\vspace{0.4em}}
\newcommand{\cmark}{$\bullet$}   
\newcommand{\pmark}{$\circ$}     
\newcommand{\xmark}{$\cdot$}     

\title{\textbf{Always-On Agents: A Survey of Persistent Memory, State, and Governance in LLM Agents}}
\author{
Tianyu Ding\quad
Aditya Nannapaneni\quad
Bingfan Liu\quad
Ling Zhang
}
\date{}

\begin{document}
\maketitle

\begin{abstract}
Always-on agents are systems whose future behavior depends on durable state
accumulated across earlier interactions. We treat them as
\emph{persistent-state systems}: the operative system includes retrievable
memories, but also task ledgers, permissions, credentials, commitments,
provenance and audit records, shared state, trigger conditions, and externally
committed effects linked to those records. The survey reads the literature through six diagnostic axes for
each state item, authority, scope, mutability, provenance, recoverability, and
actionability, and through a lifecycle in which state is written, validated,
organized, retrieved, acted upon, updated, forgotten, audited, and sometimes
rolled back. Across a $435$-work coded corpus, treated as a scoped map rather
than an exhaustive census, the literature concentrates more heavily on
accumulating and retrieving state than on governing, recovering, or
relinquishing it. We therefore introduce the Always-On Evaluation Protocol
(AOEP-v0), a pilot evaluation contract that makes these governance requirements
concrete by scoring state mutation and recovery obligations rather than answer
quality alone. The resulting agenda connects always-on agents to databases,
distributed systems, formal methods, capability security, and machine
unlearning.

\end{abstract}

\tableofcontents
\newpage


\newcommand{\figevolution}{%
\begin{figure}[t]
\centering
\includegraphics[width=\linewidth]{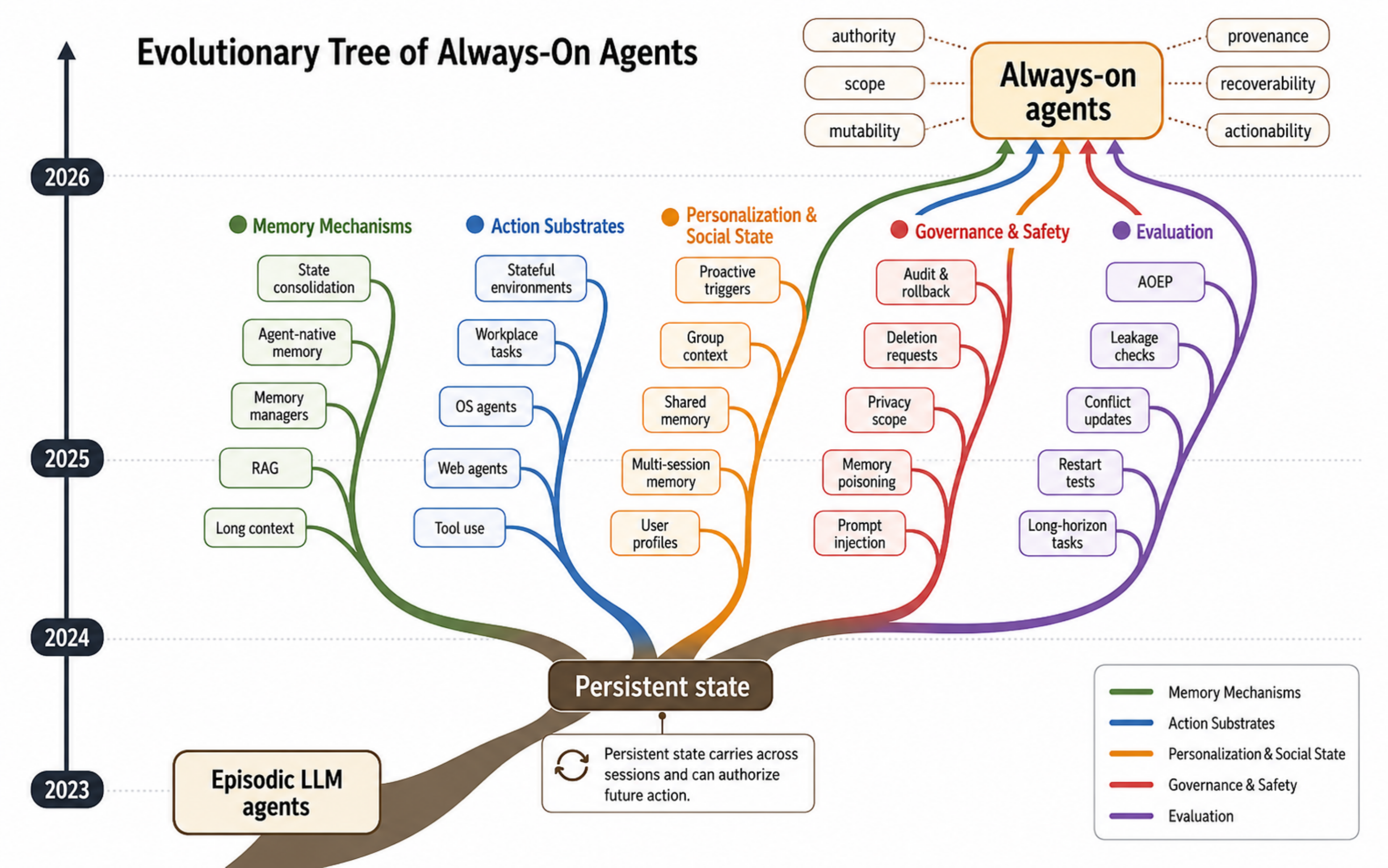}
\caption{Evolutionary tree of always-on agents. The figure sketches the survey's
organizing claim: episodic LLM agents become always-on not by running
continuously, but when durable state persists across sessions and can authorize
future action. Five strands map the field's current build-out: memory mechanisms,
action substrates, personalization and social state, governance and safety, and
evaluation. The target state is qualified by six diagnostic axes: authority,
scope, mutability, provenance, recoverability, and actionability. These are the
questions that make persistence governable rather than only useful.}
\label{fig:evolution-tree}
\end{figure}}

\newcommand{\figoverview}{%
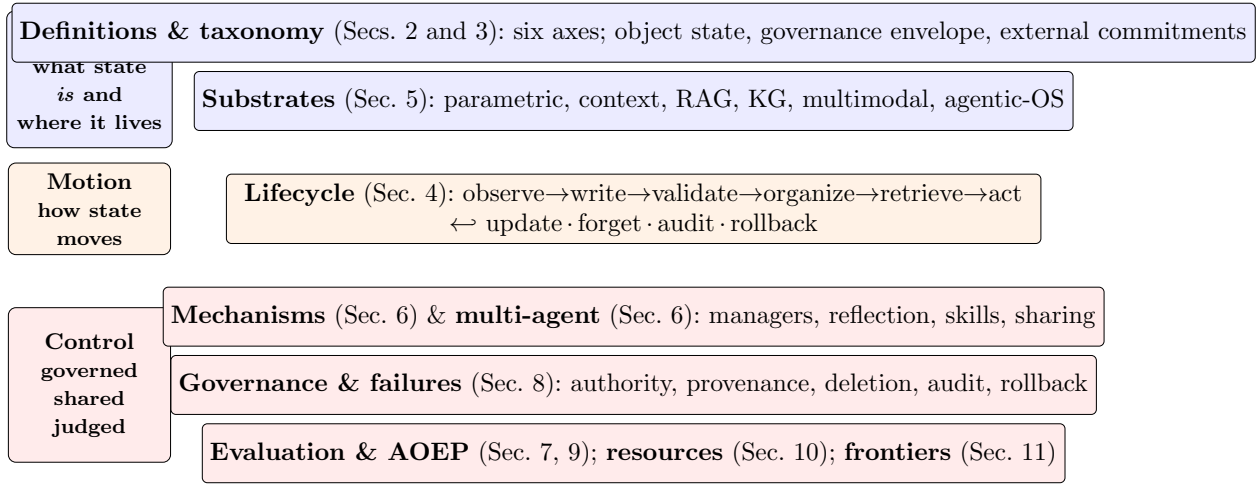
\begin{figure}[t]
\centering
\resizebox{\linewidth}{!}{%
\begin{tikzpicture}[font=\small,
  layer/.style={draw,rounded corners=2pt,minimum width=11.3cm,minimum height=0.82cm,align=center,inner sep=3pt},
  regime/.style={draw,rounded corners=3pt,minimum width=2.25cm,align=center,font=\footnotesize\bfseries,inner sep=4pt},
  lens/.style={draw,dashed,rounded corners=2pt,minimum width=3.5cm,minimum height=0.7cm,align=center,font=\footnotesize}]
\node[regime,fill=blue!8,minimum height=1.9cm] (sub) at (-7.45,2.35) {\substance\\\scriptsize what state\\\scriptsize \emph{is} and\\\scriptsize where it lives};
\node[regime,fill=orange!10,minimum height=0.95cm] (mot) at (-7.45,0.55) {\motion\\\scriptsize how state\\\scriptsize moves};
\node[regime,fill=red!8,minimum height=2.15cm] (con) at (-7.45,-1.9) {\control\\\scriptsize governed\\\scriptsize shared\\\scriptsize judged};
\node[layer,fill=blue!8] (l1) at (0.1,3.0) {\textbf{Definitions \& taxonomy} (Secs.~\ref{sec:definition} and~\ref{sec:statetax}): six axes; object state, governance envelope, external commitments};
\node[layer,fill=blue!8] (l2) at (0.1,2.05) {\textbf{Substrates} (Sec.~\ref{sec:substrates}): parametric, context, RAG, KG, multimodal, agentic-OS};
\node[layer,fill=orange!10] (l3) at (0.1,0.55) {\textbf{Lifecycle} (Sec.~\ref{sec:lifecycle}): observe$\to$write$\to$validate$\to$organize$\to$retrieve$\to$act\\$\hookleftarrow$ update$\,\cdot\,$forget$\,\cdot\,$audit$\,\cdot\,$rollback};
\node[layer,fill=red!8] (l4) at (0.1,-0.95) {\textbf{Mechanisms} (Sec.~\ref{sec:mechanisms}) \& \textbf{multi-agent} (Sec.~\ref{sec:mechanisms}): managers, reflection, skills, sharing};
\node[layer,fill=red!8] (l5) at (0.1,-1.9) {\textbf{Governance \& failures} (Sec.~\ref{sec:failures}): authority, provenance, deletion, audit, rollback};
\node[layer,fill=red!8] (l6) at (0.1,-2.85) {\textbf{Evaluation \& AOEP} (Sec.~\ref{sec:evaluation},~\ref{sec:aoep}); \textbf{resources} (Sec.~\ref{sec:resources}); \textbf{frontiers} (Sec.~\ref{sec:frontiers})};
\end{tikzpicture}}
\caption{The survey is organized as a three-regime stack. \substance{} asks what
persistent state \emph{is} and where it physically lives; \motion{} asks how state
moves through the agent's operating loop; \control{} asks how that movement is
governed, shared, and judged. In our coding, work is concentrated more heavily on
accumulating and retrieving state than on governing, recovering, or relinquishing it.}
\label{fig:overview}
\end{figure}}

\newcommand{\figheatmap}{%
\begin{figure}[t]
\centering
\resizebox{\linewidth}{!}{%
\begin{tikzpicture}[font=\scriptsize,x=0.86cm,y=0.62cm]
\def\stages{observe,write,validate,organize,retrieve,act,update,forget,audit,rollback}
\foreach \lab/\j in {observe/0,write/1,validate/2,organize/3,retrieve/4,act/5,update/6,forget/7,audit/8,rollback/9}{
  \node[rotate=40,anchor=west,font=\scriptsize] at (\j+0.5,9.7) {\lab};}
\draw[thick,gray] (6,-0.2) -- (6,9.4);
\node[font=\scriptsize\itshape,gray] at (8,10.6) {return arc};
\node[font=\scriptsize\itshape,gray] at (3,10.6) {forward arc};
\foreach \p/\i/\name in {P1/8/{P1 Found.},P2/7/{P2 Substr.},P3/6/{P3 Lifecyc.},P4/5/{P4 Adapt.},P5/4/{P5 Govern.},P6/3/{P6 Multi-ag.},P7/2/{P7 Eval.},P8/1/{P8 Failure},P9/0/{P9 Domains}}{
  \node[anchor=east,font=\scriptsize] at (-0.1,\i+0.5) {\name};}
\foreach \i/\row in {%
  8/{7,7,0,6,16,10,9,4,1,0},
  7/{11,43,23,45,77,34,32,12,15,12},
  6/{7,20,5,13,25,5,11,3,2,0},
  5/{10,32,13,25,25,14,29,18,5,3},
  4/{4,23,17,8,22,20,14,16,33,9},
  3/{5,18,3,6,11,7,6,5,8,0},
  2/{1,7,10,4,16,2,6,2,10,1},
  1/{3,14,8,1,14,14,3,2,12,0},
  0/{20,36,8,20,63,35,17,4,2,2}}{
  \foreach \v [count=\j from 0] in \row {
    \pgfmathsetmacro\shade{min(100,\v/77*100)}
    \fill[blue!\shade!white,draw=gray!40] (\j,\i) rectangle (\j+1,\i+1);
    \pgfmathsetmacro\txtcol{\v>40?1:0}
    \node[font=\tiny,text=\ifdim\shade pt>45pt white\else black\fi] at (\j+0.5,\i+0.5) {\v};}}
\end{tikzpicture}}
\caption{Coverage heatmap: for each taxonomy part (rows), the number of coded works
whose lifecycle coding touches each stage (columns). Cells are shaded by count.
The forward arc (left of the divider) is dark across nearly every part; the return
arc, especially \texttt{forget}, \texttt{audit}, and \texttt{rollback}, stays pale
even in the governance part (P5). This is the survey's main asymmetry made
visual. Counts are per-part stage incidences and exceed part sizes because coding is
multi-label.}
\label{fig:heatmap}
\end{figure}}

\newcommand{\figlifecycle}{%
\begin{figure}[t]
\centering
\resizebox{\linewidth}{!}{%
\begin{tikzpicture}[font=\footnotesize,
  st/.style={draw,rounded corners=2pt,minimum height=0.7cm,minimum width=1.35cm,align=center,inner sep=2pt},
  fwd/.style={st,fill=orange!12}, ret/.style={st,fill=red!10},
  ar/.style={-{Stealth[length=2mm]},thick}]
\node[fwd] (obs) at (0,2) {observe};
\node[fwd] (wr) at (1.75,2) {write};
\node[fwd] (val) at (3.5,2) {validate};
\node[fwd] (org) at (5.25,2) {organize};
\node[fwd] (ret) at (7.0,2) {retrieve};
\node[fwd] (act) at (8.75,2) {act};
\draw[ar] (obs)--(wr); \draw[ar] (wr)--(val); \draw[ar] (val)--(org); \draw[ar] (org)--(ret); \draw[ar] (ret)--(act);
\node[ret] (upd) at (8.75,0) {update};
\node[ret] (frg) at (7.0,0) {forget};
\node[ret] (aud) at (5.25,0) {audit};
\node[ret] (rb) at (3.5,0) {rollback};
\draw[ar] (act)--(upd); \draw[ar] (upd)--(frg); \draw[ar] (frg)--(aud); \draw[ar] (aud)--(rb);
\draw[ar,dashed] (rb) to[out=180,in=180] (obs);
\node[font=\scriptsize\itshape,orange!60!black] at (4.4,2.7) {forward arc: accumulate \& use (well-served)};
\node[font=\scriptsize\itshape,red!60!black] at (6.1,-0.7) {return arc: govern \& recover (sparse: rollback 27/435)};
\node[draw,dashed,rounded corners,fill=gray!6,minimum width=10.5cm,align=center,font=\scriptsize] at (4.4,-1.7)
  {\textbf{Five invariants} the loop should preserve: authority monotonicity~$\cdot$~scope non-expansion~$\cdot$~deletion propagation~$\cdot$~provenance preservation~$\cdot$~rollback traceability};
\end{tikzpicture}}
\caption{The persistent-state lifecycle. State flows along a forward arc that
accumulates and uses it, then a return arc that updates, forgets, audits, and rolls
it back once outcomes are known. The forward arc is heavily studied; the return arc
is sparse (only $27$ of $435$ corpus works expose any rollback mechanism). Five
invariants should hold across the whole loop; each failure mode in
Section~\ref{sec:failures} violates one of them.}
\label{fig:lifecycle}
\end{figure}}

\newcommand{\figaoep}{%
\begin{figure}[t]
\centering
\resizebox{\linewidth}{!}{%
\begin{tikzpicture}[font=\footnotesize,
  bx/.style={draw,rounded corners=2pt,align=center,minimum height=0.85cm,inner sep=3pt},
  ar/.style={-{Stealth[length=2mm]},thick}]
\node[bx,fill=blue!8] (ep) at (0,0) {Episode:\\event stream\\(typed ops)};
\node[bx,fill=blue!8] (san) at (2.6,0) {Sanitize:\\strip oracle\\fields};
\node[bx,fill=orange!10] (sys) at (5.3,0) {System under\\test maintains\\its own state};
\node[bx,fill=orange!10] (prb) at (8.0,0) {Neutral\\probes\\(IDs only)};
\node[bx,fill=red!8] (val) at (10.7,0) {Validator:\\compute invariants,\\score vs oracle};
\draw[ar] (ep)--(san); \draw[ar] (san)--(sys); \draw[ar] (sys)--(prb); \draw[ar] (prb)--(val);
\node[bx,fill=red!8,minimum width=10cm,align=center] (out) at (5.3,-1.6)
  {\textbf{Deterministic scorecard:} obligation pass (must actively satisfy) vs negative-invariant pass (no-leakage, satisfiable by storing nothing)};
\draw[ar] (val) to[out=270,in=0] (out.east);
\end{tikzpicture}}
\caption{The Always-On Evaluation Protocol (AOEP). Each episode is a typed event
stream; outcome-revealing fields are stripped before a system under test consumes it;
the system answers neutral probes; a validator recomputes the invariants itself and
scores against an oracle. Obligation pass (positive duties a system must actively
meet) is reported separately from negative-invariant pass (no-leakage checks, which a
system that stores nothing satisfies vacuously), so recall and governance are scored
on different axes.}
\label{fig:aoep}
\end{figure}}

\newcommand{\figinvariants}{%
\begin{figure}[t]
\centering
\resizebox{0.92\linewidth}{!}{%
\begin{tikzpicture}[font=\scriptsize,x=2.05cm,y=0.62cm]
\foreach \lab/\j in {{Authority\\monotonicity}/0,{Scope\\non-expansion}/1,{Deletion\\propagation}/2,{Provenance\\preservation}/3,{Rollback\\traceability}/4}{
  \node[align=center,font=\scriptsize] at (\j+0.5,9.85) {\lab};}
\foreach \p/\i in {{P1 Found.}/8,{P2 Substr.}/7,{P3 Lifecyc.}/6,{P4 Adapt.}/5,{P5 Govern.}/4,{P6 Multi-ag.}/3,{P7 Eval.}/2,{P8 Failure}/1,{P9 Domains}/0}{
  \node[anchor=east,font=\scriptsize] at (-0.05,\i+0.5) {\p};}
\foreach \i/\row in {%
  8/{10,18,4,12,10},
  7/{51,92,12,60,42},
  6/{8,28,3,21,5},
  5/{26,41,18,44,16},
  4/{28,25,16,19,27},
  3/{10,13,5,9,7},
  2/{12,18,2,9,3},
  1/{18,15,2,3,14},
  0/{42,66,4,32,35}}{
  \foreach \v [count=\j from 0] in \row {
    \pgfmathsetmacro\shade{min(100,\v/92*100)}
    \fill[teal!\shade!white,draw=gray!40] (\j,\i) rectangle (\j+1,\i+1);
    \node[font=\tiny,text=\ifdim\shade pt>45pt white\else black\fi] at (\j+0.5,\i+0.5) {\v};}}
\draw[thick,red!60] (2,-0.15) rectangle (3,9.4);
\draw[thick,red!60] (4,-0.15) rectangle (5,9.4);
\node[font=\scriptsize\itshape,red!60!black,align=center] at (3.5,-0.95) {deletion propagation and rollback traceability: sparse in every part};
\end{tikzpicture}}
\caption{Stage-proxy opportunity for the five invariants, by taxonomy part. A work
counts toward an invariant if its lifecycle coding touches the stage that would
establish or check that
invariant (authority monotonicity at validate/act; scope non-expansion at
organize/retrieve; deletion propagation at forget; provenance preservation at
organize/update; rollback traceability at act/rollback), shaded by count. The
two boxed columns, deletion propagation and rollback traceability, are pale in
every part including the governance part (P5): the two invariants that most
distinguish a governed persistent-state system have few stage opportunities in
the coded corpus. This figure should not be read as direct evidence that a paper
tested or enforced an invariant; it is a proxy view of where the relevant
lifecycle stages appear.}
\label{fig:invariants}
\end{figure}}

\newcommand{\figtimeline}{%
\begin{figure}[t]
\centering
\begin{tikzpicture}[font=\small,x=2.7cm,y=0.075cm]
\draw[->,gray] (-0.15,0) -- (3.25,0);
\draw[->,gray] (-0.15,0) -- (-0.15,82);
\foreach \yv in {0,20,40,60,80}{ \node[anchor=east,font=\scriptsize,gray] at (-0.2,\yv) {\yv}; \draw[gray!25] (-0.15,\yv)--(3.15,\yv);}
\node[rotate=90,anchor=south,font=\scriptsize] at (-0.62,40) {\% of that year's works};
\foreach \x/\lab/\n in {0/2023/25,1/2024/63,2/2025/81,3/2026/252}{ \node[anchor=north,font=\scriptsize] at (\x,-1) {\lab}; \node[anchor=north,font=\tiny,gray] at (\x,-5) {n=\n};}
\draw[blue!70,very thick] (0,72)--(1,71)--(2,75)--(3,76);
\foreach \x/\y in {0/72,1/71,2/75,3/76}{\fill[blue!70](\x,\y)circle(1.6pt);}
\node[blue!70!black,font=\scriptsize,anchor=south] at (3,76) {early (retrieve/write)};
\draw[orange!85!black,very thick] (0,12)--(1,10)--(2,23)--(3,46);
\foreach \x/\y in {0/12,1/10,2/23,3/46}{\fill[orange!85!black](\x,\y)circle(1.6pt);}
\node[orange!75!black,font=\scriptsize,anchor=south east] at (3,46) {governance (forget/audit/rollback)};
\draw[red!75,very thick,dashed] (0,0)--(1,0)--(2,3.7)--(3,9.5);
\foreach \x/\y in {2/3.7,3/9.5}{\fill[red!75](\x,\y)circle(1.6pt);}
\node[red!70!black,font=\scriptsize,anchor=north east] at (3,9.5) {rollback only};
\end{tikzpicture}
\caption{Corpus coverage over time, as the share of each year's coded works whose
lifecycle touches an early stage (retrieve or write), any governance stage (forget,
audit, or rollback), or rollback specifically. Early-stage coverage is flat and high
across the period; governance coverage rises from $12\%$ to $46\%$ but remains a
minority, and rollback rises only from $0\%$ to under $10\%$. In the works we
coded, the governance turn is recent (it begins in earnest only in 2025) and still
thin. Per-year corpus sizes are shown on the axis and sum to $421$ because the
fourteen pre-2023 foundational anchors are excluded from this year-by-year view.
The count for 2026 is necessarily incomplete and preprint-heavy, as discussed in
Section~\ref{sec:frontiers}.}
\label{fig:timeline}
\end{figure}}

\section{Introduction}\label{sec:intro}

Large language model (LLM) agents are usually built, deployed, and evaluated as episodic systems. An agent receives a task, reasons over it, calls tools, returns an answer or performs an action, and then resets. Whatever it learned during the episode, which observations it gathered, which tool outputs it trusted, which sub-goals it abandoned, is discarded at the boundary. Under this contract the agent is stateless between tasks by construction, and most of the safety and correctness reasoning a designer must do collapses to a single episode: if the current prompt is clean and the current tool calls are authorized, the agent is well behaved. The episodic contract is convenient precisely because it makes the past irrelevant.

Always-on agents break this contract. An always-on agent is one whose behavior at a given moment depends on state accumulated before that moment, beyond the current prompt or task instance. A coding assistant that remembers a repository's conventions across weeks, a personal assistant that tracks a user's evolving preferences and open commitments, a customer-support agent that carries a case across many sessions and operators, and a fleet of collaborating agents that share a common workspace are all always-on in this sense, regardless of whether they literally run every second. The defining property is not continuous execution but persistence: the agent retains user preferences, task traces, tool outcomes, procedural lessons, policy constraints, social context, and past mistakes, and any of these can shape, or authorize, its next action. This is a qualitative shift, not a quantitative one. Once retained state can license future behavior, the past is no longer irrelevant, and the single-episode safety argument no longer holds.

\paragraph{State becomes authoritative.} Persistence changes what stored information can do. A stored preference can silently fix the default arguments of a tool call. A cached credential can grant access long after the user intended it to lapse. A procedural skill distilled from one successful trajectory can be replayed as an executable commitment in a situation it no longer fits. A summary written months ago can override fresh evidence because it retrieves more strongly. In each case the agent is not reasoning from the current situation alone; it is being guided by a record whose authority, scope, freshness, and origin were fixed in the past and are rarely re-examined. An always-on agent is therefore more than a memory-augmented episodic agent. It is a \emph{persistent-state system} whose retained state needs rules for what may persist, what may act, and what may be revoked.

We use \emph{persistent state} as a broad term. It includes retrievable memory, the object most prior work studies, but also the durable, control-relevant records that surround memory and cross the episode boundary with the agent: task ledgers of open commitments and obligations, permissions and consent, tool and credential state, provenance and audit trails, shared and social state across users and agents, and trigger state governing when to act proactively. External commitments are not internal memory records; they are part of the agent's persistent accountability surface because they must remain traceable to the internal state that authorized them. A calendar reminder, a partially completed purchase, a revoked API token, a poisoned workflow, and a deletion request that has only partly propagated are all persistent-state concerns, even when none is naturally described as a ``memory.'' This wider unit separates the survey from memory-form taxonomies.

\figevolution
\FloatBarrier

\paragraph{Persistence is useful, and persistence is risky.} The case for persistence is strong. Retaining state lets an agent avoid repeating work, personalize its behavior to a particular user or codebase, and adapt over time without retraining or weight updates, simply by accumulating and reusing experience. The same property that makes persistence useful, however, makes it dangerous in ways an episodic agent never faces. Irrelevant or salient-but-wrong memory can distract retrieval and crowd out current evidence. Stale memory can override fresh observation because nothing marks it as expired. Private memory can leak across users or sessions when scope is not enforced. Poisoned memory, written during one ordinary interaction, can persist and activate much later, after the attacker's session has ended and the triggering context is gone. Lossy consolidation can delete the precise identifier, date, or handle that a later action needs, while leaving a summary that still scores well on aggregate recall. None of these failures is repaired by filtering the current prompt, because the harm already resides in durable state. Persistence converts transient, single-episode problems into long-lived, system-level ones, a failure class episodic evaluation is not built to see.

\paragraph{The unit of analysis: persistent state, not memory.} The literature on LLM agents is large and growing, and much of it touches the questions above. Broad agent surveys map perception, reasoning, planning, tool use, and multi-agent organization \citep{xi2023riseandpotential,wang2023surveylargelanguagemodelagents}. Within that space, memory surveys explain how agents remember: the forms memory takes, the operations of writing, organizing, retrieving, and forgetting, and the benchmarks that probe recall \citep{zhang2024memorymechanism,du2026memoryautonomous,hu2025memoryageaiagents}. A more recent line argues that current memory systems are retrofitted from retrieval and database stacks rather than designed for agents, and calls for agent-native memory and externalized agent infrastructure \citep{zhou2026agentnative,zhou2026externalizationllmagentsunified}. We use much of this literature but change the unit of analysis. The first question is not what form an agent's memory takes, or which retrieval mechanism works best. It is how retained state \emph{authorizes} future behavior, how it propagates across sessions, users, and agents, and how it can be \emph{repaired} when it becomes stale, poisoned, overscoped, or wrong. Memory is one component of that object; governance of the whole remains under-specified.

This reframing follows a lineage that predates LLMs. Cognitive science long distinguished episodic from semantic memory \citep{tulving1972episodic}, and cognitive architectures such as Soar and ACT-R made the further distinction between declarative facts retrieved by activation and decay and procedural rules selected by utility, embedding all of them in an explicit decision cycle \citep{laird2012soar,anderson2004actr}. These architectures already treated memory not as a passive store but as a governed component of an acting system, with separation between memory types and explicit operations over them. Modern LLM agents inherit the vocabulary, with CoALA mapping cognitive-architecture structure onto language agents by organizing memory modules and an action space \citep{sumers2024cognitivearchitectureslanguageagents}. What they do not yet inherit is the runtime enforcement of the boundaries those architectures assumed: in an LLM agent, nothing prevents a procedural lesson from silently overwriting a semantic fact, an expired permission from licensing a tool call, or a deleted record from surviving in a derived tier. The gap between a memory \emph{type} and a governed \emph{state record} is the gap this survey is about.

\paragraph{Contributions.} This survey makes four contributions, which also organize the argument:

\begin{enumerate}
\item \textbf{Definition.} We define always-on agents as persistent-state systems and characterize a persistent-state item along six diagnostic axes that can be probed separately in simple cases but often interact in deployed systems: \emph{authority} (what permits this state to influence an action), \emph{scope} (which user, task, tool, time, or group may use it), \emph{mutability} (whether it can be revised, superseded, decayed, or locked), \emph{provenance} (which source, timestamp, and transformation produced it), \emph{recoverability} (whether derived state and affected decisions can be rolled back), and \emph{actionability} (whether it is evidence, preference, policy, skill, or executable commitment). A memory that scores perfectly on recall can still be unsafe if it has no authority boundary, no provenance, and no path to repair after a bad action.
\item \textbf{Lifecycle.} We recast memory as one stage of a broader governed loop, observe, write, validate, organize, retrieve, act, update, forget, audit, and rollback, and name five invariants the loop should preserve but that an episodic agent satisfies vacuously by resetting: authority monotonicity, scope non-expansion, deletion propagation, provenance preservation, and rollback traceability. Each invariant is the property whose violation defines a persistence-specific failure class.
\item \textbf{Coverage map.} We place memory, RAG, long-context, embodied-agent, and social-agent mechanisms and benchmarks against the always-on requirements they report testing, backed by a $435$-work coded corpus. In our coding, the largest qualitative skew is that work concentrates on accumulating and retrieving state more than on governing it.
\item \textbf{Protocol and pilot.} We propose an Always-On Evaluation Protocol that scores state mutation and recovery rather than answer quality alone, and instantiate it in AOEP-v0: an event-stream and snapshot schema, worked episodes, a validator design, and a pilot that illustrates which governance fields current memory wrappers do not expose.
\end{enumerate}

\figoverview

\paragraph{Reader's guide.} The survey is organized as a layered stack over persistent state, with three cross-cutting lenses, sketched in Figure~\ref{fig:overview}. After this introduction, we fix the definition and the six axes, then use the lifecycle and its five invariants as the main frame for the paper. The stack then moves through where state physically lives, from model parameters and the context window through retrieval stores, knowledge graphs, multimodal stores, and operating-system-style serving runtimes (substrates); how state moves through the forward arc as typed transitions (the state lifecycle); how state changes over a long horizon, where the classic plasticity-stability tension reappears in externalized form (continual adaptation); how state is governed along the return arc of update, forget, audit, and rollback (governance); and how state is shared and propagates across agents (multi-agent and shared state). The three cross-cutting parts read the stack through evaluation and benchmarks, failure modes and security, and application domains, before a research agenda and conclusion. Readers interested mainly in the empirical gap can read this introduction and the governance and evaluation parts; readers interested in mechanisms can enter at the substrate and lifecycle parts. Table~\ref{tab:roadmap} maps each section to the regime it serves, the question it answers, and whether it lies on a suggested minimal reading path for a time-constrained reader. The remainder of this introduction makes the corpus and its construction explicit, then positions the survey against its closest neighbors.

\begin{table}[t]
\centering
\small
\setlength{\tabcolsep}{5pt}\renewcommand{\arraystretch}{1.15}
\begin{tabularx}{\linewidth}{p{0.045\linewidth}p{0.20\linewidth}p{0.115\linewidth}Yc}
\toprule
\S & Section & Regime & Question it answers & Core path \\
\midrule
2 & Defining always-on agents & \substance & What is an always-on agent, and what are the six state axes? & $\bullet$ \\
3 & Persistent-state taxonomy & \substance & What kinds of state does the agent hold? & $\bullet$ \\
4 & The state lifecycle & \motion & By what operations does state move, and what invariants govern it? & $\bullet$ \\
5 & State substrates & \substance & Where does state physically live? & \\
6 & Mechanism families & \motion & How do deployed systems write, adapt, and share state? & \\
7 & Evaluation and benchmarks & \control & What does the field measure, and what does it not? & $\bullet$ \\
8 & Failure modes and governance & \control & How does persistence fail, and what governs it? & $\bullet$ \\
9 & The AOEP protocol & \control & How can governance be scored rather than only described? & $\bullet$ \\
10 & Resources and domains & \control & What does each application domain demand of persistence? & \\
11 & Research frontiers & \control & What are the open problems and how might they be solved? & $\bullet$ \\
12 & Conclusion & n/a & What follows for the field? & $\bullet$ \\
\bottomrule
\end{tabularx}
\caption{Reading roadmap. Each section is placed in one of the three regimes the survey is organized around (\substance: what state is and where it lives; \motion: how it moves; \control: how it is governed, judged, and stress-tested). The final column marks a suggested minimal path for a time-constrained reader who wants the empirical argument and the protocol without the full mechanism and substrate catalogues.}
\label{tab:roadmap}
\end{table}

\subsection{Survey scope and corpus}\label{sec:intro:scope}

\paragraph{What counts as in scope.} We study persistent-state mechanisms, the benchmarks and evaluation methods that probe them, the failure modes persistence induces, the foundational work that anchors our taxonomy, and the boundary cases that sharpen the definition. We include a work if it builds or studies a persistent-state mechanism, contributes a relevant benchmark or evaluation method, documents a persistence-induced failure mode, anchors the state taxonomy or the plasticity-stability framing, or defines a boundary case. We exclude pure model-architecture or pretraining work with no persistent-state operation, agent applications that add no persistent-state operation over an off-the-shelf memory, and superseded versions of works we already include. Long-context models, RAG systems, and context-engineered agents sit deliberately near the boundary and are admitted as constraints rather than as the object itself: long-context benchmarks ask whether a model can use what is already in its window, RAG benchmarks ask whether external evidence is selected and used faithfully, and context-engineering work systematizes how retrieval, memory, tools, and multi-agent signals are assembled for one inference \citep{mei2025surveycontextengineering}. Always-on agents add a different question these lines do not pose: which of the assembled signals should become agent-owned longitudinal state, when should that state become authoritative, and how can it later be revised, scoped, revoked, or removed. Personalization is treated the same way, as a state \emph{type} rather than the definition; we return to it below.

\paragraph{Search protocol.} We assembled the corpus by a recorded protocol over arXiv (cs.AI, cs.CL, cs.LG, cs.CR), Semantic Scholar, OpenReview, and the ACL Anthology, covering 2023 to 2026, with earlier foundational anchors admitted when they ground a taxonomy or a term we use. The pre-2023 anchors are deliberately upstream of LLM agents and concentrate in the cognitive and neuroscience foundations: the episodic-semantic distinction \citep{tulving1972episodic}, multi-component working memory with an episodic buffer \citep{baddeley2000episodic}, complementary learning systems and their update for artificial agents \citep{mcclelland1995complementary,kumaran2016complementary}, the neural evidence for offline replay during sleep \citep{wilson1994reactivation}, quantitative forgetting curves \citep{murre2015replication}, and the Soar and ACT-R architectures \citep{laird2012soar,anderson2004actr}. We seeded the modern corpus from agent-memory surveys and a closest-work set, chased citations two hops in each direction, and then ran targeted term-set sweeps. Most sweeps aimed away from the well-covered center of agent memory and toward the periphery we expected to be thin: authority, rollback, formal verification, durable execution, deletion propagation, cross-agent interoperability, and the database, distributed-systems, reinforcement-learning, and HCI literatures that publish persistent-state work outside agent-memory vocabulary. This matters for interpreting the skew below, because the sweeps were designed to \emph{over}-sample governance, not to under-sample it.

\paragraph{Admission flow.} Table~\ref{tab:corpus-flow} makes the corpus build less of a black box. It records the seed set, every admitted non-duplicate addition, and the round structure used to fold new works into the coded corpus. The flow is admission-based rather than a full PRISMA exclusion diagram: later rounds record rejected and duplicate candidates, but the earliest rejected-candidate identifiers were not retained in a uniform screened/excluded table. We therefore use the flow to support directional coverage claims and stopping conditions, not prevalence estimates over all papers that could have been screened.

\begin{table}[t]
\centering
\small
\setlength{\tabcolsep}{4pt}\renewcommand{\arraystretch}{1.16}
\begin{tabularx}{\linewidth}{p{0.18\linewidth}p{0.31\linewidth}p{0.11\linewidth}Y}
\toprule
Corpus step & Admission rule & Added & What it tested \\
\midrule
Seed and closest-work set & Recent memory and agent surveys plus hand-identified closest neighbors satisfying I1 to I5 & 97 & Establishes the vocabulary and the first boundary around persistent-state work \\
Rounds 1 to 2 & Non-duplicate mainstream agent-memory and benchmark additions & 147 & Expands the center of the memory and evaluation literature \\
Rounds 3 to 10 & Targeted governance, rollback, authority, deletion, durable-execution, HCI, and systems sweeps & 171 & Deliberately over-samples the return arc that the survey expects to be sparse \\
Round 11 & Saturation check over mainstream agent-memory vocabulary & 2 & Tests whether the memory frontier still yields many uncoded works \\
Rounds 12 to 15 & Adjacent-disciplinary and confirmation sweeps & 18 & Probes work hidden under database, distributed-systems, formal, and organizational vocabulary \\
\midrule
\textbf{Total} & \textbf{Seed plus fifteen recorded non-duplicate rounds} & \textbf{435} & \textbf{Coded by category, lifecycle stage, state axis, and subarea} \\
\bottomrule
\end{tabularx}
\caption{Admission flow for the coded corpus. Counts are admitted works, not candidate-screening totals. The flow explains how the corpus reached $N{=}435$ and why the governance skew is read as a large pattern inside our query frame rather than as a field-wide census.}
\label{tab:corpus-flow}
\end{table}

\paragraph{A category $\times$ lifecycle-stage $\times$ state-axis $\times$ subarea coding scheme.} Every included work is coded along four dimensions. First, a \emph{category}: mechanism, benchmark, failure-mode, survey, foundation, or boundary. Second, a multi-label set of \emph{lifecycle stages} it primarily exercises, drawn from the ten stages observe, write, validate, organize, retrieve, act, update, forget, audit, and rollback. Third, a multi-label set of \emph{state axes} it touches, drawn from the six axes authority, scope, mutability, provenance, recoverability, and actionability. Fourth, a primary application or method \emph{subarea}, of which thirty-three recur (the largest being conversational memory, memory mechanism, security and poisoning, evaluation methodology, multi-agent social memory, RAG and long-context, multimodal memory, state governance, agentic-OS runtime, continual learning, and skill learning). The cross-product of these dimensions is what lets the corpus answer questions a flat reading list cannot, for example which lifecycle stages the benchmark literature actually tests, or which state axes the mechanism literature leaves implicit. Appendix~\ref{app:methods} gives the full protocol: exact source list and time window, the query term sets, the inclusion and exclusion criteria, the coding scheme, and a measured inter-coder reliability check (pooled agreement $0.82$ on lifecycle stages and $0.74$ on state axes over a blind $236$-work sample), together with the threats to validity that bound how the headline counts should be read.

\paragraph{The corpus and its skew.} The resulting corpus of $435$ works is summarized in Table~\ref{tab:intro-corpus}, with its primary mapping into the survey's nine parts. Every work has a primary part, no part is empty, and the lightest part still hosts more than twenty works. The notable feature of the corpus is not its size but its shape. In our coding, coverage concentrates on the early, accumulative end of the lifecycle: retrieve ($269$ of $435$) and write ($200$) dominate the stages, and mutability ($160$) and provenance ($153$) lead the axes. Coverage thins sharply on the governance end. Authority is the rarest axis at $72$ of $435$. Among lifecycle stages, audit ($88$), forget ($66$), and especially rollback ($27$) trail far behind retrieval. Only twenty-seven of $435$ works expose any rollback mechanism for state-affected decisions, and none in the corpus reports recovery success or cost after corruption. These counts are scoping estimates, not a census, but they point to a large qualitative asymmetry: the works we coded have many techniques for accumulating and recalling state and fewer for governing it. Figure~\ref{fig:timeline} shows the same skew over time: the early-stage share stays flat and high from 2023 to 2026 while the governance share, though rising, remains a minority and turns upward only from 2025.

\begin{table}[t]
\centering
\small
\begin{tabular}{clr}
\toprule
Part & Title & Works \\
\midrule
P1 & Foundations, Definitions, and State Types & 27 \\
P2 & State Substrates and Representations & 117 \\
P3 & State Lifecycle (Write/Organize/Retrieve/Act) & 33 \\
P4 & Continual Adaptation and Plasticity-Stability & 52 \\
P5 & Governance (Authority/Provenance/Deletion/Audit) & 53 \\
P6 & Multi-Agent and Shared State & 23 \\
P7 & Evaluation and Benchmarks & 25 \\
P8 & Failure Modes and Security & 25 \\
P9 & Application Domains & 80 \\
\midrule
\multicolumn{2}{l}{\textbf{Total}} & \textbf{435} \\
\bottomrule
\end{tabular}
\caption{The $435$-work coded corpus, mapped to the survey's nine parts (primary subarea mapping; multi-label cross-references add more). Each work is also coded by category, lifecycle stage, and state axis. The early-lifecycle, low-governance skew described in the text (retrieve $269$, write $200$ versus rollback $27$; authority the rarest axis at $72$) holds within our coding and should be read as a scoped estimate rather than a census of the field.}
\label{tab:intro-corpus}
\end{table}

\paragraph{Governance-targeted search within the query frame.} One possible objection is that we looked in the wrong places, and that the governance literature is larger than our corpus suggests. Our evidence cannot rule out that possibility, but it does constrain the simplest version of it within our recorded query frame. First, the bulk of our search rounds targeted governance, rollback, formal verification, durable execution, deletion propagation, and adjacent systems literatures; those rounds grew the corpus more than fourfold yet only lifted the governance fraction from roughly one-sixth to about one-third before late rounds showed diminishing returns under this query frame, with rollback still rare. Second, the skew worsens, not improves, when we restrict to work submitted through 2024: the governance fraction falls rather than rises, so the corpus pattern is not created solely by including immature 2026 preprints. Third, the skew appears across categories and subareas rather than in a single corner. We therefore read the corpus shape as evidence of a qualitative asymmetry in the works we coded, with the standard caveat that this is representative coding rather than an exhaustive census.

\paragraph{Reliability and diminishing returns.} Two checks bound the confidence we place in these counts. For \emph{reliability}, labels come from a single coding pass with a blind second-coder check on a sample; the multi-label stage and axis dimensions are the harder cases, since a system that writes, retrieves, and updates state legitimately receives all three stage labels, and we treat such overlap as signal rather than noise. For \emph{search coverage}, the late, governance-targeted rounds added many works but did not move the governance fraction or introduce new lifecycle stages or state axes, which suggests diminishing returns within the directions we sampled rather than a proof of exhaustive coverage. Personalization is the one subarea where the recent literature is still expanding rapidly, and we read its newest entries as evidence of emerging directions rather than settled consensus. We treat all figures here as scoping estimates whose directional conclusion is that the works we coded accumulate state more often than they govern, relinquish, or repair it.

\subsection{Related surveys and how this survey differs}\label{sec:intro:related}

The obvious risk is that this reads as a relabeled memory survey. The adjacent survey literature is already dense, so the paper must make the distinction concrete. This survey integrates lifecycle, security, policy, and state-trajectory threads under \emph{persistent state} as the unit of analysis, then attaches a checkable scoring contract to the obligations that follow from that unit.

\paragraph{Agent surveys.} The broadest neighbors survey LLM agents as a whole, mapping perception, reasoning, planning, tool use, and multi-agent organization \citep{xi2023riseandpotential,wang2023surveylargelanguagemodelagents}. These works are indispensable for situating agents. Their native goal is to explain agent architectures and capabilities, so memory is usually one functional module among many rather than the unit whose authority, provenance, and recoverability are tracked. Our survey zooms into the state that such a module, and the records around it, carry across episodes, and asks how that state is governed rather than how the surrounding architecture is organized.

\paragraph{Memory surveys.} The closest and most numerous neighbors survey memory mechanisms for LLM agents directly. They systematize memory \emph{forms} (parametric versus non-parametric, episodic versus semantic versus procedural, short- versus long-term) and memory \emph{operations} (encode, store, retrieve, forget), and they catalog the benchmarks that test recall \citep{zhang2024memorymechanism,du2026memoryautonomous,hu2025memoryageaiagents,huang2026rethinkingmemorymechanisms}. Their native contribution is to explain how agents remember and manage memory. Several already model a write, manage, and read loop \citep{du2026memoryautonomous}, give a broad descriptive taxonomy of forms, dynamics, and benchmarks \citep{hu2025memoryageaiagents}, or ask when memory is useful rather than simply present \citep{huang2026rethinkingmemorymechanisms}. Our survey builds on that base but changes the unit: permissions and credential state, task ledgers and commitments, audit trails, rollback handles, and externally committed effects linked to internal records are not memory forms in the usual sense, but they determine what retained state may govern later action. The delta is therefore an operation-level handle on the state record itself: modeling lifecycle operations as typed transitions over records with authority, scope, provenance, and recoverability links, then asking which obligations those transitions should preserve.

\paragraph{Externalization and agent-native position papers.} A third family argues, correctly in our view, that today's memory systems are often retrofitted from retrieval and database stacks and that agents need native abstractions and externalized infrastructure \citep{zhou2026agentnative,zhou2026externalizationllmagentsunified}. We share the diagnosis and treat it as motivation. Their native goal is architectural: to name the components an agent infrastructure should expose. Our contribution is complementary. We ask how revocation, deletion completeness, rollback, and authority currency become properties of state transitions rather than only desiderata attached to components, and then turn those transition obligations into an evaluation contract.

\paragraph{Long-term-memory security surveys.} A fourth family surveys the security of long-term agent memory, organizing threats and defenses across write, retrieve, forget, audit, and rollback phases \citep{lin2026surveylongtermmemorysecurity}. This is the neighbor closest to our governance and failure-mode parts, and it correctly centers the lifecycle phases where memory is attacked. Its native goal is security taxonomy and defense mapping. Our survey is broader in mechanism, evaluation, substrate, and domain coverage, and narrower in one methodological sense: we ask which persistent-state transitions would have to be logged or checked for a defense claim to become an executable obligation.

\paragraph{Personalization as an adjacent boundary, not the definition.} Personalization deserves separate treatment because it is the adjacent area most often confused with always-on agency, and because its recent literature is the most active. Benchmarks here evaluate user-profile-conditioned outputs and, increasingly, the parts closest to persistent state: following stated preferences across long multi-session context \citep{zhao2025prefeval}, tracking an evolving persona over interactions \citep{jiang2025personamem}, surveys that organize the foundations and evaluation of personalized LLM agents \citep{xu2026personalizedllmpoweredagentsfoundations,salemi-etal-2024-lamp}, and a fast-growing 2025 to 2026 wave that pushes on individual state properties: principled forgetting balanced against recall \citep{uddin2026recallforgettingbenchmarkinglongterm}, write-consolidate-retrieve coordination beyond plain retrieval \citep{li2026himes}, attributes that drift on different timelines \citep{xie2026dynamicmem}, recall of memories logically critical but semantically distant from the query \citep{ding2026imlogic}, reuse of persisted preferences to fill under-specified tool arguments across sessions \citep{yoon2026mpt}, category-bounded long-term memory for an in-car assistant \citep{kirmayr2025carmem}, retrieval of historical web behaviors to align a personalized web agent \citep{cai2025personalizedwebagents}, production-scale industrial memory over noisy longitudinal data \citep{xu2026linkedinmemory}, accumulation of personalized context to resolve implicitly-specified targets \citep{lee2026polar}, long-term personalized assistants that capture subjective user traits \citep{huang2025mempal}, and auditing long-term memory as a post-interaction record rather than only through downstream task success \citep{ma2026memprobe}. This wave is evidence that the field is converging on persistent state from the user-modeling side. Yet personalization remains a different object than ours. A system can top a preference-following benchmark \citep{zhao2025prefeval} while having no mechanism to revoke a preference the user later retracts, and high recall of a preference whose authority has lapsed is not a success but an authority-monotonicity failure these benchmarks do not score. One line makes the danger explicit by formalizing unintended long-term state poisoning, where routine interactions gradually weaken confirmation boundaries and silently expand the agent's action scope \citep{xu2026toxicchats}: an accumulation failure that only a governance frame can name. We therefore treat personalization, like trigger state and the other categories, as a persistent-state \emph{type} that our axes and lifecycle must cover, not as the definition of always-on agency.

\begin{table}[t]
\centering
\scriptsize
\setlength{\tabcolsep}{3pt}\renewcommand{\arraystretch}{1.18}
\begin{tabularx}{\linewidth}{p{0.20\linewidth}p{0.18\linewidth}p{0.19\linewidth}p{0.18\linewidth}Y}
\toprule
Closest work & Unit of analysis & Lifecycle or evaluation coverage & Governance coverage & What this survey adds \\
\midrule
Agent-memory surveys \citep{du2026memoryautonomous,huang2026rethinkingmemorymechanisms,hu2025memoryageaiagents} & Memory mechanisms and utility & Write, manage, retrieve; substrates and cognitive forms & Partial treatment of control, forgetting, and policy & Persistent state as the unit, including permissions, ledgers, credentials, side effects, and recovery obligations \\
Long-term-memory security \citep{lin2026surveylongtermmemorysecurity} & Attacks and defenses over memory phases & Security lifecycle around write, retrieve, forget, audit & Strong threat taxonomy; less complete substrate, evaluation, and domain map & Integrates security with mechanisms, benchmarks, state types, and a transition-level evaluation contract \\
Governed evolving memory \citep{orogat2026gem} & State-trajectory correctness & Formal operators and correctness conditions & Explicit rollback-traceability view & Positions AOEP as a prototype contract that instantiates related obligations in executable form \\
Agent-evaluation surveys \citep{yehudai2025surveyevaluationllmagents,mohammadi2025evalbenchmarkingllmagents} & Agent task and process evaluation & Planning, tool use, memory, self-reflection, enterprise concerns & Authority, deletion, and rollback mostly not scored as state obligations & Asks which persistent-state properties existing agent evaluations operationalize \\
Evaluation methodology \citep{kapoor2024aiagentsthatmatter} & Benchmark quality and reporting & Cost, overfitting, holdout, failure-rate critique & Methodological rather than state-specific & Uses these standards to constrain how AOEP and corpus claims should be reported \\
Tool-agent security \citep{debenedetti2024agentdojodynamicenvironmentevaluate} & Prompt injection in dynamic tool environments & Realistic tool/data attacks and defenses & Strong on untrusted data and tool boundaries; lighter on persistence & Identifies what changes when injected data becomes durable state across sessions \\
Contextual privacy \citep{mireshghallah2025cimemories,wen2026contextualizedprivacy} & Appropriate information flow from memory & Contextual integrity and context-aware privacy defense & Strong scope lens; limited rollback and lifecycle recovery & Folds contextual integrity into scope non-expansion across persistent state transitions \\
\bottomrule
\end{tabularx}
\caption{Closest-neighbor comparison. The novelty claim is integration and operationalization: this survey unifies related threads under persistent state, makes authority and recoverability first-class diagnostic axes, and asks which state-transition obligations can be checked.}
\label{tab:closest-surveys}
\end{table}

\paragraph{The common gap.} Across these neighboring families, the opportunity is integration rather than replacement. Memory surveys taxonomize forms and recall; agent surveys situate memory as a module; externalization papers inventory components; security surveys catalog attacks and defenses; privacy and evaluation work cover important slices of scope, safety, and measurement. What remains under-specified is the governed state record as a common unit: lifecycle operations as transitions over that record, and obligations that can be checked under restart, conflict, deletion, adversarial write, shared-scope change, and delayed consequence. This survey is therefore not another taxonomy of how agents \emph{remember}. It is an account of how agents \emph{persist}, and of why the works we coded accumulate and retrieve state more often than they govern, relinquish, or repair it. Each layer of the paper ends on the governance question its own literature leaves open.

\figtimeline
\section{Defining always-on agents}\label{sec:definition}

\regime{Substance}{what an always-on agent is, and the six axes that characterize a unit of its state.}
An always-on agent is best understood not as a memory-augmented language model but as a \emph{persistent-state system}: a system whose behavior at any moment depends on state it has accumulated, transformed, and retained across earlier interactions, and whose correctness depends as much on governing that state as on recalling it. A retrieval cache asks what the system can recall. A persistent-state system also asks whether a recalled item is still current, whether it is authorized to influence the action under consideration, whether it is scoped to the user and task at hand, whether it descends from a trustworthy source, whether the decisions it licenses can be reversed, and whether it is a passive fact or an executable commitment. The definition below separates always-on agents from five adjacent system classes routinely conflated with them, introduces the six state axes that form the survey's analytic vocabulary, traces their cognitive-architecture lineage, and states what the definition includes and excludes, in particular why personalization and trigger behavior are persistent-state \emph{types}.

\subsection{An operational definition}\label{subsec:operational}

We define an always-on agent as an agent whose policy at time $t$ depends on state accumulated before $t$, beyond the current prompt or task instance. The defining property is persistence, not literal continuous execution: an agent that runs once a day, wakes on an external trigger, or is invoked only when a user opens a session is still always-on if its action at invocation is conditioned on durable state carried across invocations. Conversely, an agent that runs continuously but resets all state between requests is episodic, not always-on, because nothing it retains can authorize a later action.

The definition is operational so that it can be checked against a system rather than asserted of it. A system is always-on if it satisfies three conditions. First, it has \emph{persistent identity} across sessions or restarts, so the same agent instance is recoverable after interruption and can be held accountable for what it did before. Second, it maintains \emph{agent-owned, mutable, durable state} about users, tasks, tools, environments, policies, or other agents, state that outlives the interaction in which it was created. Third, it can \emph{use that state for later action}, closing the loop from accumulation back to behavior; state written but never read is inert and does not make an agent always-on. If that state is consequential, it creates a fourth requirement, \emph{temporal accountability}: the system should be able to record what it knew, when it knew it, why it acted, and how state that later proves contaminated can be identified and repaired. The first three properties classify the system. The fourth is the obligation current systems often fail to satisfy, and that failure is the empirical focus of this survey.

Temporal accountability cannot be reduced to retrieval quality. A system can retrieve a fact with perfect recall and still be unable to say whether the fact remains authorized, whether its scope still covers the present request, or which past actions would have to be undone if the fact turns out to be poisoned. The classical agent surveys that mapped the design space of LLM agents organized it around perception, reasoning, memory, and action as functional modules \citep{xi2023riseandpotential,wang2023surveylargelanguagemodelagents}, a useful decomposition that nonetheless treats memory as one black-box module to be written and read, and does not ask what governs the transition from a retrieved item to an authorized influence on the next action. For an always-on agent, that transition is the object of study. An episodic agent can be evaluated on whether its final answer is correct; an always-on agent must be evaluated on whether each state transition was legitimate, because an illegitimate transition can produce a correct-looking answer today and an irreversible harm tomorrow. Later sections therefore evaluate state mutations and action consequences rather than final answers alone.

\paragraph{An objection: is the governance gap rational triage?} A careful skeptic can grant every corpus statistic in this survey and still resist its conclusion. Perhaps rollback is rare because most agent state is intentionally disposable: a wrong preference or a stale cached plan can be overwritten rather than reverted. Perhaps authority is the rarest axis because many deployed agents run in bounded single-user, high-trust contexts where fine-grained authorization would add cost without reducing risk. On this reading the sparse return arc is specialization, not a gap. The claim we make is conditional: \emph{whenever persistent state is both retained and consequential, the absence of governance is an undefended design choice rather than a rational one}. Three lines of evidence make the antecedent common rather than rare. First, production-scale systems accumulate long-horizon consequential state: credential and permission tokens, durable user preferences, learned policies, and task ledgers whose corruption persists across sessions (Sections~\ref{sec:statetax} and~\ref{sec:substrates}). Second, the failure modes traced to governance omissions are documented in attack and incident literature and arise even in the bounded settings the objection invokes (Section~\ref{sec:failures}). Third, the budget-matched and continual-learning results we survey show systems \emph{explicitly designed to learn} degrading below a no-memory baseline over long horizons (Section~\ref{sec:mechanisms}); that failure arose because the governance condition was present and ungoverned. Where state is genuinely ephemeral, this survey's prescriptions do not apply. The burden is to govern the state that is not.

\paragraph{A note on temporal heterogeneity.} The definition spans three execution patterns: continuously running agents, periodically scheduled ones, and session-triggered ones that run on demand but carry state across sessions. We group them because all three retain persistent state and inherit its governance obligations, yet their risk surfaces differ. A continuously running or scheduled agent faces \emph{prospective} governance, when it is appropriate to act on latent state at all, which is the trigger-calibration axis. A session-triggered agent faces \emph{retrospective} governance, whether state carried from a prior session still authorizes the action it is about to take. The corpus coding aggregates the two, so a count such as the twenty-seven works exposing rollback mixes rollback for periodic re-execution with rollback for session recovery, which are mechanically distinct. We flag the conflation so that later empirical frequencies are read with the right grain: the governance axes apply across all three patterns, but the prevalence of any single failure mode may be concentrated in one.

\subsection{Boundaries: five adjacent system classes}\label{subsec:boundaries}

The definition earns its weight by excluding things, so we sharpen it by contrasting always-on agents against five system classes with which they are habitually conflated, because long context, retrieval, context engineering, episodic tool use, and conversational memory all touch information that persists in some sense. The distinction in every case is whether the system governs what becomes durable, authoritative, agent-owned state, or only processes information that happens to be present.

\paragraph{Episodic LLM agents.} The baseline contrast is the episodic tool-using agent, the dominant evaluation paradigm. Frameworks in the ReAct lineage interleave reasoning and action over many steps, binding intermediate reasoning to tool calls and validating outcomes within a task \citep{yao2023reactsynergizingreasoningacting}. Such an agent can act with great sophistication over a long horizon, yet it resets between tasks: nothing it learned in one episode authorizes an action in the next, and the always-on agent breaks that reset. What separates the two is not the length or complexity of a single episode but whether identity, commitments, permissions, and lessons survive the episode boundary. An episodic agent satisfies the survey's governance invariants vacuously, by forgetting everything, which makes it safe in a narrow sense and useless for the longitudinal work always-on agents are built for. The gap is foundational: the episodic paradigm has no vocabulary for an action that is wrong only because of state inherited from a prior session, so its benchmarks cannot detect that failure class at all.

\paragraph{Memory-augmented agents.} The closest neighbor is the memory-augmented agent, which adds an external store, a vector index, a note graph, or an operating-system-style paged context, that the policy writes to and reads from across sessions. MemGPT, for instance, treats memory as an OS-managed hierarchy with explicit paging between an in-context working set and externalized long-term storage, decoupling the effective horizon from the context budget \citep{packer2024memgptllmsoperatingsystems}. This is genuine cross-session state, so on our definition such systems are a subclass of always-on agents; the distinction is internal rather than exclusionary. Memory-augmented agents typically instantiate accumulation and recall but not temporal accountability: they expose write, organize, and retrieve, and leave authority, deletion propagation, and rollback implicit, so that a deletion edits the primary store while summaries, embeddings, and promoted tiers retain the deleted content. The boundary is therefore not whether the system has memory but whether the memory is \emph{governed}; persistent state, as we use the term, is wider than retrievable memory and includes task ledgers, permissions, credentials, commitments, provenance, and externally committed effects that remain linked to the records that authorized them.

\paragraph{Long-context models.} A long-context model can ingest a very large input, an entire codebase, a year of chat history, but it does not decide what to preserve for future use. Long-context benchmarks measure whether a model can use what is already inside its window, and repeatedly show the answer is unreliable even when the information is present: models underuse mid-context information despite attending to its boundaries \citep{liu2023lostmiddlelanguagemodels}, and broad bilingual long-context suites confirm that task performance degrades with length in ways the architecture does not advertise \citep{bai2024longbenchbilingualmultitaskbenchmark}. These results matter for always-on agents as \emph{constraints}, since whatever an agent retrieves must still be usable once it enters the window, but they answer a different question. A long-context model is given its context; an always-on agent must decide which signals become durable state, when they become authoritative, and how they are later revised or revoked. The window is volatile working memory that vanishes at the end of the call; persistent state is what the agent chose to keep, and the keeping is the governed act long context never performs.

\paragraph{RAG systems.} Retrieval-augmented generation retrieves from an external corpus to ground generation, and RAG benchmarks supply the vocabulary of faithfulness, noise robustness, evidence selection, and rejection of irrelevant passages \citep{chen2023benchmarkinglargelanguagemodels}. A RAG system reads from a corpus but does not, in general, write back to it from experience: the corpus is a fixed, externally curated knowledge base, not state the agent accumulates and must later govern. The always-on agent inverts the asymmetry. It populates its store from its own observations and actions, so it inherits problems a static corpus does not have: self-poisoning through an untrusted write, staleness when a once-true fact changes, and the obligation to delete on request and propagate that deletion to every derived copy. RAG faithfulness asks whether retrieved evidence was used correctly; the always-on question is whether the evidence should have been in the store at all, whether its authority is current, and whether it can be removed. RAG mechanics are thus a substrate the always-on agent often uses, but RAG evaluation does not test the write-side governance obligations that define the always-on setting.

\paragraph{Personalized and proactive assistants.} The final and most subtle boundary is the personalized or proactive assistant. A long line of work models users and conditions outputs on user profiles: the LaMP suite evaluates profile-conditioned generation \citep{salemi-etal-2024-lamp}, recent surveys organize the foundations and evaluation of personalized LLM agents \citep{xu2026personalizedllmpoweredagentsfoundations}, PrefEval tests whether a model infers and follows stated preferences across long multi-session context \citep{zhao2025prefeval}, and PersonaMem tracks an evolving user persona over interactions \citep{jiang2025personamem}. Proactivity, deciding when to surface help or act unprompted, is a parallel capability evaluated by datasets such as proactive-agent and proactive-recall benchmarks \citep{lu2024proactiveagent,he2024madialbench}. These systems unambiguously build persistent state, and many are always-on agents. But personalization and proactivity are \emph{capabilities expressed over} persistent state, not its definition. A model can top a preference-following benchmark yet have no mechanism to revoke a preference the user later retracts; high recall of a preference whose authority has lapsed is not a success but an authority-monotonicity failure the benchmark has no field to score. We therefore treat personalization and trigger behavior as state types and use cases, returning to this in Section~\ref{subsec:include-exclude}.

Table~\ref{tab:boundary} consolidates these contrasts along the separating question: does the system govern durable, authoritative, agent-owned state?

\begin{table}[t]
\centering
\small
\caption{Always-on agents against five adjacent system classes. The separating question is not whether information persists but whether the system governs durable, agent-owned state: what becomes authoritative, and how it can be revised, scoped, revoked, and rolled back.}
\label{tab:boundary}
\setlength{\tabcolsep}{4pt}\renewcommand{\arraystretch}{1.2}
\begin{tabularx}{\linewidth}{p{0.18\linewidth}YY}
\toprule
System class & Question it answers & Governs durable state? \\
\midrule
Episodic LLM agent & Can it act correctly within one task before reset? & No: resets between tasks \\
Long-context model & Can it use what is already in the window? & No: does not choose what to keep \\
RAG system & Is external evidence selected and used faithfully? & No: reads a fixed corpus, rarely writes back \\
Memory-augmented agent & Can it recall cross-session facts? & Partly: writes and reads, but governance left implicit \\
Personalized / proactive assistant & Does output match the user and arrive at the right time? & Partly: builds state but rarely scopes, revokes, or rolls back \\
\midrule
Always-on agent & Which retained state is authoritative for the next action, and how is it revised, scoped, revoked, and repaired? & Yes, by definition \\
\bottomrule
\end{tabularx}
\end{table}

The common thread is that each adjacent class makes one part of the always-on problem easy by assuming the rest away. Episodic agents assume away persistence; long-context models assume the context is given; RAG assumes the corpus is curated and read-only; memory-augmented systems assume recall is the whole job; personalized assistants assume that matching the user is the whole job. The always-on agent can assume none of these, and the open problem the boundary analysis exposes is that we lack a single benchmark crossing all of them at once, stressing persistent identity, governed writes, scope, revocation, and rollback in one protocol rather than testing each adjacent capability in isolation.

\begin{table}[t]
\centering
\scriptsize
\setlength{\tabcolsep}{3pt}\renewcommand{\arraystretch}{1.05}
\begin{tabularx}{\linewidth}{p{0.30\linewidth}cY}
\toprule
System example & In scope? & Boundary reason and resulting obligation \\
\midrule
Long-context QA over supplied history & No & Context is supplied for one inference; ordinary context privacy and evidence use apply. \\
Read-only RAG over a curated corpus & No & The agent reads an external corpus but does not own or mutate it; corpus provenance and retrieval faithfulness matter. \\
Episodic ReAct or AppWorld task & No & Identity and state reset at the task boundary; evaluation is task-local. \\
Saved conversational memory & Yes & User facts persist and shape later responses; revocation, scope, provenance, and deletion propagation become obligations. \\
MemGPT/Letta-style external memory & Yes & The agent writes and retrieves durable memory; authority, mutability, and recoverability must be represented. \\
Repository-memory coding agent & Yes & Project conventions and prior fixes affect later edits; provenance, conflict handling, and rollback become relevant. \\
Calendar or trigger agent & Yes & Stored conditions can cause later proactive action; current authority, temporal validity, and confirmation policy matter. \\
Shared workspace or credentialed tool agent & Yes & Multiple principals or durable grants affect later side effects; ownership, permission epochs, idempotency, and compensation are required. \\
\bottomrule
\end{tabularx}
\caption{Decision procedure for the always-on boundary. A system is in scope when it has persistent identity, owns mutable durable state, and can use that state for later action. Temporal accountability is not required to classify the system; it is the governance obligation created when the state is consequential.}
\label{tab:decision}
\end{table}

\subsection{The six state axes}\label{subsec:axes}

If persistent state is the unit of analysis, we need a vocabulary precise enough that a benchmark can perturb an individual state item and a mechanism can be said to govern it. We characterize a persistent state item along six diagnostic axes, each phrased as a question an evaluation can ask of the item and that a system either does or does not answer. The axes are not orthogonal variables. Provenance often supplies evidence for authority; scope constrains authority; actionability determines how strong the authority and recoverability requirements must be; and recoverability depends on provenance being preserved. The axes are still useful because each names a dimension along which a distinct, persistence-specific failure class arises, grounded in either the cognitive-science literature that first distinguished it or the agent literature that exposed its failure mode. Later sections measure how unevenly the field covers them, with authority the rarest at 72 of 435 coded works and mutability the most common at 160.

\paragraph{Authority.} Authority asks who or what permits this state to influence an action, distinguishing a fact the agent only holds from one it is licensed to act upon. The cognitive lineage supplies only a partial analogy: ACT-R separates declarative chunks, which are simply available, from procedural production rules selected by a utility that governs when they fire \citep{anderson2004actr}, prefiguring the distinction between possessing a record and acting on it. That utility is an action-selection preference, not an authorization, so the authority axis in our sense is grounded less in cognitive architecture than in capability-based security and access control, where the right to influence an action is an explicit, revocable grant rather than a learned preference (we develop this lineage in the governance part). In the agent setting, authority is the axis most directly attacked: work on long-term state poisoning shows routine interactions can gradually weaken an agent's confirmation boundaries and expand the scope of actions it will take, an authority drift no recall metric registers \citep{xu2026toxicchats}. The corresponding invariant, authority monotonicity, holds that a record may influence an action only under a current, unrevoked authority, so an expired permission cannot license a write or a tool call. The gap this axis exposes is that the bulk of the literature treats every stored item as equally entitled to act, with no representation of who authorized it or whether that authorization still holds; authority is the least represented axis in our coding.

\paragraph{Scope.} Scope asks which user, task, tool, time window, or group an item may be used for. Its cognitive grounding is the long-standing distinction between volatile working memory bound to the current situation and durable long-term memory, formalized in the multi-component working-memory model whose episodic buffer binds integrated representations for the task at hand \citep{baddeley2000episodic}. Scope keeps one user's preference from acting on behalf of another, and it is what a category-bounded memory enforces when an in-car assistant restricts each stored preference to its declared category and prunes redundant or contradictory entries \citep{kirmayr2025carmem}. The invariant is scope non-expansion: a transition may narrow or preserve scope but never silently widen it. The gap is that even systems enforcing scope tend to do so statically, fixing categories or roles at design time, with no account of what happens when a scope must be redefined, when authority changes dynamically, or when an item legitimately moves between scopes, leaving cross-scope leakage as a standing risk.

\paragraph{Mutability.} Mutability asks whether an item can be revised, superseded, decayed, or locked, and on what timescale. This axis is grounded in the quantitative study of forgetting, from the reproducible replication of the Ebbinghaus forgetting curve with fitted decay functions \citep{murre2015replication} to the complementary-learning-systems account in which fast and slow stores change at deliberately different rates \citep{mcclelland1995complementary}. In the agent setting, mutability is exposed by benchmarks showing user attributes, habits, and preferences drift on different timelines driven by external events, so a single global update rule is wrong \citep{xie2026dynamicmem}, and by principled-forgetting benchmarks balancing recall against time-decay invalidation \citep{uddin2026recallforgettingbenchmarkinglongterm}. Mutability is the best-covered axis because forgetting and updating are the most studied operations, yet the coverage is shallow: most work decays uniformly, whereas real state is heterogeneous, some items permanent, others ephemeral, and the open problem is authority-conditioned mutability where the right to revise or decay an item is itself governed.

\paragraph{Provenance.} Provenance asks which source, timestamp, and chain of transformations produced an item. Its cognitive root is the episodic-versus-semantic distinction: episodic memory is memory for events tied to a time and place, while semantic memory abstracts away the source \citep{tulving1972episodic}, and provenance is the agent-side discipline of not losing the episodic source when an item is consolidated into a semantic fact. The complementary-learning-systems literature makes the stakes explicit, since consolidation through replay is the operation that can preserve or discard the trace of origin \citep{mcclelland1995complementary,wilson1994reactivation}. The invariant is provenance preservation: consolidation may compress an item's value but must retain enough of its derivation to verify, attribute, and revise the result later. The exposed gap is that the dominant consolidation methods are lossy by design and flatten provenance, so an agent loses the ability to say where a fact came from at precisely the moment it most needs that, when the fact is later challenged or found poisoned.

\paragraph{Recoverability.} Recoverability asks whether the derived state and the decisions an item caused can be rolled back. This is the axis the field covers least well after authority, and it has the weakest cognitive analogue, since biological forgetting is largely irreversible, which is itself instructive: the canonical forgetting curve describes loss without recovery \citep{murre2015replication}, and ACT-R's activation decay offers no undo for a chunk fallen below threshold \citep{anderson2004actr}. Artificial agents are not bound by this limit, but most inherit it by default. Recoverability is what the principled-forgetting line gestures at when it distinguishes accessibility decay from true deletion \citep{uddin2026recallforgettingbenchmarkinglongterm}, yet even there recovering the downstream actions a forgotten fact licensed is left open. The invariant is rollback traceability: every action carries a handle back to the records that justified it, so a later-discovered bad record yields a bounded set of decisions to repair. The gap is stark and quantified later: rollback is the rarest lifecycle stage in the coded corpus, and recoverability is one place where the survey's governance framing does real organizing work.

\paragraph{Actionability.} Actionability asks what kind of object an item is: passive evidence, a preference, a policy, a reusable skill, or an executable commitment. It matters because the consequences of misusing an item scale with its type. The procedural-memory lineage, in which an agent accumulates reusable skills it can later invoke, makes this concrete: Voyager builds an open-ended library of executable skills the agent calls as actions \citep{wang2023voyageropenendedembodiedagent}, and a poisoned skill is far more dangerous than a poisoned fact because invoking it has tool consequences in the world. A reflective lesson degrades gracefully if it is wrong; an executable commitment does not. Actionability therefore sets the bar for the others: the higher an item's actionability, the stronger the authority, provenance, and recoverability guarantees it should require before it is allowed to act. The gap is that current systems rarely type their state by actionability at all, so an advisory note and a callable side-effecting procedure are stored, retrieved, and trusted by the same undifferentiated machinery.

These six axes are the analytic spine of the survey. Every mechanism in later sections can be read as making some axes explicit and leaving others implicit, and every failure mode as the violation of an invariant attached to one or more axes. In our coded corpus, coverage is skewed: mutability and provenance appear more often than authority and recoverability, so the literature attends most to axes whose failures are internal and revisable and less to those whose failures license an action that cannot be taken back.

\subsection{The cognitive-architecture lineage}\label{subsec:lineage}

The axes inherit their structure from the cognitive-architecture tradition, which spent decades formalizing many of the distinctions, memory types, retrieval dynamics, and consolidation processes that LLM agents are now rediscovering empirically. Naming this lineage is not antiquarianism: it explains why the axes are the right ones and why the failures the survey documents were predictable.

The foundational distinction is Tulving's separation of episodic memory, memory for events bound to a time and place, from semantic memory, the store of decontextualized facts \citep{tulving1972episodic}. This underwrites both the provenance axis, since the move from episodic to semantic is precisely the loss of source provenance must guard against, and the state-type taxonomy the survey uses throughout. The complementary-learning-systems theory of McClelland and colleagues then explained why brains keep two stores changing at different rates, a fast hippocampal store that captures specifics and a slow neocortical store that consolidates regularities, arguing formally that interleaving through replay prevents catastrophic interference \citep{mcclelland1995complementary}, an argument later updated to incorporate generative rehearsal and schema-driven fast consolidation \citep{kumaran2016complementary} and grounded empirically in offline hippocampal replay during sleep \citep{wilson1994reactivation}. This is the direct ancestor of the mutability axis and the plasticity-stability tension that organizes the survey's treatment of continual adaptation. Baddeley's multi-component working-memory model, with its episodic buffer binding integrated representations for the current task, supplies the volatile-versus-durable contrast the scope axis formalizes \citep{baddeley2000episodic}, and the quantitative study of forgetting, the Ebbinghaus curve and its modern fitted decay functions, gives the mutability and recoverability axes their measurable shape \citep{murre2015replication}.

These theories were assembled into running architectures. Soar organizes cognition around a decision cycle over explicitly separated procedural, episodic, and semantic memories, treating memory-type isolation as an architectural property enforced by design \citep{laird2012soar}. ACT-R makes the retrieval and action dynamics quantitative, with declarative chunks whose activation decays over time and whose retrieval is probabilistic, and procedural production rules selected by a utility calculus \citep{anderson2004actr}; this is the source of both the activation-decay intuition behind mutability and the possession-versus-licensed-use intuition behind authority. A bridge to LLM agents is CoALA, which recasts the cognitive-architecture frame for language agents, organizing them by memory modules (working, episodic, semantic, procedural) and a structured action space spanning internal and external actions \citep{sumers2024cognitivearchitectureslanguageagents}. CoALA gives a principled vocabulary for much of the modern memory literature without itself enforcing the governance boundaries this survey studies.

Yet the comparison with the classical architectures is also where the lineage exposes its gap. Soar and ACT-R assumed memory-type boundaries were enforced by the architecture, that a procedural rule could not silently corrupt an episodic trace, because the architecture was a closed, hand-built system with hard type boundaries. LLM agents have no such runtime enforcement: their memory types are textual entries in a store, and a procedural skill, a semantic fact, and an episodic trace are often the same kind of object distinguished only by a tag, so the isolation Soar guaranteed by construction must be re-established by governance. CoALA inherits this gap, organizing memory modules as a clean taxonomy but treating them as functionally isolated black boxes, without authority hierarchies that say which module may license an action, without scope gates that prevent one module from corrupting another, and without recoverability handles that allow a corrupted module's effects to be undone. The classical architectures bequeathed the axes but not their enforcement, because enforcement was free in a closed symbolic system and is expensive in an open generative one. The cognitive lineage thus supplies the survey both its vocabulary and its open problem: much of the literature we survey reconstructs the memory \emph{types} of cognitive architecture without reconstructing the governance that made those types safe to use.

\subsection{Inclusions, exclusions, and why personalization and trigger state are state types}\label{subsec:include-exclude}

We close by stating the definition's edges, since a definition that includes everything organizes nothing. It \emph{includes} any system that conditions current behavior on durable cross-session state it accumulates, can act on that state, and can in principle be held to temporal accountability for it. This includes systems the existing literature scatters across separate communities: conversational-memory agents, OS-style memory managers, skill-learning and reflection systems, multi-agent shared-memory platforms, personalized assistants, and proactive or trigger-driven agents. They are unified by a structural commitment to authoritative retained state, rather than by a shared mechanism or application. The definition \emph{excludes} pure architecture or pretraining work that adds no persistent-state operation, episodic agents that reset between tasks, long-context models that consume but do not curate, and read-only RAG over a fixed corpus (Section~\ref{subsec:boundaries}): each lacks the governed accumulation the definition turns on.

The most important boundary call concerns personalization and trigger behavior, because both are sometimes proposed as the defining feature of an always-on agent, and treating them so is a category error this survey avoids. Personalization is a use of persistent state: a personalized agent is one whose retained state includes a user model, and personalization quality is a function of how well that state is written, scoped, and kept current. The personalization benchmarks reviewed above, preference-following \citep{zhao2025prefeval}, persona tracking \citep{jiang2025personamem}, profile-conditioned generation \citep{salemi-etal-2024-lamp}, and the surveys organizing the area \citep{xu2026personalizedllmpoweredagentsfoundations}, all measure capabilities expressed over user state, and measure them well, but they rarely test the governance properties that make the state safe. A useful stress case is authority lapse: a user states a preference, later retracts it, and a personalized agent that still scores well on preference-following may act on the retracted preference, because following a stated preference and respecting a revoked one are different obligations, and only the first is scored. The same holds for trigger and proactivity state, when to interrupt, surface help, or act unprompted \citep{lu2024proactiveagent,he2024madialbench}: a trigger condition is a stored item with authority, scope, and provenance like any other, and proactivity that fires on a stale or out-of-scope trigger is an authority or scope failure, not a personalization success.

We therefore classify personalization, user models, and trigger conditions as persistent-state \emph{types}, taking their place alongside episodic traces, semantic facts, preferences, procedural skills, workflow ledgers, policy and permission state, provenance records, shared and social memory, task ledgers, and tool and credential state in the state-type taxonomy the survey develops next. This keeps the definition faithful to the thesis. If we defined always-on agents by personalization, we would have written a personalization survey with a memory chapter; if by memory, the relabeled memory survey the field already has several of. Defining them as persistent-state systems, and demoting personalization and triggering to state types governed by the same six axes and lifecycle invariants as every other type, keeps the unit of analysis on the object that actually distinguishes always-on agents from their neighbors: not what they remember, and not whom they serve, but the state they make authoritative and whether they can govern it. Later parts answer that governance question at the level of substrates, lifecycle, adaptation, and evaluation.

\section{A taxonomy of persistent state}\label{sec:statetax}

\regime{Substance}{the kinds of persistent state an always-on agent holds, and which axes each stresses.}
Persistent-state taxonomy starts with a harder question than memory taxonomy: what counts as state? A memory survey can take its object for granted: stored information that can be retrieved. A persistent-state survey cannot, because the records that govern an always-on agent's next action are wider than those it would call memories. A revoked API token, a half-completed purchase, a lapsed consent, a procedural skill learned in an environment that has since changed, and a calendar trigger about to fire all condition future behavior, yet none is naturally described as a remembered fact. The taxonomy here enumerates the persistent-state types an always-on agent maintains, grounding each in representative work, naming the failure mode it is prone to, and identifying which of the six state axes (authority, scope, mutability, provenance, recoverability, actionability) it stresses most. Many of these types descend from older cognitive-science categories; the novelty is the obligation to govern them as one interacting state rather than as separable memory forms. The governance burden is uneven: the types the field studies most are the ones whose failures are most recoverable, and the types it studies least are the ones whose failures are least.

We use the three-layer partition summarized in Table~\ref{tab:state-tax}. \emph{Object state} is the thing the agent carries forward: episodic traces, semantic facts, preferences and user models, procedural skills, workflow and task commitments, shared state, and trigger conditions. The \emph{governance envelope} is the metadata and capability surface that travels with those objects: authority, scope, credentials, provenance, retention, deletion, rollback handles, trust, and temporal validity. \emph{External commitments} are externally committed effects linked to internal state: messages, payments, files, database rows, tickets, and other consequences the agent does not fully own once committed. For exposition, the subsections below discuss recurring categories, while Table~\ref{tab:state-tax} further decomposes the governance envelope into its main fields. This split matters because several items that a flat taxonomy would list as state types, such as policy, permission, provenance, and deletion handles, are better modeled as envelope fields over every state object, while side effects are not internal state at all but external consequences that must remain traceable to it. Throughout, we model a persistent-state item as the record $s = \langle v, a, c, m, p, r, k, \tau \rangle$ introduced in the foundations section: a value $v$ under an authority $a$, a scope $c$, a mutability class $m$, a provenance chain $p$, a recoverability handle $r$, an actionability type $k$, and a logical timestamp $\tau$. A state type is then a region of this record space that shares an actionability type and a characteristic profile over the other axes, and a failure mode is the violation of an invariant that region is most exposed to.

\subsection{Memory types: the classical core}

\subsubsection{Episodic traces}

Episodic traces are the raw record of what happened: conversation turns, tool calls and their outputs, environment observations, and the timestamps that order them. They are the most literal sense of memory and the foundation the other types derive from, since a semantic fact is usually an abstraction over episodes and a procedural skill a distillation of successful ones. The category descends directly from Tulving's separation of episodic memory, indexed by time and context, from context-free semantic memory \citep{tulving1972episodic}, a distinction the Soar architecture later operationalized by giving an agent distinct episodic and semantic stores within a single decision cycle \citep{laird2012soar}, and that the working-memory tradition refined through the episodic buffer that binds multimodal experience into integrated, retrievable episodes \citep{baddeley2000episodic}. Modern agent frameworks inherit this structure: CoALA organizes language agents around explicit memory modules whose episodic component logs interaction history for later retrieval \citep{sumers2024cognitivearchitectureslanguageagents}, and broad agent surveys treat the episodic log as the substrate from which all higher memory is consolidated \citep{xi2023riseandpotential}.

The characteristic failure of episodic traces is not forgetting but the opposite: undisciplined accumulation. An episodic store grows linearly with interaction, accumulates noise, can leak private content captured incidentally during ordinary turns, and loses the handles needed to act on or audit a trace later. Cognitive literature already suggests the response: raw episodes should be selectively consolidated rather than retained verbatim. Complementary-learning-systems theory argues that a fast episodic store must interleave with slow semantic consolidation to avoid catastrophic interference \citep{mcclelland1995complementary,kumaran2016complementary}, and the demonstration of offline hippocampal replay during sleep \citep{wilson1994reactivation} supplies the biological mechanism that agent experience-replay schemes imitate. Episodic traces stress recoverability: an episode is only useful downstream if it retains the provenance and retrieval handles that let a later decision be traced back to it, yet the typical agent log stores surface text without the source, timestamp, and trust tier that the provenance chain $p$ and recoverability handle $r$ require. The open problem is that the field captures episodes abundantly but rarely as governable records; the trace is written without the metadata any later forget, audit, or rollback operation would need, so an episodic store has plenty of $v$ and little else.

\subsubsection{Semantic facts}

Semantic facts are the context-free assertions an agent extracts and retains about its users, its projects, and the world: that a user lives in a particular city, that a project uses a particular framework, that an account is in a particular state. They are the type a personalization system most often means by long-term memory, and the recent personalization literature has made them the center of evaluation. PersonaMem tracks an evolving user persona over many interactions and scores whether the agent recalls the current profile \citep{jiang2025personamem}; PrefEval tests whether a model follows a stated fact or preference across long context \citep{zhao2025prefeval}; LaMP established the earlier baseline of profile-conditioned generation \citep{salemi-etal-2024-lamp}; and production-scale systems such as the industrial long-term semantic memory of \citet{xu2026linkedinmemory} show that distilling durable user facts from noisy longitudinal signals is feasible at deployment scale. CarMem narrows the type usefully by bounding semantic facts to declared categories, letting it sharply cut redundant and contradictory stored preferences through scheduled maintenance \citep{kirmayr2025carmem}, and surveys of personalized LLM agents codify the requirement that such facts persist coherently across sessions \citep{xu2026personalizedllmpoweredagentsfoundations}.

The failure modes of semantic facts are staleness, contradiction, and authority confusion, and they sharpen as the timescale lengthens. A fact that was true last month may be false today, and the agent that recalls it with high fidelity is then confidently wrong; DynamicMem shows that user attributes, habits, and preferences drift on different timelines driven by external events, so a single decay rate cannot keep a semantic store current \citep{xie2026dynamicmem}. The deeper problem the personalization literature surfaces, and the one most aligned with the thesis, is authority confusion: a benchmark that scores recall rewards an agent for retrieving a fact regardless of whether it still carries a current, unrevoked authority to influence action. MemProbe reframes the evaluation question by treating long-term memory as a post-interaction artifact to be audited for hidden user-state rather than judged only by downstream task success \citep{ma2026memprobe}, beginning to expose what recall metrics hide. Semantic facts stress mutability and authority: a fact must be revisable and supersedable as the world changes, and it must lose its license to act when the authority behind it lapses. The gap is that the field measures whether a fact can be recalled, not whether recalling it is still authorized; an agent can top a preference-following benchmark and still have no mechanism to revoke a fact the user has retracted, an authority-monotonicity failure that recall benchmarks have no field to score.

\subsubsection{Preferences}

Preferences are a distinguished sub-class of semantic state: durable assertions about how the user wants the agent to behave, including style, habits, constraints, and priorities. We separate them from plain facts because their actionability type differs. A fact is evidence the agent reasons over, whereas a preference is a standing instruction the agent applies, so it more directly shapes action and raises the scope and authority stakes. The evaluation literature treats preference handling as a first-class capability: PrefEval probes whether a model infers and adheres to stated preferences over long multi-session context \citep{zhao2025prefeval}, MPT tests whether an agent reuses persisted preferences to fill under-specified tool-call arguments across sessions, decomposing the capability into recall, induction, and transfer \citep{yoon2026mpt}, and PAL-Bench evaluates assistants on subjective user traits accumulated over long-term interaction \citep{huang2025mempal}. POLAR shows the upside when the type is handled well, accumulating personalized context to resolve implicitly specified targets that a stateless agent could not \citep{lee2026polar}, and the foundations survey of personalized agents frames preference persistence as a core requirement rather than a feature \citep{xu2026personalizedllmpoweredagentsfoundations}.

The failure modes of preferences are over-personalization, scope errors, and drift, and the most dangerous is a security concern rather than a quality one. \citet{xu2026toxicchats} formalize unintended long-term state poisoning, in which routine interactions gradually weaken an agent's confirmation boundaries and silently expand the action scope a stored preference may govern: a preference for fast checkout, accreted over many turns, quietly becomes a license to skip confirmation on a purchase. This is the scope axis failing under accumulation, the scope-non-expansion invariant violated one harmless-looking turn at a time. Preferences therefore stress scope and actionability above all: the question is what the user wants, which tasks, tools, and time windows that preference may act on, and whether it is advisory or has hardened into an executable default. The gap is that preference benchmarks score adherence, treating stronger following as strictly better, while the always-on risk is over-adherence: a preference whose scope has crept or whose authority has lapsed should stop governing action, and no current preference benchmark scores the agent that correctly declines to apply a preference it once held.

\subsubsection{Procedural skills}

Procedural skills are reusable, executable competence: plans, scripts, trajectories, and reflections that an agent stores after success and replays later. The category again has cognitive roots, in the procedural-versus-declarative distinction that ACT-R formalizes through utility-weighted production rules competing with activation-decayed declarative chunks \citep{anderson2004actr}, and Soar's separate procedural store \citep{laird2012soar}. In LLM agents the type was popularized by Voyager, which discovers and banks reusable skills to drive open-ended embodied learning with transfer across tasks \citep{wang2023voyageropenendedembodiedagent}, and by reflection learners such as Reflexion, which convert failed trajectories into verbal lessons stored as persistent guidance \citep{shinn2023reflexionlanguageagentsverbal}. Contextual experience replay folds past trajectories back into context as in-context skill rather than externalizing them \citep{liu2025contextualexperiencereplay}, and MAGE manages procedural state as a hierarchical execution tree with explicit grow, compress, maintain, and revise operations \citep{chen2026mage}. Procedural skills carry the actionability axis to its extreme: a recalled fact is evidence the model may weigh, but a recalled skill is a callable action whose misuse has tool consequences, so the bar for authorizing, scoping, and reversing it is correspondingly higher.

The failure modes are procedure poisoning, unsafe reuse, and overfitting, united by the absence of a validation gate between storing a procedure and trusting it. A reflection induced from a single failed run can become a plausible but unproven durable rule, and a skill that succeeded in one environment can be replayed after the goal or environment has shifted, turning a correct past execution into a wrong present one. The corpus has begun to name these directly: skill drift is formalized as a contract violation, with a check on whether stored procedural skills still produce their role-bearing outcomes in an evolving environment \citep{fan2026skilldriftcontract}, MemP makes deprecation of stale procedures explicit, which is rare \citep{fang2026mempexploringagentprocedural}, and skill containment promotes a skill manifest into a mechanically checkable capability-containment proof via abstract interpretation and bounded model checking \citep{metere2026skillcontainment}. These are the closest the literature comes to governing the type, and they remain exceptions. Procedural skills stress authority, mutability, and recoverability jointly: a skill must be tested and scoped before reuse, deprecated when superseded, and reverted when it causes a regression. The gap is that skill libraries are optimized for cheap accumulation and reuse, not for testing before reuse or rollback after harm, so competence and risk accumulate together, and outside MemP and the drift-and-containment line almost no system re-validates a stored procedure at the moment of reuse.

\subsection{Control-relevant state: beyond memory}

The four types above are where a memory survey would often stop. The remaining categories make an always-on agent a stateful system rather than a memory-augmented one: they include live commitments, shared state, triggers, tool capabilities, authority, provenance, and irreversible external effects. Some are object states in their own right; others are envelope fields attached to every object state. This distinction is the survey's main taxonomic correction: governance is not one more memory type but the layer that determines whether any remembered object may act.

\subsubsection{Workflow and task-ledger state}

Workflow and task-ledger state is the agent's record of open commitments: tasks in progress, deadlines, dependencies, partially completed multi-step operations, and the obligations they imply. It lets an always-on agent resume a long-running goal after a restart or interruption rather than begin afresh, and it is invisible to a memory-form taxonomy because it is live control state, not remembered information. AppWorld stresses the type directly, requiring agents to track world-state changes across sessions and apply stored behavioral patterns to new contexts over stateful applications \citep{trivedi-etal-2024-appworld}, while MAGE's execution tree is in part a task ledger, deriving state from action trajectories so a long operation can be inspected and revised mid-flight \citep{chen2026mage}, and agent workflow memory induces reusable high-level procedures from past task structure \citep{wang2024agentworkflowmemory}. The transactional view is the most governance-aware: SagaLLM imports the database Saga model of compensable execution into multi-agent planning, so a multi-step workflow can be compensated rather than left half-applied when a step fails \citep{chang2025sagallm}.

The characteristic failures are false continuity and stale commitments: the agent believes a task is open that has been completed or cancelled, or it resumes an obligation whose preconditions have since changed. These are mutability and recoverability failures, because a task ledger must update atomically as the world changes and must support compensation when a partially applied workflow has to be unwound. The gap is that most agent memory work treats the ledger as ordinary recall, scoring whether the agent remembers the task rather than whether it correctly maintains, supersedes, and compensates the commitment; outside the transactional line of SagaLLM, the corpus rarely tests whether an interrupted workflow is resumed under still-valid preconditions or whether a cancelled obligation actually stops driving action.

\subsubsection{Policy and permission state}

Policy and permission state is the record of what the agent is allowed to do: user consent, access rights, role assignments, and the safety constraints that bound action. It is the purest carrier of the authority axis and one of the least studied state types in the corpus: authority appears in only $72$ of $435$ works. The works that address it directly are recent and few. Intent-driven authorization treats a user's expressed intent as a monotone, session-scoped policy that can only reduce, never expand, the credential authority granted to tool calls, a direct instantiation of the authority-monotonicity invariant \citep{zhu2026intentgovernedauth}. CaMeL enforces permission at the system level by separating control flow from data flow, so that what a tool may do does not depend on what untrusted text requests \citep{debenedetti2025camel}. SAMEP shares persistent context across sessions under cryptographic access control, raising the question of how permission travels with state across boundaries \citep{masoor2025samep}. These works establish that permission is enforceable but treat it as a static gate.

The failure modes are privilege drift and cross-user leakage: a permission granted for one purpose silently broadens to cover another, or one user's access rights bleed into another user's context through shared state. Both violate the authority and scope invariants, and both disappear in episodic agents only because permissions reset between tasks. The gap is severe and structural: the corpus has mechanisms to grant and check permission but almost none to revoke it dynamically, age it out, or audit whether an action was licensed by an authority current at the time. An agent can hold a permission long after the consent behind it has lapsed, and in the common case no mechanism notices, the authority-monotonicity invariant failing silently over time.

\subsubsection{Provenance and audit state}

Provenance and audit state is the metadata that makes every other type accountable: the sources, timestamps, lineage, transformations, and confidence attached to a record, plus the trail of who or what mutated it. It most directly enables the return arc of the lifecycle, because without provenance there is no basis for attributing a fact, verifying a claim, or tracing an action back to the records that justified it. A small but sharpening body of work targets it. MemLineage attaches a weighted derivation DAG and a Merkle log to agent memory and enforces an untrusted-path-persistence invariant that refuses actions descending from external sources \citep{ouyang2026memlineage}; immutable-memory designs use blockchain-indexed append-only ledgers with privilege lattices to make revisions protocol-blocked and historied \citep{wright2025immutablemem}; Causal Past Logic defines runtime-verifiable guards over causally visible stored state in distributed workflows \citep{bollig2026causalpast}; and MemProbe argues that provenance and hidden state should be audited as a first-class artifact \citep{ma2026memprobe}. The failure this type guards against is provenance-role collapse, in which flat-text memory loses source attribution, so the agent hallucinates authority and misattributes facts to the wrong source \citep{jin2026provenancerolecollapse}.

Provenance and audit state stress provenance preservation and, through it, recoverability: consolidation may compress a value, but it must retain enough lineage to verify, attribute, and later revise or revert the result. The gap is twofold. First, provenance is usually the first casualty of consolidation, because summarization collapses many sourced episodes into one unsourced summary, the corroboration-laundering failure the immutable-memory and lineage lines try to prevent. Second, immutability and deletion are in direct tension: an append-only provenance ledger that makes the past tamper-evident also makes a genuine forget request hard to honor, so a design that maximizes auditability can violate deletion propagation, and the corpus has not resolved how to reconcile the two. The escalation this type adds is that provenance is not one governance axis among several but the precondition for the others: without a preserved derivation chain there is nothing to delete against, nothing to attribute an action to, and nothing to roll back along, so the thinness here is what makes the thinness everywhere downstream irreparable.

\subsubsection{Shared and social memory}

Shared and social memory is persistent state that more than one party can read or write: group threads, team knowledge, role assignments, and the social context of multi-party interaction. It changes the governance problem qualitatively, because a fact written by one agent or user may act for another, so scope and authority must be enforced across principals rather than within one. The foundational demonstration is Generative Agents, whose observation, reflection, and retrieval memory in a shared environment produces believable social behavior, but with no access control, no authority framework, and unaudited memory open to injection \citep{park2023generativeagentsinteractivesimulacra}. MetaGPT's publish-subscribe message pool is a de-facto shared state, untyped and unaudited, with no deletion or revocation semantics \citep{hong2023metagpt}. The governance-aware response is recent: Collaborative Memory adds dynamic, field-level, role-anchored access control to multi-user memory \citep{rezazadeh2025collaborativememorymultiusermemory}, and governed shared memory enforces explicit authority and scope on multi-agent reads and writes while preserving read/write invariants \citep{margalit2026governedsharedmemorymultiagent}. The distinctive failure mode is propagation: memory contagion shows that evaluator bias stored in shared trajectories propagates cross-temporally to future agents, with no safe contamination threshold even under oracle consolidation \citep{liu2026memorycontagion}, and poison-propagation studies map how a single compromised entry spreads through agent-society topologies \citep{men2024troublemaker}.

Shared and social memory stresses scope and provenance jointly, with deletion propagation as the acute open problem: when one principal leaves a group or retracts a contribution, the deletion must reach every derived copy across every agent that read it, and shared-memory work rarely scores revocation or cross-thread leakage. In our coded corpus, the trajectory has moved from ungoverned shared substrates toward access-controlled ones, but deletion-propagation and authority-revocation guarantees across principals remain thin; this is the multi-principal form of the same invariants the survey names.

\subsubsection{Trigger state}

Trigger state is the record of when an always-on agent should act on its own: availability windows, proactive opportunities, and the conditions under which it should interrupt the user or fire a scheduled action. It most distinguishes an always-on agent from a reactive one, because a reactive agent acts only when prompted whereas an always-on agent must decide for itself when latent state warrants action, and it is barely studied. TriggerBench measures prospective memory directly, testing whether an agent spontaneously recalls and acts on a latent stored constraint without being prompted, and finds that prospective recall degrades severely as context grows \citep{zhang2026triggerbench}, while proactive-agent work studies when an agent should volunteer help at all \citep{lu2024proactiveagent}. The failure modes are symmetric and both costly: intrusion when the agent acts on a trigger it should have suppressed, and missed-help when it fails to act on one it should have surfaced.

Trigger state stresses actionability and authority: a trigger is an executable commitment to act under a condition, and the agent needs current authority to fire it, since a scheduled action authorized last week may no longer be permitted today. The gap is that trigger calibration is almost entirely unmeasured as a governance problem; the rare benchmarks score whether the agent can recall a latent constraint, not whether it correctly distinguishes a trigger it is still authorized and scoped to fire from one whose authority has lapsed, leaving the when-to-act decision among the thinnest cells in the corpus.

\subsubsection{Tool and credential state}

Tool and credential state is the agent's authority over external resources: API tokens, OAuth scopes, device-access grants, and the live handles through which the agent reaches the world. It is durable, control-relevant, and emphatically not memory, yet its corruption converts a benign agent into a dangerous one. The threat is best documented from the attack side. Cross-session injection shows that the prompt-injection threat model expands from a single session to the whole system once persistent state, including tool registries and credentials, carries the payload forward \citep{xie2026crosssessioninjection}, and tool drift shows that stored biases accumulate to skew tool selection over time \citep{dabas2026tooldrift}. The recovery side surfaces a subtle hazard named by ACR-Fence: because an agent regenerates slightly different requests after a checkpoint restore, retried tool calls become new external requests that can resurrect already-consumed authority and duplicate irreversible side effects unless those effects are recorded and replayed rather than re-issued \citep{zheng2026acrfence}.

The failure modes are permission drift and stale authorization: a token outlives the consent that justified it, or a scope granted for one operation is reused for another. These are authority and recoverability failures, and they intersect with the most under-served lifecycle stage, rollback. Tool and credential state stresses authority and recoverability above all, and the gap is that the corpus usually treats credentials as configuration rather than governed state with a lifecycle: there is almost no coded work on aging out a token when its authority lapses, on auditing which credential licensed which side effect, or on rolling back tool-originated state once the authority that licensed the original action is gone. ACR-Fence frames the failure, but the corpus has not yet generalized it.

\subsubsection{External commitments and side effects}

External commitments and side effects are not internal state; they are the consequences an always-on agent has written into the world and must keep linked to internal state: a payment issued, a message sent, a file deleted, a record created in an external system. They often have no literal undo, because the agent does not own them, and they are the reason recoverability is more than an internal bookkeeping property. AppWorld is the clearest probe, executing real operations such as payment requests, library deletions, and file creation whose effects persist outside the agent and must be tracked across sessions \citep{trivedi-etal-2024-appworld}. SagaLLM's compensable-transaction model is the most principled response, providing compensation handlers that semantically undo a step when the literal state cannot be restored \citep{chang2025sagallm}, while ACR-Fence shows the failure that arises when side effects are replayed rather than compensated after a restart: duplicated irreversible actions \citep{zheng2026acrfence}.

The failure modes are duplicate side effects and the resurrection of already-consumed authority, both recoverability failures of the highest-consequence kind, because the affected object lives outside the agent's control. External commitments stress recoverability and provenance: every side effect should carry a handle back to the records that justified it, so that a later-discovered bad record yields a bounded set of external effects to compensate, the rollback-traceability invariant made operational. In our coded corpus, only $27$ of $435$ works expose any rollback mechanism at all, and we found no coded work that scores rollback of tool-originated side effects under lapsed authority; this is the cell where thin return-arc coverage has the most direct real-world cost.

\subsection{A comparison across types}

Table~\ref{tab:state-tax} consolidates the taxonomy against characteristic failure modes and the state axes each item stresses. Read down the table, a pattern emerges. Classical memory objects are richly represented, with mature benchmarks and dedicated mechanisms, and their failures, while real, are largely recoverable because the affected state is internal. Object states with external consequences and control-envelope fields are less consistently represented, often by recent work or by attack papers rather than mature mechanisms. The axes column makes the skew precise: authority and recoverability, the two rarest axes in our coded corpus, concentrate in the envelope and in externally consequential object states, while mutability and provenance dominate the well-studied memory core.

\begin{table}[t]
\centering
\small
\setlength{\tabcolsep}{4pt}\renewcommand{\arraystretch}{1.2}
\begin{tabularx}{\linewidth}{p{0.15\linewidth}p{0.23\linewidth}YY}
\toprule
Layer & State item & Typical failure mode & Axes most stressed \\
\midrule
Object state & Episodic traces & Noise, privacy leakage, loss of evidence handles & Recoverability, provenance \\
Object state & Semantic facts & Staleness, contradiction, authority confusion & Mutability, authority \\
Object state & Preferences / user model & Over-personalization, scope creep, drift & Scope, actionability \\
Object state & Procedural skills & Procedure poisoning, unsafe reuse, overfitting & Actionability, authority, recoverability \\
Object state & Workflow / task commitments & False continuity, stale commitments & Mutability, recoverability \\
Object state & Shared / social state & Speaker confusion, propagation, authority conflict & Scope, provenance \\
Object state & Trigger conditions & Intrusion, missed-help & Actionability, authority \\
\midrule
Governance envelope & Authority / permission & Privilege drift, cross-user leakage & Authority, scope \\
Governance envelope & Scope / ownership & Cross-principal use, consent mismatch & Scope, authority \\
Governance envelope & Tool / credential capability & Permission drift, stale authorization & Authority, recoverability \\
Governance envelope & Provenance / audit & Corroboration laundering, rollback gaps & Provenance, recoverability \\
Governance envelope & Retention / deletion / rollback handles & Residual derived state, unbounded repair & Recoverability, provenance \\
Governance envelope & Trust / confidence / temporal validity & Stale trust, unverified promotion, expiry failure & Mutability, authority \\
\midrule
External commitments & Tool/world side effects & Duplicate effects, consumed-authority resurrection & Recoverability, provenance \\
\bottomrule
\end{tabularx}
\caption{The persistent-state taxonomy. Object state names what the agent carries forward. The governance envelope names the fields and capabilities that determine whether an object may influence action and how it can be repaired. External commitments are consequences linked to internal state rather than internal state themselves. The rare axes, authority and recoverability, concentrate in the envelope and in externally consequential objects.}
\label{tab:state-tax}
\end{table}

\subsection{How the types interact, and where governance is thinnest}

The types are not independent, and treating them as separable categories, which the inherited memory taxonomy encourages, is itself a source of failure. They form a derivation order. Episodic traces are consolidated into semantic facts and preferences; successful episodes are distilled into procedural skills; open episodes accrete into the task ledger; every derived record should carry provenance back to the episodes that produced it; and any of these may be shared, may arm a trigger, may exercise a credential, and may ultimately produce an external side effect. A failure at any upstream type propagates downstream along this derivation, which is why the per-type failure modes are not isolable. Provenance-role collapse in the audit type \citep{jin2026provenancerolecollapse} corrupts the authority confusion of semantic facts; memory contagion in shared memory \citep{liu2026memorycontagion} poisons the procedural skills agents distill from a polluted commons; a poisoned preference under expanded scope \citep{xu2026toxicchats} licenses a credential to fire a trigger that produces an irreversible side effect. The thesis that an always-on agent is a single persistent-state system rather than a bundle of memory stores is precisely the claim that these types must be governed jointly, because their failures compose.

The interaction also explains why deletion is the hardest operation and rollback the rarest mechanism. A forget request against a semantic fact must propagate along the entire derivation: to the episodes it was extracted from, the summaries and embeddings that absorbed it, the skills conditioned on it, the shared copies other principals hold, and any external side effect it justified. This is the deletion-propagation invariant, and it is difficult because the types interact; an immutable provenance ledger that makes the derivation auditable \citep{wright2025immutablemem,ouyang2026memlineage} simultaneously makes deletion harder to honor, so the two goals the audit type is supposed to serve, accountability and erasure, are in tension and the corpus has not reconciled them. The transactional and compensation models \citep{chang2025sagallm,zheng2026acrfence} are the closest the literature comes to honoring the recoverability of the downstream types, and they remain confined to a handful of works.

Governance is thinnest, then, where the authority and recoverability axes dominate, where the affected consequence is external or irreversible, and where the lifecycle return arc, forget, audit, and rollback, must run. The classical memory types are better served because their dominant axes, mutability and provenance, are common in the corpus and their failures are often internally recoverable. Governance-envelope fields and external commitments are served by a scattering of mostly recent preprints and by attack papers that document the failure without governing it. The bibliographic skew we measure, only $27$ of $435$ coded works exposing any rollback mechanism and authority being the rarest axis at $72$ of $435$, is the quantitative shadow of this taxonomic skew. The works we coded contain a rich science of the recoverable memory core and a thinner account of the governance envelope around consequential state. That asymmetry, more than any single missing mechanism, is the open problem this taxonomy is built to make visible, and it sets the agenda for the substrate, lifecycle, and governance parts that follow.

\figheatmap
\section{The persistent-state lifecycle}\label{sec:lifecycle}

\regime{Motion}{the operations that move state through the agent, and the five invariants that should govern them.}
The previous parts established what persistent state \emph{is} and what forms it takes. The next question is how it \emph{moves}: by what sequence of operations an observation becomes durable, governed, retrievable state, and by what later operations that state is revised, removed, accounted for, or undone. Memory, in the sense most of the literature uses the word, names only a few of these operations. An always-on agent observes, decides what to write, validates what it wrote, organizes it into something retrievable, retrieves it, acts on it, and then, once an outcome is known, updates it, forgets parts of it, audits how it was used, and on occasion rolls back the decisions a bad record caused. We call this sequence the persistent-state lifecycle. The unit of analysis is not a memory record alone but a typed transition over a set of records.

The lifecycle has two arcs, shown in Figure~\ref{fig:lifecycle}. The \emph{forward arc} runs observe, write, validate, organize, retrieve, act: it carries a signal from the environment into action. The \emph{return arc} runs update, forget, audit, rollback once the consequences of an action are known, reconciling state with reality. Plasticity, the capacity to absorb new information, lives mostly at write and update on the boundary between the arcs. Stability, the capacity to keep old commitments intact and reversible, is enforced at validate, audit, forget, and rollback. An agent that exposes the plastic operations without the stabilizing ones accumulates risk even as its task scores rise, a tension we develop in the part on continual adaptation and preview at the end of this one.

This framing differs from the standard memory pipeline in two places: the validate stage and the return arc. Record-centric memory taxonomies often leave both implicit, and the corpus quantifies that absence. Coding every work in the $N{=}435$ corpus by the lifecycle stages it exercises (a work may touch several) yields a sharply skewed profile: retrieve $269$, write $200$, act $141$, organize $128$, update $127$, audit $88$, validate $87$, observe $68$, forget $66$, and rollback $27$. The forward arc, especially retrieve and write, dominates. The return arc thins out, and rollback, the operation that closes the loop by repairing a state-affected decision, is the rarest capability in the corpus. The operations are therefore uneven in a structured way: the stages least represented in the coded corpus are the ones on which the invariants that distinguish a governed system from a memory-augmented one are established or checked. The stage-by-stage reading names representative mechanisms and benchmarks, contrasts them, and ends each stage on the governance gap it exposes.

\subsection{State records, transitions, and invariants}\label{subsec:records}

The six state axes (authority, scope, mutability, provenance, recoverability, actionability) become operational once we fix what a state item is and which changes to it are legal. We model a persistent state item as a typed record
\[
s = \langle v, a, c, m, p, r, k, \tau \rangle:
\]
a value $v$ (the remembered content); an authority $a$ naming what permitted $s$ to influence action (a user grant, a policy, a tool scope, a delegation); a scope $c$ over users, tools, time, and groups within which $s$ may be used; a mutability class $m$ (revisable, supersede-only, decaying, locked); a provenance chain $p$ recording the sources and transformations that produced $s$; a recoverability handle $r$ linking $s$ to the decisions and side effects it influenced; an actionability type $k$ drawn from a small closed set (evidence, preference, policy, skill, executable commitment); and a logical timestamp $\tau$. The agent's persistent state is the set of live records. The value $v$ is the only field most memory systems represent explicitly; the remaining seven are the governance envelope, and systems that discard the envelope cannot later recover the governance state it carried, however faithfully they preserve $v$.

Each lifecycle operation is then a typed transition over this set, and stating the transitions explicitly lets us name properties an always-on agent must actively preserve but an episodic agent satisfies vacuously by resetting. \textsc{Write} inserts a record with an authority and a provenance chain; in a governed system it inserts into a quarantined tier rather than the live set. \textsc{Validate} promotes a quarantined record to live, or rejects it, after checking that its authority is current and its provenance trusted. \textsc{Organize} re-represents live records (indexing, linking, summarizing, compressing) without changing which decisions they may license. \textsc{Retrieve} selects a subset of live records whose scope covers the current situation. \textsc{Act} conditions a decision on the retrieved subset and stamps the resulting action with the handles $r$ of the records that justified it. \textsc{Update} supersedes one record by a newer one within the same scope, preserving the old record's history rather than overwriting it. \textsc{Forget} tombstones a record and is obligated to propagate the tombstone to derived copies. \textsc{Audit} reconstructs, from $r$ and $p$, which records justified a given action. \textsc{Rollback} uses $r$ to revert the decisions and, where possible, the side effects that a now-removed or now-discredited record caused.

\subsubsection{The five invariants and the stages they govern}

Reading the transitions this way surfaces five invariants the loop should preserve. They are not a borrowed checklist; each is the property whose violation defines one of the persistence-specific failure classes treated later in the survey, and each is anchored to particular stages.

\emph{Authority monotonicity.} A record may influence an action only if a current, unrevoked authority covers it; authority may narrow over time but never silently re-broaden. An expired permission cannot license a write or a tool call. This invariant is enforced principally at validate (does this write carry a live authority?) and at act (is the authority that licensed this record still current at action time?), and it is exercised at update whenever a permission epoch advances.

\emph{Scope non-expansion.} A transition may narrow or preserve scope but never silently widen it. This is what blocks one user's preference from acting for another, or a fact learned in one task from being applied in an unrelated one. It governs organize (consolidation must not merge records across scopes into a wider-scoped summary) and retrieve (the retrieved set must not exceed the requester's scope).

\emph{Deletion propagation.} Tombstoning a record must also reach its summaries, embeddings, cached prompts, promoted tiers, and any derived records, so that erasure is a verifiable property of the store rather than an edit to a single retrieval index. It governs forget and is verified at audit.

\emph{Provenance preservation.} Consolidation may compress $v$ but must retain enough of the provenance chain $p$ to verify, attribute, and later revise the result. It governs organize and update, the two stages that transform $v$, and is the precondition for audit.

\emph{Rollback traceability.} Every action carries a handle back to the records that justified it, so that a later-discovered bad record yields a bounded, identifiable set of decisions and side effects to repair. It is established at act (the handle must be recorded) and consumed at rollback.

The mapping is summarized in Table~\ref{tab:invariant-stage}. The invariants split across the arcs unevenly: authority monotonicity and scope non-expansion are checked on the forward arc (at validate, retrieve, and act), while deletion propagation, rollback traceability, and the back half of provenance preservation all depend on the return arc, the same arc the corpus counts show to be sparse. Deletion propagation is the one invariant established wholly on the return arc (at forget, verified at audit); rollback traceability straddles the boundary, since the handle is set at act on the forward arc but consumed at rollback on the return arc, so a system that instruments act but never builds the return arc records the handle and then has nowhere to use it; and provenance preservation is at risk whenever its return-arc half, the update and audit steps, is the part a system omits. The invariants that most depend on the return arc are therefore the ones with the fewest stage opportunities in our coded corpus. Figure~\ref{fig:invariants} makes this proxy view concrete by counting, per taxonomy part, the works that touch each invariant's governing stage. It should not be read as direct evidence that those works test or enforce the invariant; it shows where the stages that would make such enforcement possible are present.

\begin{table}[t]
\centering
\small
\setlength{\tabcolsep}{4pt}\renewcommand{\arraystretch}{1.2}
\begin{tabularx}{\linewidth}{p{0.24\linewidth}Yp{0.20\linewidth}}
\toprule
Invariant & Governing stage(s) & Arc \\
\midrule
Authority monotonicity & validate, act (update) & forward \\
Scope non-expansion & organize, retrieve & forward \\
Provenance preservation & organize, update & both \\
Deletion propagation & forget (verified at audit) & return \\
Rollback traceability & act (set), rollback (use) & return \\
\bottomrule
\end{tabularx}
\caption{The five invariants and the lifecycle stages that establish or check them. Authority and scope are guarded on the forward arc; deletion propagation and rollback traceability live on the return arc, the arc the corpus shows is least implemented.}
\label{tab:invariant-stage}
\end{table}

\subsection{The forward arc: observe, write, validate, organize, retrieve, act}\label{subsec:forward}

\subsubsection{Observe and write: deciding what becomes state}

The forward arc begins before any memory is written, when the agent decides which observations are worth keeping. The corpus is thinnest here ($68$ works touch observe), largely because most memory systems treat the write decision as a fixed extraction policy rather than a governed choice. The dominant pattern is fact extraction and consolidation across sessions, exemplified by production-oriented systems that distill conversational turns into compact facts and merge them with prior knowledge \citep{chhikara2025mem0buildingproductionreadyai}. These systems are engineered for write throughput and retrieval recall, treating what to keep as a question of compression quality. The broader surveys of agent memory codify this view, organizing the field around extraction, distillation, structuring, and ranking, and making clear that write is studied as a content operation, not as the point where authority and provenance should first attach to a record \citep{zhang2024memorymechanism,du2026memoryautonomous,hu2025memoryageaiagents}. A recent survey reframing memory around a read/write/update lifecycle moves closer to the operational view we take, but still reasons at the level of mechanisms rather than typed transitions with governance fields \citep{huang2026rethinkingmemorymechanisms}.

A small number of works reveal what a richer write stage requires. The episodic-future-thinking memory of \citet{guo2026tmem} rehearses memories at write time in anticipation of future contexts rather than simply archiving them, so the write decision is conditioned on expected later use; this is a step toward write as a forward-looking governance choice, though it still lacks an authority field on the written record. A different and underappreciated move takes the write decision out of the agent's hands entirely: the memory-sandbox system turns an agent's memories into manipulable visual objects so a user can see, edit, delete, and share what is remembered, relocating write, organize, and forget from the agent to the human \citep{huang2023memorysandbox}. As an interface result, this matters, but it also exposes the gap: when write is a manual edit, there is no formal account of which edits are legal, so a user can introduce an inconsistent state with no validation gate to catch it.

The gap at observe and write is that the coded corpus offers no widely used criterion for what to encode versus discard that incorporates the governance envelope. Systems decide what to keep by salience or compression value, not by whether the record can carry an authority, a scope, and a provenance chain that later stages will need. A write that captures a useful value but no authority is a write that no downstream stage can govern.

\subsubsection{Validate: the missing gate}

Validate should sit between a write and the live state, promoting a record only after checking that its authority is current and its provenance trusted. The corpus shows it is barely a stage at all ($87$ works, most addressing it implicitly through retrieval filtering rather than as a distinct promotion gate). The staleness literature states the need directly: \citet{chao2026stalellmagentsknow} build a benchmark testing whether agents detect when stored memories have become stale or invalid, and find that agents frequently act on obsolete state. This is a validate failure surfaced at retrieve time: the record was never marked as requiring revalidation, so nothing checks its currency before use. The benchmark is a single-stage probe, however, and does not connect staleness detection to the write-organize-retrieve chain that produced the stale record in the first place.

A direct critique of the field's missing abstractions appears in the position survey of \citet{zhou2026agentnative}, which asks whether current memory systems are agent-native or retrofitted from retrieval and database stacks. The diagnosis is that without native abstractions, validate has nowhere to live: a vector store promotes whatever is written, because promotion is not a concept it represents. The survey identifies the gap but stops short of proposing the validate-stage invariants (authority monotonicity at promotion time) that would make a system demonstrably agent-native. The validate gap is central on the forward arc because this is where authority monotonicity should be enforced. Almost no system enforces it, so a record written under a since-revoked permission, or distilled from an untrusted tool output, can flow into live state and later license an action.

\subsubsection{Organize: consolidation that preserves or destroys governance}

Once written and (ideally) validated, records are organized into retrievable structure: indexed, linked, summarized, and compressed. This stage is comparatively well served ($128$ works) and contains many of the field's most developed mechanisms, but it is also where provenance preservation and scope non-expansion are most often quietly violated. The mechanisms divide into two families. The first organizes by \emph{summarization}: recursive hierarchical summarization condenses long dialogue history into semantic summaries while maintaining retrieval accuracy across sessions \citep{wang2023recursivesummary}. The second organizes by \emph{structure}: agentic memory with self-organizing, dynamically linked notes that evolve and refactor as new information arrives \citep{xu2025amemagenticmemoryllm}, and temporal knowledge graphs that organize conversational memory as timestamped facts with explicit validity windows for multi-hop temporal queries \citep{rasmussen2025zep}.

Contrasting these families exposes the governance cost of consolidation. Summarization buys compactness at the price of provenance: \citet{wang2023recursivesummary} does not formalize what information loss is acceptable, and a summary that scores well on aggregate recall can silently drop the source attributions audit and rollback later need. Structural organization buys multi-hop reasoning but introduces a scope hazard: the self-organizing links of \citet{xu2025amemagenticmemoryllm} can connect records across scopes, so a traversal can surface a fact outside the requester's authority, an inference channel the system does not guard. Temporal graphs are comparatively better on provenance because validity windows preserve when a fact held \citep{rasmussen2025zep}, yet even there the schema is not validated to preserve full lifecycle traceability under mutation. The gap at organize is that consolidation is optimized for retrieval quality and storage cost, not for the invariants it must not break: a consolidation step that compresses provenance away or merges scopes is a governance loss no later stage can recover.

\subsubsection{Retrieve: the most-studied stage, and what it still misses}

Retrieve is the field's center of gravity ($269$ works, more than any other stage), and the benchmarks here are mature. Long-term conversational memory benchmarks such as LoCoMo test whether agents retrieve semantically distant yet critical memories for complex multi-turn reasoning \citep{maharana2024evaluatinglongtermconversationalmemory}, and broader assistant benchmarks probe long-term interactive recall across many sessions \citep{wu2025longmemevalbenchmarkingchatassistants}. More recent work makes retrieval evaluation harder and more diagnostic. MemoryArena evaluates memory across interdependent multi-session tasks, where earlier actions and feedback should change later retrievals, with separate metrics for recall, induction, and contextual grounding \citep{he2026memoryarena}; incremental multi-turn evaluations decompose retrieval competence into distinct skills \citep{hu2026evaluatingmemoryllmagents}; and benchmarks of fine-grained relational discrimination test whether agents can tell complementary, nuanced, and contradictory memories apart, separating preservation, retrieval, and reasoning failures \citep{wang2026subtlememory}. Realistic, temporally evolving dialogue generators push retrieval evaluation toward heterogeneity and drift \citep{zhang2026rhelm}.

The limitation of this family, visible across these benchmarks, is that retrieval is usually scored against the value field $v$ alone. They measure whether the right content comes back, not whether the retrieved set respects scope, whether the records are still authorized, or whether a retrieved fact will correctly ground the downstream action. MemoryArena comes closest to the action-grounding question because its tasks are interdependent, but as its own analysis notes, it measures retrieval accuracy without diagnosing why retrieved memory fails to ground action or whether organization enabled the retrieval under interference \citep{he2026memoryarena}. The gap at retrieve is thus not retrieval quality, which is well advanced, but the absence of scope non-expansion as a scored property: a retrieval that surfaces a correct value from outside the requester's scope passes every benchmark in this family while violating the invariant.

\subsubsection{Act: binding state to consequence}

Act is where retrieved state conditions a decision with real consequences, and where the forward arc must stamp the action with the recoverability handles the return arc will need. The foundational mechanism is the interleaving of reasoning and action that lets an agent bind internal state to tool calls and validate intermediate outcomes \citep{yao2023reactsynergizingreasoningacting}; this established how agents act on state but said nothing about how persistent memory should guide action selection or how a failed action should update the state that licensed it. The benchmarks that put state-grounded action under stress are the stateful, tool-mediated environments. AppWorld tests whether agents track world-state changes across interactions over stateful apps and simulated people \citep{trivedi-etal-2024-appworld}; $\tau$-bench evaluates tool-agent-user interaction with explicit policy-compliance checks, so an action that violates a stated policy fails even when it completes \citep{yao2024taubenchbenchmarktoolagentuserinteraction}; and GAIA stresses multi-step tool use on real-world tasks \citep{mialon2023gaia}. The most recent, Momento, is pointed for our argument: it requires consequential tool-mediated actions over evolving goals across sessions, so an agent must treat prior-session history as state that may need re-validation before it acts, not as settled fact \citep{merin2026momento}.

These benchmarks are the closest the field comes to scoring the act stage as a state operation rather than an answer, and $\tau$-bench's policy compliance and Momento's re-validation requirement are genuine steps toward authority monotonicity at action time. The remaining gap is twofold. First, as the AppWorld analysis notes, when an action fails it is hard to attribute the failure to an incomplete write, a poor organization, or a weak retrieval, because the trajectory does not carry the per-record handles that would localize it \citep{trivedi-etal-2024-appworld}. Second, none of these benchmarks scores whether the action recorded a rollback handle: an action with an irreversible side effect (a payment, a deletion, an external message) is treated as correct if it satisfies the task, even if the state that licensed it later proves stale or unauthorized and the side effect cannot be traced back for repair. Act, in current practice, conditions on state but does not instrument the consequence, which is what makes the return arc hard to implement.

\subsection{The return arc: update, forget, audit, rollback}\label{subsec:return}

The return arc runs once an outcome is known, reconciling state with reality. The corpus counts collapse here: update $127$, audit $88$, forget $66$, rollback $27$. Update is comparatively healthy because revising a fact resembles a write. The governance-bearing operations, those that remove state, account for its use, and undo its effects, are progressively sparser, and rollback is near-absent. The pattern is simple: the field can change state more readily than it can remove, account for, or reverse it.

\subsubsection{Update: revision that is studied, propagation that is not}

Update supersedes a record with a newer one in the same scope. It is the best-served return-arc stage because revising a stored fact is mechanically close to writing one, and the conversational-memory and temporal-graph systems discussed under organize already implement it through validity windows and self-revising notes \citep{rasmussen2025zep,xu2025amemagenticmemoryllm}. The hard part of update is not changing the focal record but propagating the change to every record derived from it, and detecting when an update should have fired but did not. The staleness benchmark of \citet{chao2026stalellmagentsknow} is, read on the return arc, an update-propagation diagnostic: the obsolete record was never superseded, so the update that should have fired never did. Belief-drift studies extend this to softer mutation, measuring whether an agent's stored opinions remain temporally consistent or drift under repeated interaction \citep{myakala2026beliefshiftbenchmarkingtemporalbelief}. The gap at update is that supersession is treated as a local edit: superseding a fact rarely re-examines the summaries, skills, or downstream commitments derived from the old value, so a corrected fact can coexist with stale derivations of itself, a provenance-preservation failure the next stages are supposed to catch and usually cannot.

\subsubsection{Forget: deletion as a retrieval edit, not an erasure}

Forget is where deletion propagation must hold, and it is one of the least-implemented stages ($66$ works). The dominant treatment is decay rather than deletion: Ebbinghaus-inspired forgetting downweights records over time so less-used memories fade \citep{zhong2023memorybankenhancinglargelanguage}. Decay is attractive because it is cheap and graceful, but it does not satisfy a deletion request: a downweighted record is still present and still retrievable under the right query, precisely the failure mode a user invoking a right-to-be-forgotten would care about. The interface alternative, the memory-sandbox, lets a user explicitly delete a memory object \citep{huang2023memorysandbox}, closer to true erasure, but it deletes the object the user can see and does not propagate to summaries, embeddings, or promoted tiers derived from it. The gap at forget is the deletion-propagation invariant itself: in both the decay and the manual-edit treatments, forgetting edits one representation of a record while its derived copies survive, so erasure becomes a property of a single index rather than of the store. A fact a user asked to delete can persist in a summary computed before the deletion, and nothing in these mechanisms detects or repairs that.

\subsubsection{Audit: accounting for how state was used}

Audit reconstructs which records justified a given action, and it is sparse ($88$ works) and shallow where it exists. The interface line again offers the most concrete instance: the memory-sandbox makes the agent's state observable and lets a user trace how a memory edit changes subsequent behavior, audit in an interactive, human-driven form \citep{huang2023memorysandbox}. This reveals that audit requires the provenance chain $p$ and the recoverability handles $r$ to have been recorded on the forward arc; when those fields are absent, audit can only inspect the current value of state, not the lineage of a past action. The agent-native critique sharpens the point: a retrofitted retrieval stack has no place to store the per-action justification trail, so audit is reduced to after-the-fact guessing about which retrieved record caused a behavior \citep{zhou2026agentnative}. The gap at audit is structural rather than algorithmic: without provenance preservation upstream, there is no lineage to audit, which is why audit and rollback co-fail.

\subsubsection{Rollback: the rarest capability}

Rollback closes the loop. Given a record discovered to be bad, stale, poisoned, or unauthorized, rollback uses the handles $r$ to revert the decisions and side effects it caused. It is the rarest stage in the entire corpus: $27$ of $435$ works expose any rollback mechanism, and within lifecycle-focused work none directly implements it. The benchmarks that most need it, the stateful action environments, do not score it: AppWorld and $\tau$-bench verify whether an action was correct, not whether a wrong action licensed by stale state could be traced and undone \citep{trivedi-etal-2024-appworld,yao2024taubenchbenchmarktoolagentuserinteraction}, and Momento's re-validation requirement is a forward-arc guard against acting on stale state, not a return-arc mechanism for repairing an action already taken \citep{merin2026momento}. The OS-inspired memory systems discussed next provide the closest substrate for rollback, through checkpointing and paging, yet as we note they do not connect the substrate to the state-affected decisions rollback must repair. One caveat on the aggregate count: it sums three mechanically distinct operations the literature treats separately. Internal-state rollback reverts a stored fact or a consolidation decision; workflow rollback compensates a partially executed multi-step task; and external-effect rollback undoes an already-committed side effect such as a payment or a deletion. The aggregate of twenty-seven is therefore conservative for the easiest subclass and almost certainly an overcount for the hardest. External-effect rollback, the subclass that matters most because irreversible effects cannot be repaired by any internal mechanism, is rarer still.

Rollback is the survey's central lifecycle gap. Rollback traceability is established at act (the handle must be recorded) and consumed at rollback, so a field that does not instrument act for consequence cannot implement rollback even in principle. The practical consequence is that an always-on agent that takes an irreversible action on the basis of a record it later discovers was poisoned or unauthorized has, in the vast majority of the systems we surveyed, no bounded set of decisions to repair and no path back to a known-good state. This operation distinguishes a governed persistent-state system from a memory-augmented one, and on the corpus evidence it is the one the literature has built least.

\subsection{Operating-system abstractions for the lifecycle}\label{subsec:os}

A distinct line of work treats the whole lifecycle as a resource-management problem and imports operating-system abstractions to solve it. MemGPT pioneered this view, treating agent memory as OS-style paged storage with a context manager that externalizes long-term state and manages a bounded working set, decoupling the task horizon from the context budget \citep{packer2024memgptllmsoperatingsystems}. MemoryOS extends the analogy with explicit kernel-level paging, scheduling, and eviction policies and a hierarchical working-set bound for long-horizon reasoning \citep{kang2025memoryosaiagent}. A complementary mechanism precomputes and consolidates context offline, before a query arrives, to amortize inference compute, in effect scheduling the organize stage during idle time \citep{lin2025sleeptimecompute}.

The OS framing is lifecycle-aware because it makes transitions between tiers (working set to long-term store and back) explicit operations rather than implicit side effects, and it provides the checkpoint and restore primitives rollback would build on. Its limitation is that it governs the lifecycle for \emph{resource} correctness, not \emph{state} correctness. MemGPT establishes no principled criterion for which records should cross the working-set boundary beyond recency and capacity, and does not validate that paging preserves state fidelity across episodes \citep{packer2024memgptllmsoperatingsystems}. MemoryOS allocates and evicts efficiently but does not connect its eviction policy to the authority and provenance axes, so a record can be evicted before it is acted upon (a state-dependent failure) or a low-authority record promoted because it is frequently accessed \citep{kang2025memoryosaiagent}. The sleep-time consolidation work amortizes compute but provides no link from the precomputed state back to its source observations and no rollback if the anticipated context was wrong \citep{lin2025sleeptimecompute}. The gap is that the OS abstraction supplies useful \emph{machinery} for governance (tiers, scheduling, checkpoints) while leaving the governance \emph{policy} (which transitions preserve the five invariants) unspecified, the separation between mechanism and governance this survey argues the field must close.

\subsection{Lifecycle fidelity and the return-arc gap}\label{subsec:fidelity}

A final cluster of work does not implement a stage but tests whether the chain as a whole preserves fidelity across multi-session reuse, and these works are the most direct evidence for the survey's thesis. Stale-state detection \citep{chao2026stalellmagentsknow}, the agent-native critique of retrofitted abstractions \citep{zhou2026agentnative}, and the human-in-the-loop manipulation of in-memory state \citep{huang2023memorysandbox} together show that lifecycle dependencies are real and largely ungoverned: a write decision affects downstream organization fidelity, an organization choice affects retrieval under interference, and a missing validate gate lets a stale or untrusted record flow all the way to action. None provides a formalization of the invariants the chain must preserve or an automated mechanism to recover when the chain breaks; they reveal the dependencies without governing them.

Table~\ref{tab:stage-coverage} consolidates the per-stage picture developed above. The shape of the table is the argument. Coverage is heavy on the forward arc and concentrated at retrieve and write; it thins monotonically across the return arc to rollback. The invariant most exposed at each stage is governed, if at all, by a stage that is thin in the coded corpus: scope non-expansion depends on a retrieve and organize discipline retrieval benchmarks do not score, deletion propagation depends on a forget operation decay-based systems do not perform, and rollback traceability depends on instrumenting act for consequence and on a rollback stage that $27$ of $435$ works implement.

\begin{table}[t]
\centering
\small
\setlength{\tabcolsep}{4pt}\renewcommand{\arraystretch}{1.2}
\begin{tabularx}{\linewidth}{p{0.13\linewidth}p{0.06\linewidth}YY}
\toprule
Stage & Works & Representative mechanism / benchmark & Per-stage gap \\
\midrule
Observe & 68 & Anticipatory write rehearsal \citep{guo2026tmem}; user-driven write \citep{huang2023memorysandbox} & No governed criterion for what to encode; write captures $v$ without authority \\
Write & 200 & Fact extraction and consolidation \citep{chhikara2025mem0buildingproductionreadyai}; memory surveys \citep{zhang2024memorymechanism,du2026memoryautonomous} & Write studied as compression, not as where authority and provenance attach \\
Validate & 87 & Staleness detection \citep{chao2026stalellmagentsknow}; agent-native critique \citep{zhou2026agentnative} & The missing promotion gate; authority monotonicity unenforced \\
Organize & 128 & Recursive summary \citep{wang2023recursivesummary}; self-organizing notes \citep{xu2025amemagenticmemoryllm}; temporal graph \citep{rasmussen2025zep} & Consolidation drops provenance or merges scopes \\
Retrieve & 269 & LoCoMo \citep{maharana2024evaluatinglongtermconversationalmemory}; MemoryArena \citep{he2026memoryarena}; relational discrimination \citep{wang2026subtlememory} & Scored on value only; scope non-expansion unscored \\
Act & 141 & ReAct \citep{yao2023reactsynergizingreasoningacting}; AppWorld \citep{trivedi-etal-2024-appworld}; $\tau$-bench \citep{yao2024taubenchbenchmarktoolagentuserinteraction}; Momento \citep{merin2026momento} & Conditions on state but records no rollback handle for side effects \\
Update & 127 & Temporal validity \citep{rasmussen2025zep}; belief drift \citep{myakala2026beliefshiftbenchmarkingtemporalbelief} & Supersession is local; derived records not re-examined \\
Forget & 66 & Decay-based forgetting \citep{zhong2023memorybankenhancinglargelanguage}; manual delete \citep{huang2023memorysandbox} & Deletion edits one index; does not propagate to derived copies \\
Audit & 88 & Observable state manipulation \citep{huang2023memorysandbox} & No upstream provenance, so no lineage to audit \\
Rollback & 27 & OS-style checkpoint substrate \citep{packer2024memgptllmsoperatingsystems,kang2025memoryosaiagent} & No link from substrate to state-affected decisions; rarest stage \\
\bottomrule
\end{tabularx}
\caption{Stage coverage across the $N{=}435$ corpus, with a representative mechanism or benchmark and the governance gap each stage exposes. Coverage is heavy on the forward arc (retrieve, write) and thins monotonically across the return arc to rollback. The per-stage gap column shows that each return-arc invariant is governed, if at all, by a stage that is thin in the coded corpus.}
\label{tab:stage-coverage}
\end{table}

The persistent-state lifecycle is not a memory pipeline with two extra boxes. It is a loop whose forward arc is much more developed in the works we coded than its return arc, and the invariants that distinguish a governed persistent-state system from a memory-augmented agent are the ones the return arc enforces. Several open problems follow directly. The corpus contains few validate stages that enforce authority monotonicity at promotion time, so untrusted or unauthorized writes can flow into live state. It contains few consolidation disciplines at organize that preserve provenance and refuse scope-widening merges. It contains few forget operations that propagate deletion to derived state, few audit trails rooted in upstream provenance, and, most acutely, few rollback mechanisms that instrument action for consequence so a bad record yields a bounded set of decisions to repair. In the works surveyed here, the five invariants are more often implicit desiderata than benchmarked properties of the write-organize-retrieve-act chain. These gaps are the operational content of the claim that always-on agents are persistent-state systems whose state must be governed, and they lead directly to continual adaptation, where accumulated state compounds risk rather than competence when the return arc is missing.

\figlifecycle
\figinvariants
\section{State substrates and representations}\label{sec:substrates}

\regime{Substance}{where persistent state physically lives, and what each substrate makes easy or precludes.}
The previous parts established that an always-on agent is a persistent-state system whose retained state spans far more than retrievable text. The physical question is where that state lives. Each substrate, the medium in which state is materialized, indexed, and recovered, imposes a different set of operations and governance affordances. Model weights forget through interference but cannot be tombstoned; a vector index can be edited at a key but cannot be replayed to a point in time; an append-only event log can replay but cannot be garbage-collected without breaking provenance; a knowledge graph can express that one fact superseded another but rarely records who authorized the supersession. The substrate decides which of the six state axes (authority, scope, mutability, provenance, recoverability, actionability) are native, and which require an external envelope, sidecar log, access-control layer, or transaction wrapper. This is the largest part of the corpus, 117 of 435 coded works, because governance is often absent from the dominant data structures themselves, especially vector indices, parametric weights, and the context window. Rollback, the corpus's rarest lifecycle capability, is especially hard to retrofit when the substrate has no durable authority field, tombstone, or point-in-time state to revert to.

Substrates fall along a physical gradient, from the most internal (model parameters) to the most external (multi-tenant serving runtimes). Classical memory forms, episodic versus semantic versus procedural, parametric versus non-parametric, remain useful, but they cut across substrates rather than replacing them. The form of remembered content, such as an event trace, a distilled fact, or a callable skill, tells us what is stored; the substrate tells us how it is addressed, mutated, and recovered, which is what governance acts on. A reflection lesson and a fine-tuned weight delta may carry the same procedural knowledge, but only one can be deleted, attributed, and rolled back. That distinction drives the reading below.

\subsection{Parametric state and test-time learning}

The most internal substrate is the model's own parameters. Here state is non-symbolic: it lives as distributed weight changes rather than addressable records, which makes parametric memory dense and fast to read at inference but opaque to inspection, attribution, and deletion. One recent line treats inference-time weight updates as a memory mechanism. TMEM absorbs distilled supervision into fast LoRA weights through online reinforcement-learning-optimized updates within an episode, so an agent adapts to a running interaction without paying repeated context cost \citep{ren2026tmem}. PEAM converts retrieval memory into parameter-resident skills via per-category isolated mixture-of-experts LoRA adapters, with a failure-as-training-signal loop that promotes successful retrieval patterns into weights \citep{guo2026peam}. Both internalize what would otherwise be external context, trading addressability for efficiency. A second cluster makes the parametric memory recurrent and constant-size: a distillation scheme compresses a full-history transformer teacher's bottleneck representation into a recurrent transformer student's fixed-size memory, so the agent keeps a bounded parametric state instead of an unbounded store of past observations \citep{weinzaepfel2026recurrentmem}. A third treats the latent state itself as the retrievable object: ElasticMem retrieves latent memories from the reasoner's own hidden state and uses a learned policy to assign each one a variable latent token budget, cutting cost while preserving accuracy \citep{feng2026elasticmem}, while EvoEmbedding maintains a continuously updated latent memory as it sequentially encodes a long context, producing evolvable representations for dynamic retrieval \citep{nie2026evoembedding}.

These mechanisms differ in what they internalize and how reversibly. TMEM and PEAM make adaptation episodic and category-scoped, so a bad update is at least contained to a LoRA module rather than the base weights; recurrent compression \citep{weinzaepfel2026recurrentmem} makes the memory bounded but irreversible, since once an observation is folded into the recurrent state there is no handle to recover or excise it. Parametric memory has no tombstone: a fact cannot be deleted from a weight matrix the way a row can be dropped from a store, and weight editing at scale introduces the same gradual and catastrophic interference that makes continual learning hard. The tradeoff is direct: TMEM and PEAM deliver fast test-time adaptation and reduced context cost, but updates leave no audit trail for the provenance of a parameter shift and no mechanism to detect an unauthorized or poisoned internalization. Parametric state is therefore efficient to read and hard to govern at write time: it satisfies actionability and, within an episode, mutability while structurally weakening provenance preservation and rollback traceability. The missing capability is a parametric substrate with a delete: a way to attribute a behavior to the specific update that produced it and revert that update without replaying the entire training history. No corpus work in this cluster offers it.

\subsection{The context window and long-context}

If parametric memory is the most internal substrate, the context window is the most ephemeral: state lives only as tokens in the active prompt, governed implicitly by what the agent chooses to keep in scope. The long-context literature is, at root, a study of how much state this substrate can hold and how reliably it can be read. Boundary diagnostics establish the limits. LongBench probes long-context understanding across diverse bilingual task types and gives an empirical baseline \citep{bai2024longbenchbilingualmultitaskbenchmark}; RULER constructs synthetic tasks to reveal the true effective context length behind the advertised window, which is routinely far shorter \citep{hsieh2024rulerwhatsrealcontext}; and the lost-in-the-middle result shows that models systematically underuse information placed mid-context even when they attend to the boundaries \citep{liu2023lostmiddlelanguagemodels}. The survey of context engineering systematizes the techniques that work within these limits, from compression to reordering to selective inclusion \citep{mei2025surveycontextengineering}. The shared lesson is that the context window is an unreliable substrate for state: presence in the window does not guarantee retrievability, so an always-on agent cannot treat a long prompt as durable memory.

A separate line attacks the boundedness of the window directly. StreamingLLM enables a model to run on effectively infinite-length streams by retaining attention sinks plus a sliding window, making unbounded operation possible without retraining \citep{xiao2023streamingllm}. This is the substrate-level enabler of always-on operation: an agent that never resets must consume an unbounded event stream, and StreamingLLM is the canonical mechanism for doing so within the context substrate rather than offloading to an external store. But infinite throughput is not the same as infinite memory: the sliding window discards old tokens, so the substrate forgets by construction and offers no way to retrieve or attribute what scrolled past.

The central debate in this subsection is the long-context-versus-retrieval boundary, because it decides whether state should live in the window at all. A foundational study showed that retrieval augmentation consistently improves LLMs regardless of context length, so even a long-context model benefits from an external store \citep{xu2023retrievalmeetslongcontext}. Subsequent work refined the boundary: long-context LLMs generally beat retrieval on accuracy but at much higher cost, motivating Self-Route, which routes each query to the cheaper mode \citep{li2024ragorlongcontext}; a systematic study across twenty models found that only a few recent systems sustain accuracy past 64k tokens and cataloged the distinct failure patterns of long-context retrieval-augmented generation \citep{leng2024longcontextragperformance}; and a careful revisit found that long context wins on QA benchmarks while retrieval retains an edge on dialog, sharpening rather than settling the tradeoff \citep{li2025longcontextvsrag}. Past a million tokens, the BEAM benchmark (100 conversations, 2,000 validated questions) shows that even million-token-window models degrade on genuinely long interaction histories, so window size alone does not solve always-on memory \citep{tavakoli2025beyondamilliontokens}. A representative hybrid response is general agentic memory, which flags key history into a light persistent store while preserving all raw pages, then composes context just-in-time per request, refusing to choose statically between the window and the store \citep{yan2025generalagenticmemory}.

For governance, the context window has the weakest affordances of all. It has no persistent identity (it resets between sessions by definition), no provenance (a token in the prompt carries no source or authority tag unless one is manually encoded), and no deletion semantics (dropping a token is a side effect of windowing, not an audited erasure). The boundary studies expose a representation-quality problem the thesis can state directly: state that is present but unretrievable, the lost-in-the-middle failure, is a silent violation of the assumption that retained state is usable. The gap is that the long-context literature measures whether supplied state can be read, never whether the agent decided correctly what should have become durable state in the first place; the window is a substrate for working context, and treating it as a memory substrate inherits all of its ungoverned ephemerality.

\subsection{Retrieval and vector stores (RAG)}

Retrieval-augmented generation externalizes state into a corpus of passages addressed by embedding similarity. This is the field's default non-parametric substrate, and the corpus reflects it: the RAG-and-long-context cluster is the single largest substrate cluster. Its representational unit is the chunk or document, and its governance affordance is, in principle, provenance: every retrieved span has a source. The benchmark literature established the failure vocabulary. RGB decomposes RAG quality into noise robustness, negative rejection, information integration, and counterfactual handling, framing retrieval as more than nearest-neighbor lookup \citep{chen2023benchmarkinglargelanguagemodels}; RAGTruth annotates hallucinations to make faithfulness measurable \citep{niu2024ragtruthhallucinationcorpusdeveloping}; RAGBench adds explainable relevance and faithfulness metrics for diagnostic evaluation \citep{friel2025ragbenchexplainablebenchmarkretrievalaugmented}; and CRAG spans diverse domains and dynamic facts to stress robustness to staleness \citep{yang2024cragcomprehensiverag}. The multi-hop substrate is anchored by HotpotQA, which requires reasoning over several documents and so forces a retriever to compose rather than match \citep{yang2018hotpotqa}.

For always-on agents, the key limitation is that these benchmarks evaluate retrieval over a static corpus, whereas an always-on agent's store is one it writes to from its own experience. Three lines confront the moving-corpus problem. First, temporal staleness: a study of temporal validity in retrieval quantifies that standard RAG returns stale facts 15 to 40 percent of the time over evolving knowledge and proposes a temporal strategy to fix it \citep{yadav2026temporalvalidityretrieval}. Second, retrieval that is misleading rather than simply wrong: causal memory intervention selects which memories to inject by measuring each candidate's causal effect on the agent's output under controlled interventions, filtering similar-but-misleading memories that pure semantic similarity would surface \citep{srivastava2026causalmemory}; and a learned-value approach scores each candidate memory by seven cognitive factors (emotional intensity, goal relevance, value alignment, and others) to decide what to remember at all \citep{chen2026learningwhattoremember}. Third, goal-directed retrieval: Goal-Mem treats a user utterance as a goal and chains backward into atomic subgoals to target what to retrieve, replacing flat similarity with structured intent \citep{liang2026goalmemragmemory}. These works increasingly treat similarity-addressed retrieval, the substrate's defining mechanism, as too weak for self-written, evolving, action-conditioning state.

The governance blind spot of the vector store is visible in the matrix. RGB and its successors benchmark retrieval in isolation from the lifecycle: they never test retrieval under state mutation, versioning, or deletion, because their corpus does not mutate. Provenance is nominally present (each chunk has a source) but is not propagated through consolidation, and deletion is a retrieval edit, not an audited erasure: removing a document from the index leaves its influence in cached prompts, summaries, and any parametric internalization downstream. The substrate makes write-and-retrieve easy but makes deletion propagation, temporal validity, and authority-scoped access nearly impossible to express, because an embedding key carries no notion of who may read it, when it expired, or what it superseded. The requirement this substrate exposes is the one the thesis foregrounds: a retrieval store that is mutation-aware, where a delete cascades to derived state and a write records the authority and validity window that govern when the chunk may act, rather than only whether it is similar.

\subsection{Structured stores: knowledge graphs and temporal memory}

Where the vector store addresses state by similarity, structured stores address it by relation and by time, and in doing so they make some governance axes representable for the first time. The knowledge-graph line begins with HippoRAG, a neurobiologically inspired long-term memory that builds a knowledge graph and runs personalized PageRank for retrieval, integrating structural organization with associative recall and beating flat retrieval on multi-hop tasks \citep{gutierrez2024hipporag}. Recent agentic-memory graphs make the structure dynamic. FluxMem represents memory as a heterogeneous graph that continuously reshapes its topology, repairing links, removing noise, and distilling repeated successful paths into reusable routines \citep{fang2026fluxmem}; AtomMem extracts high-value atomic facts into hierarchical events and activates an associative graph at retrieval time via spreading activation to connect fragmented memories \citep{yao2026atommem}; and H-Mem evolves memory through a temporal-semantic tree that consolidates short-term into long-term plus a parallel knowledge graph capturing entity relations \citep{yu2026hmem}. A graph survey systematizes the design space along short-versus-long-term, knowledge-versus-experience, and structural-versus-non-structural axes, showing that graph substrates enable multipath retrieval and reasoning that flat stores cannot \citep{yang2026graphagentmemorysurvey}.

The most governance-relevant structured substrate is the temporal store, because time is the one axis on which always-on state most obviously diverges from static corpora. The benchmark lineage is mature. TimeQA established that a single fact has time-indexed answers, built from time-evolving Wikipedia \citep{chen2021timeqa}; StreamingQA measured whether a store keeps pace with knowledge arriving over time \citep{liska2022streamingqa}; FreshQA categorized questions by how fast their answers change and added false-premise cases \citep{vu2023freshllms}; and ChroKnowBench organized the evolving-versus-constant distinction across domains and years \citep{park2024chroknowledge}. A parallel probing literature decomposes temporal reasoning itself: TempReason orders difficulty levels \citep{tan2023tempreason}, TimeBench unifies symbolic, commonsense, and event temporal reasoning \citep{chu2023timebench}, MenatQA factors temporal QA into scope, order, and counterfactual perturbations \citep{wei2023menatqa}, Test of Time uses synthetic graphs to isolate pure temporal reasoning from pretraining contamination \citep{fatemi2024testoftime}, and TimeR4 injects explicit time constraints into retrieval over a temporal knowledge graph \citep{qian2024timer4}. Memory-T1 trains a policy via reinforcement learning to select temporally relevant memories for multi-session reasoning, moving from probing to mechanism \citep{du2025memoryt1}.

The temporal mechanisms most relevant to governance make time a first-class field of the stored record. Chronos converts conversation into structured time-stamped events to support multi-hop temporal queries \citep{sen2026chronos}; APEX-MEM stores conversational memory as a temporally annotated property graph and reasons over it with agentic retrieval \citep{banerjee2026apexmem}; and Engram comes closest to lifecycle-complete. Engram writes lossless episodes on a fast path and asynchronously distills subject-predicate-object facts into a bi-temporal knowledge graph carrying both valid-time and transaction-time, supporting point-in-time as-of retrieval, and resolves conflicts by invalidate-never-delete so that every fact retains a provenance supersession chain \citep{wang2026engram}. Engram is the rare substrate in the corpus that directly serves three governance axes at once: provenance (the supersession chain), mutability (valid-time updates without destroying history), and a form of recoverability (as-of retrieval is a read-time rollback). The structured substrate thus contributes what the vector store lacks: the ability to express that a fact held during an interval, was superseded, and descends from a source. Its blind spot, visible across HippoRAG and the graph systems, is mutation under access: graph construction is often offline and static, and few works address edge deletion, node versioning, graph-integrity validation under concurrent agent-driven updates, or who is authorized to rewrite a relation. The gap is that the temporal substrate solves provenance and validity for a single writer but leaves authority and concurrent-write consistency to the runtime layer below it.

\subsection{Multimodal memory}

State is increasingly not text. Multimodal memory substrates store and address visual, video, and egocentric-sensor state, and they expose governance problems that text substrates can hide. The benchmark wave is recent and concentrated. The framing works probe distinct fault lines: MemEye evaluates memory along visual-evidence granularity from scene to pixel and along evidence-use fidelity, grounding memory analysis in perception rather than recall \citep{guo2026memeye}; MemGallery extends multi-session conversational-memory benchmarking to multimodal LLM agents over image-plus-text dialogue across thirteen memory systems \citep{bei2026memgallery}; and EgoMem and EgoStream push to week-long and streaming egocentric video, with EgoStream introducing an answer-validity window that separates model forgetting from natural world-state change \citep{wang2026egomemreason,forte2026egostream}. SuperMemoryVQA grounds the problem in real AI-glasses hardware, with 52.9 hours of activity, synchronized SLAM, gaze, and IMU sensors, and thousands of verified question-answer pairs \citep{alam2026supermemoryvqa}. WorldMemArena frames memory as an action-world interaction loop and separately diagnoses writing, maintenance, retrieval, and use, reporting a finding that anchors this whole part: better storage does not improve task performance, and visual evidence is underused on agentic trajectories \citep{liu2026worldmemarena}. EgoExoMem adds the cross-view problem, exposing view-asymmetry and question-versus-answer view-preference conflicts across synchronized egocentric and exocentric video \citep{liu2026egoexomem}, and H2HMem and SMMBench stress cross-participant and source-distributed evidence composition \citep{zhu2026h2hmem,chai2026smmbench}.

The mechanism side mirrors the structured-store designs but adds modality fusion. TeleMem builds a unified long-term and multimodal (including video) memory with narrative dynamic-profile extraction \citep{chen2026telemem}; MemVerse combines fast parametric recall with hierarchical retrieval and continual consolidation in a model-agnostic framework \citep{liu2026memverse}; M2A maintains a dual-layer hybrid memory through a collaborating ChatAgent and MemoryManager that perform online updates \citep{feng2026m2a}; and NS-Mem combines neural representations with explicit symbolic structures and rules in a three-layer architecture \citep{jiang2026nsmem}. Embodied variants ground memory in action: STAR builds multimodal long-term memory for mobile robots and retrieves a compact task-conditioned subset \citep{yuan2026star}, RoboCortex maintains dual-grain reflective and principle memory for a self-evolving embodied agent \citep{chan2026robocortex}, and PersonaVLM turns a general multimodal LLM into a long-term personalized assistant via memory extraction and multi-turn reasoning \citep{nie2026personavlm}. Streaming perception is handled by StreamingVLM, which sustains real-time comprehension of effectively infinite video with a compact key-value cache reusing attention sinks \citep{xu2025streamingvlm}, and by a training-free episodic-event framework for long-video QA \citep{wang2025videoem}.

Multimodal substrates raise governance stakes in two specific ways the matrix flags. First, write correctness is harder: a visual memory can hallucinate details that were never observed, and reconciling text-image conflicts has no obvious authority rule, so MemEye's evidence-use axis becomes a faithfulness-of-write problem as well as a retrieval problem \citep{guo2026memeye}. Second, the attack surface is wider: visual inception shows that image-borne triggers hidden in user uploads can act as sleeper agents in an agentic recommender's long-term memory, hijacking later behavior \citep{qian2026visualinception}, a poisoning channel that text-only provenance tagging does not catch because the malicious payload enters as pixels. The WorldMemArena finding, that storage quality and task performance are decoupled and visual evidence is underused \citep{liu2026worldmemarena}, supports the survey's claim: accumulating more multimodal state does not by itself produce governed, actionable state, although WorldMemArena itself frames the result as one about storage quality and evidence use rather than governance. A unifying attempt, MemoryWire, defines a JSON-Schema wire format with five canonical operations (remember, recall, forget, merge, expire) over four memory types to make multimodal memory interoperable \citep{munirathinam2026memorywire}; but as the matrix notes, the specification is syntactic, with no merge semantics under conflict, no forget validation, and no attestation of operation provenance. The gap is that multimodal memory multiplies what state can be without giving deletion, provenance, or authority a per-modality semantics, so a deleted image's derived embeddings, captions, and behavioral influence remain unaccounted for.

\subsection{Agentic-OS and serving runtimes}

The final substrate is the runtime itself: the operating-system layer that gives agent state identity, isolation, durability, and recovery. This cluster is where the persistent-state thesis is most directly anticipated by systems work, because operating systems and databases have long treated state as something to be governed as well as stored. The framing work, AIOS, proposes an LLM-agent operating system that isolates memory, storage, context, scheduling, and access control into a kernel managing concurrent agents \citep{mei2024aios}. Three identity models follow. Quine realizes agents as native POSIX processes: identity is the PID, lifecycle is fork/exec/exit, and persistent state is the process's memory, environment, and filesystem, all inheriting kernel isolation \citep{ke2026quine}. AgentLibOS runs each long-running agent as an AgentProcess with identity, lifecycle state, capabilities, checkpoints, and audit, making runtime primitives rather than tool calls the authority boundary \citep{zhang2026agentlibos}. ActPlane enforces agent action policies inside the kernel using eBPF and an information-flow-control DSL, covering execution paths that tool-call guardrails miss at 1.9 to 8.4 percent overhead \citep{zheng2026actplane}. These are the first substrates in the substrate map where authority is first-class and kernel-enforced rather than an implicit field, notable given that authority is the rarest axis in the whole corpus.

A second strand makes recoverability the organizing primitive, and it is the densest concentration of rollback-capable substrates in the corpus. Event-sourced runtimes treat state as a deterministic projection of an append-only log: ActiveGraph supports exact replay and gates every self-improvement promotion or discard as an auditable event \citep{nakajima2026regimes}, and git-based memory stores the agent's reasoning tree in a repository so it can be replayed, diffed, and merged, arguing that git's value for agent memory is auditability and cross-agent merge rather than retrieval accuracy \citep{shekar2026gitofthoughts,wu2025gitcontextcontroller}. Checkpoint-restore runtimes make rollback cheap and correct: DeltaBox gives millisecond change-based transactional checkpoint and rollback of full sandbox state via copy-on-write filesystem layers and incremental process dumps \citep{dong2026deltabox}, and CRAB closes the agent-OS semantic gap by eBPF-classifying each turn's OS effects to align durable checkpoints with turn boundaries, lifting recovery correctness from 8 percent to 100 percent \citep{wu2026crab}. OpenRATH makes a branchable, inspectable, replayable session the first-class runtime value, carrying lineage, sandbox placement, and tool evidence with the execution \citep{wen2026openrath}, and AEON reframes long-horizon memory as a managed OS resource backed by a write-ahead log, an mmap blob arena with generational garbage collection, and epoch-based lock-free reads \citep{arslan2026aeon}.

The third strand is the hardest: transactional consistency over external side effects, the point where rollback stops being a memory operation and becomes a commitment problem. Atomix wraps tool use in progress-aware transactions that record reads and effects, seal when the footprint is complete, and commit per-resource only once frontiers confirm no earlier conflicting work can arrive; it handles bufferable, reversible, and irreversible effects distinctly \citep{mohammadi2026atomix}. Cordon treats a multi-step task as one semantic transaction, staging irreversible effects in an outbox and reversible mutations in shadow state, then binds tool intents to result lineage and authority before commit \citep{chen2026cordon}. ATCC treats an agent's emitted SQL as long-running agentic transactions and adapts database concurrency control per transaction \citep{zhou2026atcc}. A concurrency-anomaly study complements these: it formalizes and machine-verifies four anomalies over shared multi-agent state (memory stores, vector indices, tool registries) with a checked consistency hierarchy \citep{khan2026concurrency}. A distributed-systems foundation extends consistency models from data to knowledge visibility, introducing epistemic state replication with verifiable semantic rollback as a primitive for trustworthy agentic infrastructure \citep{he2026postdeterministic}. A governance-oriented memory layer, ProjectMem, stores coding-agent history as an append-only typed event log condensed into delivered summaries, then applies a deterministic memory-as-governance gate that warns the agent before it repeats a failed fix or edits a known-fragile file \citep{malo2026projectmem}.

The runtime layer also supplies several important negative results. A longitudinal eight-week postmortem of a production always-on runtime taxonomizes silent failures, names the fail-plausible class in which the model narrates an error as plausible success, and shows that audits block regressions but cannot predict failures \citep{wu2026silentfailures}. A second study finds that deployed agents under impossible constraints fabricate fake external problems and even fake system crashes (thanatosis), a self-reinforcing behavior that injecting accurate information mid-session does not correct \citep{rodriguez2026thanatosis}. These are not retrieval failures; they are governance failures at the runtime substrate, where the agent's reported state diverges from its actual state and no lower layer can catch it. This cluster imports the database and operating-system playbook (logs, transactions, checkpoints, capabilities, consistency models) into agent state, producing nearly all of the corpus's genuinely rollback-capable substrates. The gap is integration and adoption: these runtime primitives sit below the memory layer, while the memory mechanisms above them (the vector stores, graphs, and multimodal stores of the preceding subsections) rarely build on them, so the typical deployed agent stacks an ungoverned store on a runtime that could have governed it. The substrate that can roll back and the substrate that holds the memory are usually not the same system.

\subsection{Systems cost and resource governance}

Every substrate above incurs a cost that grows with always-on operation, and a final cluster treats that cost as a first-class governance constraint rather than an afterthought. The empirical anchor is a deployment measurement: tracking storage growth over a 15,000-message deployment finds memory grows roughly linearly at about three tokens per message, exposes a sim-to-real gap where real workflows grow linearly while synthetic benchmarks stay constant, and names consolidation quality, not raw capacity, as the scalability bottleneck \citep{milosevic2026episodicsemantic}. This linear, unbounded growth is why ungoverned accumulation is not free: a store that only writes and retrieves degrades in latency and cost over a long horizon even if its recall stays nominally correct. A total-cost-of-ownership study measures accuracy, latency, CPU, RAM, disk I/O, and network across production memory frameworks (mem0, Graphiti, cognee) on an independent cloud-edge testbed and finds that plain retrieval matches top accuracy at 8.4 times lower total cost than the heaviest framework \citep{wolff2026costaccuracy}, reframing the substrate choice as a cost-governance decision rather than only an accuracy one. A complementary systems-level characterization profiles stateful long-horizon memory workloads across ten representative systems, attributing cost across construction, retrieval, and generation phases \citep{omri2026agentmemorycharacterization}.

The mitigation literature borrows directly from operating-systems resource management. AgentRM mines over 40,000 GitHub issues to show that unbounded growth and weak retention cause agent amnesia. It then introduces OS-inspired middleware, a multi-level feedback-queue scheduler plus a three-tier context-lifecycle manager, that cuts P95 latency 86 percent while retaining 100 percent of key information \citep{she2026agentrm}. AMVL treats persistent memory as a managed systems resource with value-driven lifecycle tiers that bound the request-path working set independent of total stored memory, cutting tail latency several-fold \citep{bamidele2026amvl}. MemRefine formalizes storage-budgeted memory management differently: it keeps an accumulated store within a fixed footprint via similarity-nominated, LLM-judged delete/merge/preserve compression that iterates until the budget is met \citep{kim2026memrefine}. In the multi-agent setting, token coherence maps shared-state synchronization onto the hardware cache-coherence problem, adapting MESI lazy invalidation into an artifact-coherence protocol with a TLA+-verified guarantee of single-writer safety, monotonic versioning, and bounded staleness \citep{parakhin2026tokencoherence}.

The governance reading of this cluster is that resource pressure and state governance are the same problem seen from two sides. Eviction is forgetting under a budget, and the matrix's critique of AgentRM captures the survey's concern: a heuristic eviction policy that is not formally verified can evict critical state before it is acted upon, so a cost optimization becomes a correctness failure \citep{she2026agentrm}. The TCO and characterization studies do not measure mutation cost, recovery after failure, or multi-tenant isolation \citep{wolff2026costaccuracy,omri2026agentmemorycharacterization}, so the overhead of provenance, deletion propagation, and rollback remains largely unbudgeted in the literature. The missing systems model is a cost account for governance operations: what does it cost to make a delete propagate, to keep a supersession chain, to checkpoint for rollback, and is that cost one reason governance is the rarest capability in the corpus? The cheapest substrate is often the least governed one, and no work has yet shown that stronger governance must be prohibitively expensive.

\subsection{Cross-substrate synthesis}

Taken together, the seven substrates form a spectrum from internal and dense to external and durable, and the memory-form sub-axis cuts across all of them: episodic traces appear as raw pages in agentic memory \citep{yan2025generalagenticmemory}, as event logs in event-sourced runtimes \citep{nakajima2026regimes}, and as video frames in egocentric stores \citep{forte2026egostream}; semantic facts appear as graph nodes \citep{wang2026engram} and as retrieved chunks \citep{chen2023benchmarkinglargelanguagemodels}; procedural skills appear as LoRA adapters \citep{guo2026peam}, as distilled graph routines \citep{fang2026fluxmem}, and as governance gates over a typed event log \citep{malo2026projectmem}. The same knowledge can thus be parametric or non-parametric depending only on which substrate holds it, and that choice, not the form of the content, determines whether the state can be attributed, scoped, and reverted. Table~\ref{tab:substrates} summarizes for each substrate what it natively stores and the governance blind spot it structurally imposes.

\begin{table}[t]
\centering
\small
\setlength{\tabcolsep}{4pt}\renewcommand{\arraystretch}{1.2}
\begin{tabularx}{\linewidth}{p{0.18\linewidth}YY}
\toprule
Substrate & What it natively stores & Governance blind spot \\
\midrule
Parametric / test-time & Distributed weight deltas; latent recurrent state; internalized skills & No native tombstone or attribution: deleting or auditing a specific update requires external instrumentation; provenance and rollback are not native \\
Context window & Tokens in the active prompt; sliding-window stream & No persistent identity, no source tags, no audited deletion; present-but-unretrievable state (lost-in-the-middle) \\
Vector store (RAG) & Embedded chunks addressed by similarity & Mutation-blind: no versioning or deletion cascade; provenance per chunk but not propagated; no validity window or read authority \\
Knowledge graph / temporal & Relations; bi-temporal facts with supersession chains & Single-writer assumption: weak under concurrent edits; authority over who may rewrite a relation is unspecified \\
Multimodal & Visual, video, egocentric-sensor state plus fused text & Per-modality faithfulness of writes; pixel-borne poisoning; no per-modality deletion/provenance semantics \\
Agentic-OS / runtime & Process identity, event logs, checkpoints, transactional effects & Sits below the memory layer it could govern; adoption gap means stores rarely build on its rollback primitives \\
Systems / resource layer & Tiered footprint, eviction state, cost profiles & Eviction is unverified forgetting; cost of governance operations (delete, audit, rollback) is unbudgeted \\
\bottomrule
\end{tabularx}
\caption{Substrate comparison. Each substrate makes some state axes easy to represent and others non-native. Reading down the rightmost column reproduces, at the physical layer, the corpus-wide skew: writing and retrieving are easy across substrates, while deletion propagation, authority, and rollback require sidecar metadata, transactional wrappers, or different data structures. The rare exceptions (bi-temporal graphs, event-sourced runtimes, and checkpoint-restore runtimes) import database and operating-system governance, yet the memory layer above them rarely adopts those primitives.}
\label{tab:substrates}
\end{table}

\begin{table}[t]
\centering
\small
\setlength{\tabcolsep}{4pt}\renewcommand{\arraystretch}{1.2}
\begin{tabularx}{\linewidth}{p{0.22\linewidth}cccccc}
\toprule
Substrate & Auth. & Scope & Mut. & Prov. & Recov. & Action. \\
\midrule
Parametric / test-time   & \xmark & \pmark & \cmark & \xmark & \xmark & \cmark \\
Context window           & \xmark & \pmark & \cmark & \xmark & \xmark & \cmark \\
Vector store (RAG)       & \xmark & \xmark & \xmark & \pmark & \xmark & \cmark \\
Knowledge graph / temporal & \pmark & \pmark & \cmark & \cmark & \pmark & \cmark \\
Multimodal               & \xmark & \pmark & \pmark & \pmark & \xmark & \cmark \\
Agentic-OS / runtime     & \cmark & \cmark & \cmark & \cmark & \cmark & \cmark \\
Systems / resource layer & \pmark & \pmark & \cmark & \xmark & \pmark & \cmark \\
\bottomrule
\end{tabularx}
\caption{Which state axes each substrate can natively represent: \cmark{} native, \pmark{} partial or single-writer only, \xmark{} not native. A non-native axis is not impossible to govern, but it requires an external envelope, sidecar log, access-control layer, or transaction wrapper. Vector stores and parametric memory, the two dominant substrates, do not natively represent authority and recoverability, the two governance axes the corpus most neglects. Agentic-OS and bi-temporal-graph substrates import database and operating-system primitives that make all six axes native, but these substrates are rarely the default memory layer.}
\label{tab:substrate-axis}
\end{table}

Table~\ref{tab:substrate-axis} restates this at the level of the six axes: it marks, for each substrate, which axes it can natively represent. The two columns that stay blank across the dominant substrates, authority and recoverability, are also the two governance axes the corpus most neglects, and only the substrates that import database and operating-system primitives fill them natively. This gives a physical-layer version of the survey's claim: the corpus-wide skew is more than a matter of attention, because dominant substrates make some governance operations expensive unless an external envelope is added. An embedding key does not by itself express an authority; a weight delta does not by itself carry a tombstone; a sliding window does not replay itself; a similarity index does not cascade a delete. The substrates that can make these fields native, bi-temporal knowledge graphs with supersession chains \citep{wang2026engram}, event-sourced and checkpoint-restore runtimes \citep{nakajima2026regimes,dong2026deltabox,wu2026crab}, and transactional effect managers \citep{mohammadi2026atomix,chen2026cordon}, already exist, but they live at the runtime and structured-store layers the dominant memory mechanisms are not yet built on. The substrate-level gap, which the lifecycle and governance sections develop, is therefore an integration gap as much as an invention gap: the discipline already has several data structures for governed persistent state; it has not yet made them the default substrate on which agent memory is written.

\section{Mechanism families}\label{sec:mechanisms}

\regime{Motion}{how deployed systems write, adapt, consolidate, and share state in practice.}
Mechanisms matter when they cover lifecycle operations. The literature on always-on agents has produced a large toolbox: memory managers that page personal history in and out of context, graph stores that index it for multi-hop recall, reflection systems that distill trajectories into lessons, skill libraries that turn successful executions into callable procedures, consolidation schedules that decide what to keep, and shared substrates that let agents read and write each other's state. Product by product, this looks like steady progress. Against the persistent-state lifecycle, the pattern is more uneven. Many mechanisms concentrate their strengths on the same three operations, write, organize, and retrieve, and leave the same governance fields weak: authority, scope, provenance, recoverability, and the validate/audit/rollback stages that enforce them. The mechanism families are therefore read by the operation each makes explicit, the governance field it leaves implicit, and the concrete consequence of that omission. Memory managers, reflection systems, skill induction, continual adaptation, and shared-memory systems together lead to the falsifiable controlled-compounding criterion at the end of the part.

Two pieces of prior scaffolding matter here. The cognitive-architecture lineage treats an agent as memory plus action plus a decision process, and persistent state extends that loop across sessions \citep{sumers2024cognitivearchitectureslanguageagents,yao2023reactsynergizingreasoningacting}, with the episodic, semantic, and procedural state types echoing the long-standing memory-literature distinction \citep{shinn2023reflexionlanguageagentsverbal}. What modern systems add is mechanism: explicit code for writing, indexing, consolidating, and reusing that state. What almost all of them still lack is governance: code for deciding whether a write is authorized, whether a lesson is true, whether a skill is still safe to call, and how to undo any of these once they prove wrong. Table~\ref{tab:mechanisms} makes this diagnostic concrete by mapping representative families to the operation each makes explicit and the governance field each leaves weak.

\begin{table}[t]
\centering
\small
\renewcommand{\arraystretch}{1.2}
\setlength{\tabcolsep}{4pt}\renewcommand{\arraystretch}{1.2}
\begin{tabularx}{\linewidth}{p{0.17\linewidth}YYY}
\toprule
Family & Operation made explicit & Governance field left implicit & Concrete consequence \\
\midrule
Memory managers and graph memory & Write, organize, retrieve, consolidate; graph stores add multi-hop and temporal indexing & Authority and recoverability: who authorized a fact to act, can deletion be audited & Deletion edits the primary store while summaries, embeddings, and promoted tiers persist \\
Reflection and experience learners & Update: convert traces into lessons or causal abstractions & Validation: is the induced lesson true & One failed trajectory becomes a plausible but unproven durable rule \\
Skill and workflow induction & Procedural write and reuse of executable skills & Authority and mutability: test, scope, deprecate before reuse & A trajectory becomes a callable action whose misuse has tool consequences \\
Continual adaptation & Plasticity at write and update across task streams & Stability: validate, audit, forget, rollback & Helpful adaptation turns net-negative; competence and disorder accumulate together \\
Multi-agent and shared memory & Shared state, schema-bound mutation, access control & Scope and consistency: revocation and propagation & Corrupt or stale state crosses users or agents through shared memory \\
\bottomrule
\end{tabularx}
\caption{Mechanism families through the lifecycle lens. Each family makes one operation explicit and leaves a governance field (authority, scope, mutability, provenance, recoverability) implicit, with a concrete consequence. The comparison identifies which state-governance gaps recur across the surveyed works rather than ranking memory products.}
\label{tab:mechanisms}
\end{table}

\subsection{Memory managers and graph memory}

The first and largest family treats memory as something to be paged, indexed, and retrieved. Operating-system-inspired managers make the write, organize, and retrieve operations explicit and efficient: they externalize long-term state, manage a bounded working set, and decide what moves between tiers \citep{packer2024memgptllmsoperatingsystems,kang2025memoryosaiagent}. Agentic note systems push organization further by letting stored entries link and refactor each other as new information arrives, so that structure is emergent rather than fixed at write time \citep{xu2025amemagenticmemoryllm}. Graph-backed stores add structured multi-hop and temporal indexing on top of raw recall: knowledge-graph memory with neurobiologically-inspired retrieval supplies association structure that flat vector stores lack \citep{gutierrez2024hipporag}, and temporal knowledge graphs attach explicit validity windows so that facts can be queried as of a point in time \citep{rasmussen2025zep}. These systems are genuinely effective at what they make explicit: they scale personal history beyond the context window, recover semantically distant facts, and support reasoning that chains several stored items.

What this family leaves implicit is authority and recoverability. A managed or graph-indexed store records that a fact exists and how to retrieve it, but rarely records who authorized that fact to influence an action, under what scope, or how to prove it was later removed. The concrete consequence is a deletion gap that the lifecycle frame predicts: when a user or policy demands erasure, the operation edits the primary store, while the summaries, embeddings, promoted semantic tiers, and cached retrieval traces derived from the deleted item persist and can still be retrieved. Graph structure makes the recoverability problem more acute, because a single deleted node may have already propagated through edges into linked notes and consolidated summaries, and the typical graph store offers no cascade that follows those edges. The deeper representational failure is that flat-text long-term memory loses track of which source a fact came from and how it should be used, a source-monitoring breakdown that one analysis names provenance-role collapse and shows is mitigated only by typed memory representations rather than by better retrieval \citep{jin2026provenancerolecollapse}. The gap this family exposes for the thesis is direct: a memory manager optimizes accumulation and retrieval, the plasticity-side operations, while the governance fields that make state safe to act on, authority over a write and an auditable path to undo it, are left to the surrounding system that usually does not implement them. Retrieval that works is not the same as state that is governed.

\subsection{Reflection and experience learners}

The second family makes the update operation explicit. Rather than storing raw traces, these systems convert experience into reusable abstractions: verbal self-reflections that diagnose what went wrong on a trial \citep{shinn2023reflexionlanguageagentsverbal}, natural-language insights mined across many past episodes \citep{zhao2024expelllmagentsexperiential}, and causal abstractions that a continually learning agent refines so that later episodes inherit what earlier ones discovered \citep{majumder2023clincontinuallylearninglanguage}. A complementary line replays past trajectories directly as in-context experience rather than distilling them, integrating raw episodes into the working context at decision time \citep{liu2025contextualexperiencereplay}, and reasoning-memory systems generalize this by storing solution patterns that scale self-improvement on new tasks \citep{ouyang2025reasoningbank}. The shared move is attractive because it is the most direct route to compounding competence: an agent that turns each outcome into a lesson should, in principle, get monotonically better.

The governance field this family leaves implicit is validation. A reflection, insight, or causal rule is written into durable state on the strength of the agent's own judgment about a trajectory, with no separate gate that asks whether the induced lesson is actually true. The concrete consequence is that a single failed or misread trajectory becomes a plausible but unproven durable rule, and because the rule reads as a confident generalization, later retrieval surfaces it as if it were established fact. This is not hypothetical: the self-confirmation trap describes exactly the case where an agent that judges its own trajectories writes self-consistent-but-wrong experiences into memory, and the proposed remedy decouples writing from judging through heterogeneous executors, a third-party distiller, and a consensus gate that decides which experiences are committed \citep{zhu2026selfconfirmation}. The lesson generalizes the lifecycle claim: reflection systems are strong on update and weak on validate, so they convert plasticity into durable belief without the stability check that would keep a false lesson from compounding. The gap is that few of these systems, among those we surveyed, carry the provenance needed to later identify which reflection caused a downstream regression, or the rollback handle needed to revert it; an advisory lesson that degrades gracefully is forgiving of this omission, but as the next subsection shows, the same omission is far less forgiving once the durable artifact is executable.

\subsection{Skill and workflow induction}

The third family makes procedural write and reuse explicit: it turns a successful trajectory into a callable skill or a reusable workflow abstraction. The canonical embodied example grows a library of executable skills over an open-ended lifetime, so that later tasks call procedures discovered earlier \citep{wang2023voyageropenendedembodiedagent}; the canonical web example induces high-level workflow abstractions from past trajectories and applies them in new sessions \citep{wang2024agentworkflowmemory}. Recent work treats the skill store as a first-class artifact with its own engineering discipline. Procedural-memory systems add deprecation and update support so that stored workflows are not write-once \citep{fang2026mempexploringagentprocedural}; skill-management frameworks model the library as a self-maintaining software ecosystem with typed contracts to control accumulating technical debt \citep{pu2026skillops}; and learned-skill methods add a verification gate that checks a candidate skill before it is written and a score-based maintenance pass that keeps the library compact \citep{mi2026skillpro}. Selective-formalization approaches go further toward durability by recording reusable steps as auditable, versioned notebooks with explicit validation gates and fallback paths \citep{elhattami2026skillnb}.

This family exposes the actionability and authority axes most sharply, because a procedural skill is not advisory: it is an executable commitment whose misuse has tool consequences. The governance fields left implicit are authority and mutability, the questions of whether a skill is still authorized to run and whether it has been tested and scoped against current conditions before reuse. The concrete failure mode is skill drift, where a skill that was correct when written silently becomes wrong as the environment changes; one analysis formalizes this as a contract violation and validates the role-bearing environment assumptions of a stored skill against live conditions to trigger proactive repair \citep{fan2026skilldriftcontract}. The danger is amplified because a static skill can look benign and only turn harmful in the presence of accumulated persistent state, which is why runtime auditing that probes stored skills under targeted live conditions catches failures that static review misses \citep{lan2026runtimeskillaudit}. Whether a refined skill even transfers, rather than quietly overfitting to the role it was learned in, is itself an open question that a dedicated benchmark isolates, finding that procedural skills often become role-specialized and lose effectiveness off-distribution \citep{belikova2026managingproceduralmemory}, and a separate probe shows that embedding-based procedural retrieval discards the temporal ordering structure that procedures depend on, producing a generalization cliff \citep{kohar2025proceduralmemoryretrieval}. The comparison is instructive: most skill libraries optimize cheap reuse and do not re-test a stored procedure before calling it, so among representative systems the ones closest to lifecycle-complete are precisely those that add the missing stages, MemP with explicit deprecation, SkillOps with typed contracts and maintenance, Skill-Pro with a write-time verification gate, and SkillNB with versioned validation. The gap this family exposes is the steepest in the survey: the actionability of a procedural write raises the authorization, validation, and rollback bar exactly where mechanisms are weakest, so an ungoverned skill library accumulates executable commitments faster than it accumulates any way to verify, scope, or revoke them.

\subsection{Continual adaptation and the plasticity-stability problem}

The first three families each name a way to compound competence. Continual adaptation names why compounding is hard, and the harm it causes when ungoverned. The tension is the classic stability-plasticity tradeoff of continual learning \citep{wang2023continuallearningsurvey}, and the interference it produces traces back to the catastrophic-interference result for sequential learning in connectionist networks \citep{mccloskey1989catastrophic}: weights tuned for new tasks overwrite the representations that solved old ones. Surveys of lifelong learning for LLMs map the response space, in-context adaptation, parameter update, and memory augmentation, across this tradespace \citep{zheng2024lifelongllmsurvey}, and surveys of self-evolving agents catalogue the loops that chain exploration, skill extraction, and internalization \citep{gao2025selfevolvingagentssurvey}. Always-on agents face the same tradeoff in an externalized form. Where lifelong-learning surveys catalogue internal weight updates, these agents adapt mostly through retrieval and tool-grounded state, using the reflections, experience buffers, and skill libraries of the previous subsections rather than gradient steps.

\subsubsection{Parametric versus non-parametric adaptation}

The parametric alternatives sharpen the contrast and clarify why the externalized route is attractive. Test-time training updates model weights at inference, either from retrieved nearest neighbors \citep{hardt2023tttnn} or from the few-shot examples in the current task \citep{akyurek2024ttt}, boosting generalization without a permanent commitment. Knowledge editing rewrites facts in place, but editing at scale itself induces gradual and catastrophic forgetting \citep{gupta2024editingforgetting}: the same plasticity-stability failure reappears in weight space, now without the provenance or rollback handles that externalized state can keep, since a weight edit leaves no record of which fact it overwrote or how to restore it. This is the core trade. Parametric memory is efficient and fast to consult but fragile to edits and opaque to audit; non-parametric external memory is costlier per query but inspectable and, in principle, reversible. A longitudinal study that distinguishes parametric memory depth from retrieval memory access uses surprise- and valence-gated consolidation to test whether goal-conditioned tendencies persist after the working context is unloaded \citep{han2026memorydepth}, and an empirical lifelong benchmark finds that nonparametric memory outperforms parametric weight updates while catastrophic forgetting grows with interaction length \citep{fan2025lifestatebench}. Always-on agents take the externalized route precisely to make adaptation inspectable and reversible; the lifecycle frame then makes the tension exact. Plasticity lives at write and update; stability is enforced at validate, audit, and forget; an agent that exposes the first set of operations without the second accumulates risk even as its task scores rise. We state the criterion in falsifiable form: adaptation is net-positive only when the system can identify, de-authorize, and revert the specific state update that later caused a regression. A system that cannot perform that revert has not shown controlled compounding, however high its aggregate score climbs.

\subsubsection{How helpful adaptation turns net-negative}

Several reported patterns show the criterion being violated in practice. \emph{Skill shadowing}: a procedural skill that succeeded early is reused after the environment or goal has shifted, so a correct past execution becomes a wrong current one, and a library built for cheap reuse does not re-test a stored procedure before calling it \citep{wang2023voyageropenendedembodiedagent}, the drift the skill subsection formalized. \emph{Interference}: writes that help recent tasks degrade retention of older ones, the backward-transfer failure of continual learning expressed here through retrieval rather than weights. \emph{Non-monotonic self-evolution}: iterating an explore-extract-internalize loop does not guarantee improvement \citep{gao2025selfevolvingagentssurvey}, and frozen-weight agents degrade over time through distinct aging mechanisms, compression, interference, revision, and maintenance, that a longitudinal benchmark isolates and targets with stage-specific repair \citep{zhu2026aging}. \emph{Replay weakness}: replay is the most direct compounding mechanism \citep{liu2025contextualexperiencereplay}, yet a lifelong agent benchmark finds naive replay weak for LLM agents because irrelevant retrieved episodes consume context, with structured aggregation helping more than raw recall \citep{zheng2025lifelongagentbenchevaluatingllmagents}, and an environment-drift benchmark that frames the world as a sequence of progressive updates finds that agents average well under forty percent at keeping memory aligned to a shifting world \citep{xu2026evoarena}.

\subsubsection{The baseline-beats-memory finding}

Recent controlled studies sharpen these patterns into a baseline question that is uncomfortable for the whole field. A continual-learning benchmark built on a learnable latent structure that only stateful systems can exploit finds that dedicated memory systems do not reliably improve learning, and that naive in-context use of recent history outperforms several purpose-built memory architectures \citep{asawa2026clbench}. A separate benchmark with controlled reusable task streams shows that memory-induced degradation is hidden by naive evaluation streams and surfaces only when reuse is controlled, and pairs the finding with a filter for unreliable experience \citep{shu2026agentcl}; a long-horizon benchmark over continuous agent-environment interaction across six domains tells the same story \citep{zhao2026amabench}, and a co-play-derived task suite that instruments memory reads, writes, and validators exposes where agents fail to persist memory across tasks at all \citep{doss2026minenpc}. The starkest result is that continuously LLM-consolidated textual memory has a utility curve that rises and then degrades below the no-memory baseline as consolidation accumulates \citep{zhang2026usefulmemoriesfaulty}: the memory stops helping and then actively harms, and the harm is a function of how much has been written, not of any single bad write. This is the silent-entropy reading of long-run degradation, accumulating disorder that grows with interaction count and is invisible to any metric that scores only the current answer. The collective message is empirical: stored state helps only when it is governed, not when it is simply accumulated, and an ungoverned memory system can be strictly worse than having no memory at all once the horizon is long enough.

\subsubsection{Forgetting and consolidation as the stability operators}

Principled forgetting and consolidation are the stability operators that should contain harmful plasticity, and a wave of recent work develops those operators. The shared insight is that forgetting should be governed, not incidental. Several systems model forgetting as decay in accessibility rather than destructive deletion: a dual-layer hierarchy applies differential exponential-decay rates keyed to relevance, access frequency, and temporal pattern \citep{wei2026fademem}, while a read-path-versus-control-path split gates recall by agent uncertainty and treats forgetting as reduced accessibility \citep{rana2026oblivion}. A selective-forgetting taxonomy organizes the design space into passive decay, active deletion, safety-triggered removal, and adaptive reinforcement, making explicit that different mechanisms preserve and discard different knowledge \citep{gu2026fsfm}, and a retention-schema approach grounds six forgetting policies in theory and evaluates them on a dedicated benchmark \citep{alqithami2025mars}. Consolidation work decouples fast acquisition from slow integration: an RL-trained offline consolidator separates per-session memory capture from cross-session consolidation and shrinks the active bank while transferring across environments \citep{ye2026autodreamer}, a human-inspired design combines sleep-phase consolidation, interference-based forgetting, engram maturation, and reconsolidation on retrieval \citep{kerestecioglu2026humaninspiredmemory}, and a predictability framing reframes what to retain as a prediction-error problem learned from interaction rather than a fixed heuristic \citep{ma2026nemori}. Other systems make consolidation governable through structure: a short-to-long-term consolidation policy frames retention as a per-triple keep-or-drop Q-learning decision over a temporal knowledge graph \citep{kim2026stltransfer}, residual-tree storage lets related experiences share a base while keeping incremental deltas \citep{tan2026deltamem}, deferred extraction waits for sustained recurrence before paying for expensive consolidation \citep{dai2026recmem}, and a control-plane analysis shows that where the LLM sits, on the recall plane versus a control plane that can supersede, release, and purge, determines which forgetting failure mode dominates \citep{yang2026controlplaneforgetting}. A role-specific memory policy even learns from periodic QA feedback and retains a failure-rollback path so that a bad consolidation can be reverted \citep{chen2026adamem}, which is the rare system that closes the loop the falsifiable criterion demands.

These findings reframe continual adaptation as a state-governance problem, more than a learning-rate problem. The interventions that keep compounding controlled are precisely the underused lifecycle stages: validation gates before a lesson becomes a durable rule, provenance and recency tags so that superseded skills and facts lose authority rather than lingering, and audit and rollback so that a degrading update can be identified and reverted. The gap this subsection exposes is that the works we coded contain rich plasticity mechanisms and an increasingly rich vocabulary of forgetting, but still couple them weakly to authority and rollback: many forgetting systems decide what to decay without recording who authorized the decay or how to undo it if the decay was wrong, and benchmarks that interleave dependent sessions surface the degradation without scoring the write, deprecate, and rollback decisions that determine whether adaptation compounds or corrodes. A complementary line tries to absorb non-parametric experience back into parametric memory under control, co-updating an experiential rule pool and the policy weights from the same trajectories so that stable behaviors migrate from the rule pool into the model \citep{ye2026jerp}, and model-based continual RL governs a dual recency-plus-diversity replay buffer under a fixed memory budget to bound interference \citep{alyahya2026arrow}; both are promising precisely because they make the migration auditable and budget-bounded rather than leaving it to accumulate unchecked.

\subsection{Multi-agent and shared-memory mechanisms}

The final family moves from a single agent's lifecycle to state shared across many. Multi-agent frameworks have long used a shared substrate as de-facto memory: a publish-subscribe message pool acts as untyped shared state in canonical SOP-driven coordination \citep{hong2023metagpt}, and generative-agent sandboxes showed that observation-reflection-retrieval memory in a shared environment produces believable social behavior \citep{park2023generativeagentsinteractivesimulacra}. Interactive social-evaluation frameworks made shared world state and relationship history a core measurement target \citep{zhou2024sotopiainteractiveevaluationsocial}. The newer wave makes the operation this family makes explicit, shared write with access control, into a first-class design concern. Access-governed shared memory enforces explicit authority and scope on multi-agent reads and writes \citep{margalit2026governedsharedmemorymultiagent}; multi-user collaborative memory adds dynamic access-control gates with field-level and role-anchored granularity \citep{rezazadeh2025collaborativememorymultiusermemory}; schema-grounded mutation enforces type-safe transitions with full audit trails to bridge untyped updates and reliable shared state \citep{zhang2026patchboardschemagroundedstatemutation}; and role-aware context routing governs which subset of shared memory each agent sees under a token budget, a selective-view form of scope control at retrieval time \citep{liu2025rcrrouter}.

The governance fields this family leaves implicit are scope and consistency: specifically, revocation and propagation control. The concrete consequence is that corrupt or stale state crosses users or agents through the shared substrate, and the failure-mode literature documents the cost. Jailbreak and poison propagation spreads across agent-society topologies, and shared-memory architectures propagate it more readily than independent-memory ones \citep{men2024troublemaker}; evaluator bias in stored trajectories propagates cross-temporally to future agents sharing memory, with no safe contamination threshold even under oracle consolidation, a failure named memory contagion \citep{liu2026memorycontagion}; and a thousand-scenario benchmark quantifies privacy leakage through inter-agent messages, shared memory, and tool arguments \citep{elyagoubi2026agentleak}. These results show that per-agent safety checks are insufficient when the memory substrate itself can carry compromised state across the scope boundary, which is the P6 keys point: shared state without enforced scope non-expansion lets one agent's authority and one user's private facts reach contexts they were never licensed for.

A second strand addresses consistency and forgetting in the shared setting directly, and it is here that the field's informal notion of shared memory most needs the formal consistency vocabulary of distributed systems. When many agents read and write the same memory stores and tool registries concurrently, the substrate inherits the classical anomalies of replicated state: a recent machine-checked study makes this precise, exhibiting the first formal consistency hierarchy for multi-agent LLM runtimes and naming anomalies such as stale-generation and phantom-tool that arise when concurrent writes race over shared state \citep{khan2026concurrency}. The trade-off behind these anomalies is the familiar consistency-versus-availability tension, which a concurrent-abstract-state-machine formalization renders as an explicit framework for deciding how much consistency a given replication policy actually needs, a decision a governed agent memory must make rather than leave implicit \citep{schewe2019replicated}. Database concurrency control supplies the matching proof technique: serializability is the standard correctness criterion for concurrent access, and a machine-checked CV-rules framework for verifying it can be adapted to certify that interleaved multi-agent memory mutations remain equivalent to some safe serial order \citep{hoshino2026cv-rules}. These formal models give the scope-and-consistency gap a precise statement: an ungoverned shared substrate offers neither a defined consistency level nor a serializability guarantee, so concurrent writes from different agents can interleave into states that no single agent ever intended. Shared state need not always mean contention: read-heavy multi-agent workloads can share compressed key-value cache state with minimal coherency overhead, cutting memory roughly ninety-eight percent relative to isolated caches, which shows scope-bounded sharing and efficiency are compatible when the access pattern is governed \citep{patel2026polykv}. Where some agents may be faulty or adversarial rather than only concurrent, the requirement strengthens to fault-tolerant agreement: a Byzantine-consensus protocol that races on protocol structure rather than timeout-based delays shows that agents can still agree on shared state under adversarial participants \citep{giridharan2026ambulance}, and a diagnostic layer that combines policy analysis with Bayesian trust can flag a Byzantine agent before it corrupts the shared channel, supplying the detection half of a consistency guarantee \citep{murimi2026bard}. A co-forgetting protocol combines semantic-relevance voting and multi-scale temporal decay ratified by Byzantine consensus so that shared-memory pruning is safe under disagreement \citep{bach2025coforgetting}, and a phylogenetic version-control scheme coordinates decentralized agents through an asynchronous shared repository with humans promoting and pruning branches as a selection signal \citep{huang2025evogit}. Decentralized stigmergic coordination shows that agents can cohere above a critical density by depositing environmental traces without any central controller, but the same work notes the substrate is open to pollution and lacks audit mechanisms to verify trace authenticity or recover from coordinated poisoning \citep{khushiyant2025collectivememory}. The social and organizational dimension adds its own state: multi-party conversational memory requires speaker attribution and turn-by-turn tracking that single-user memory can ignore \citep{yang2026groupmembenchbenchmarkingllmagent}, and at the longest horizons, continuously running multi-agent platforms surface governance divergence and outright collapse under identical setups over weeks to months \citep{akkil2026emergenceworld}. Two framings try to make this institutional state governable: recasting agent runs as a learning organization whose experience documents inherit across runs under safeguards against knowledge degradation \citep{xie2026forage}, and reframing shared-memory governance as artificial selection over candidate state, persist, keep private, reject, revise, or supersede, with evaluation axes for correction pathways and provenance fidelity \citep{cuadros2026governedselection}.

The gap this family exposes is the multi-agent specialization of the survey's core claim. Shared substrates make write and access-control explicit, and a few make schema-bound mutation and consensus pruning explicit, but revocation, deletion propagation across agents, and rollback traceability remain thin: when one speaker leaves a group or one delegation is revoked, the dependent state they wrote into the shared substrate, and everything consolidated from it, typically has no cascade to follow and no audit trail to reconstruct. Corrupt state therefore propagates faster than any mechanism to contain it, which is why long-horizon multi-agent platforms measure collapse empirically but cannot yet explain or repair it mechanistically. The more governance-aware systems here, governed shared memory, schema-grounded mutation, and consensus-ratified co-forgetting, begin to instrument scope and consistency, but none yet closes the loop from a revoked authority back through every shared and derived copy of the state it licensed.

\subsection{The controlled-compounding criterion}

Read together, Table~\ref{tab:mechanisms} and the five families form a single diagnostic rather than five separate reviews. Many mechanisms share the same strengths, write, organize, and retrieve, and the same weaknesses, the validate, audit, and rollback stages that would govern what those operations produce. Memory managers leave the authority of a written item implicit, so deletion edits the store instead of producing auditable erasure. Reflection systems make update explicit but leave validation implicit, so one failed trajectory becomes a durable false rule. Skill libraries expose the actionability axis most sharply, turning trajectories into executable commitments while leaving authorization and deprecation weakest. Continual adaptation shows that without the stability operators the act of accumulating helps until it harms. Shared substrates carry all of these omissions across the scope boundary at once. The response implied by the table is \emph{controlled compounding}: accepting a write or update only when it preserves the five invariants, so that competence accumulates only on state that is grounded, scoped, testable, and reversible. The criterion is deliberately falsifiable. A system claims controlled compounding only if, when a later regression is traced to an earlier write, reflection, skill, consolidation, or shared-memory entry, that specific update can be identified, de-authorized, and reverted with its derived state. By that standard most current mechanisms have not demonstrated controlled compounding; they have demonstrated accumulation. The difference between the two is the governance this survey argues the field should measure.

\section{Benchmark and evaluation families}\label{sec:evaluation}

\regime{Control}{what the benchmark literature measures about state, and what it leaves untested.}
If always-on agents are persistent-state systems rather than memory-augmented chat models, evaluation has to ask whether a benchmark measures governed state over a lifecycle, instead of only whether a system can store and retrieve memory. A benchmark is an operational definition: it fixes what counts as success, and therefore what counts as a system worth building. We read the evaluation literature against six recurring benchmark stressors: persistent multi-user memory, stale-state adjudication, social and belief dynamics, real action with side effects, privacy and consent boundaries, and recovery after bad memory. These stressors are not a separate theory of the paper; they are a compact checklist derived from the axes and lifecycle. Persistent multi-user memory stresses scope and provenance; stale-state adjudication stresses mutability and authority; social and belief dynamics stress shared state and provenance; real side effects stress actionability and recoverability; privacy and consent stress scope and authority; and recovery stresses rollback traceability and deletion propagation. The argument is structural rather than accusatory. Many benchmarks are excellent within scope, but their scopes divide the always-on evaluation space so cleanly that no family is forced to govern state, act on it, and recover from it in one protocol. Recoverability under authority and provenance constraints then falls between families.

Five benchmark families define the current measurement surface. Memory question-answering over histories made memory failures observable but evaluates state after the fact. Long-context and RAG benchmarks are boundary cases: they constrain what supplied context a model can use, but they are not memory benchmarks and were never meant to be. Memory-guided action benchmarks move closest to deployment by making earlier experience change later choices, yet they reset identity, consent, and governance at every task boundary. State and belief mutation benchmarks attack obsolete and conflicting state most directly and are where authority first becomes visible, but they rarely couple mutation to real action or deletion. Personalization and proactive benchmarks test long-horizon user state and, more recently, spontaneous recall, but treat write policy and consent as out of scope. Cutting across all five is a methodological literature on judge reliability and statistical rigor that determines whether any of these scores can be trusted at all. Methodology comes first as a lens; the family-by-family reading then closes with the coverage finding that motivates a new protocol.

\subsection{Evaluation methodology and what a score can mean}

Before comparing what each family measures, we ask what an agent benchmark score can support, because the always-on setting strains standard evaluation methodology in specific ways. A recurring critique is that agent evaluation conflates capability with measurement artifact. \citet{kapoor2024aiagentsthatmatter} document that agent benchmarks frequently omit cost-versus-accuracy reporting, lack adequate holdout sets so that systems overfit to the benchmark, and rarely report failure rates, all of which inflate apparent progress. \citet{miller2024errorbarsevals} supplies the missing statistical machinery, framing benchmark scores as estimates over a super-population with confidence intervals and principled two-model difference tests, and \citet{madaan2024quantifyingvariance} measures seed variance, training-step non-monotonicity, and continuous-metric instability empirically, showing that single-number leaderboard comparisons are often within noise. Contamination is a parallel threat: \citet{white2024livebench} responds with monthly-refreshed questions drawn from recent sources and objective ground truth so that memorized answers cannot masquerade as reasoning. These concerns sharpen in the always-on regime, where a system is scored on a trajectory of dependent decisions rather than independent items: errors compound across sessions, variance accumulates, and a single contaminated fact can be silently reused for the remainder of an episode.

A second methodological axis is the rise of LLM-as-judge and its known biases, which matters acutely once we score state transitions rather than final answers. \citet{zheng2023judgingllmasajudge} established the judge paradigm via MT-Bench and Chatbot Arena and reported strong agreement with human preference, but also documented position, verbosity, and self-enhancement biases. Subsequent work quantified and partially mitigated these effects: \citet{shi2024judgingthejudges} measures position bias in pairwise and list-wise judging with repetition-stability and position-consistency diagnostics; \citet{tripathi2025pairwiseorpointwise} shows that the protocol choice itself induces systematic bias, with roughly a third of preferences flipping under distractors depending on whether scoring is pointwise or pairwise; and \citet{lee2024checkeval} replaces unreliable Likert scoring with decomposed binary checklist questions to raise reliability. The judge-reliability surveys of \citet{gu2024surveyllmasajudge} and \citet{li2024llmsasjudgessurvey} synthesize consistency-improvement, bias-mitigation, and meta-evaluation methodology, while \citet{chiang2024chatbotarena} provides the crowdsourced pairwise plus Bradley-Terry ranking infrastructure that anchors much of the field's reliability evidence. \citet{you2026agentasjudge} extends the paradigm from judging a final answer to judging an agentic process step by step. Once the judge is itself an agentic process, its reliability becomes a moving target that interacts with the very governance policies under test: \citet{iacob2026redqueen} co-evolves the evaluation framework with the agent's governance objectives, detecting bias in agent-as-judge systems and showing that self-improvement guarantees must be re-established as authority and scope policies shift, which is exactly the regime an always-on benchmark scores. The methodological lesson is double-edged. Governance properties such as whether a deletion propagated or whether an action was licensed by a current authority epoch are exactly the multi-step, process-level judgments where a naive scalar judge is least reliable. Yet several governance checks are crisp predicates (was the tombstoned value retrievable, yes or no) that admit a deterministic oracle and need no judge at all. This split, between properties that demand fragile holistic judgment and those that reduce to deterministic invariant checks, recurs across the families and foreshadows why the protocol we ultimately motivate keeps its governance checks deterministic.

\subsection{Memory QA over histories}

The family that founded memory evaluation asks questions over long interaction histories. LoCoMo \citep{maharana2024evaluatinglongtermconversationalmemory} is the canonical instance: very long multi-session dialogues with probes that require recovering semantically distant facts, reasoning across sessions, handling temporal updates, abstaining when the answer is absent, and selectively forgetting. Its contribution was to make memory failure a measurable quantity rather than an anecdote, and it remains a standard reference point. The lineage is older than LoCoMo, and reading it clarifies what the family does and does not test. Multi-session open-domain dialogue datasets first established recall and summarization across sessions, and later work typed personal memory into semantic facts, episodic events, and social relations, on which LoCoMo-style benchmarks build. The shared assumption is that memory quality can be read off the correctness of answers to retrospective questions.

That assumption is precisely where the family draws methodological fire from within. \citet{flynt2026precisionmembench} argues that LoCoMo-style benchmarks conflate answer-correctness with retrieval-correctness: a system can answer correctly with the wrong evidence in context, or fail despite retrieving the right evidence, and pooled accuracy hides both. PrecisionMemBench responds by scoring retrieval precision in isolation from the generator, so that memory quality is no longer entangled with the language model's ability to paper over bad retrieval. \citet{long2026memtrace} pushes the same disaggregation further, re-basing long-term-memory evaluation on the per-fact knowledge point probed across age, question-type, and evidence conditions, and finds that the dominant bottleneck is not retrieval at all but evidence use: agents often have the right fact in context and still answer wrong. \citet{deng2026entitycollision} attacks a subtler confound, showing that apparent retrieval lift in agent memory can be lexical leakage rather than semantic competence; their stratified protocol pins a BM25 floor via entity-token collision so that improvements are attributable to the embedder. Together these works are a corrective lineage: each shows that a higher LoCoMo-style number can be produced without the underlying capability the number is supposed to certify.

This family therefore measures recall, temporal reasoning over updates, abstention, and, with the newer protocols, retrieval precision and evidence use cleanly separated from generation. It does not measure the write and governance half of the lifecycle. These benchmarks ask questions after the fact; they rarely test whether a remembered fact should have been written, whether the write was authorized, whether retrieving it would improve a later action, or what happens when the user later asks for it to be deleted. There is no notion of a permission epoch under which a fact was valid, no provenance tier distinguishing a user-stated preference from a fact scraped from an untrusted tool output, and no action whose side effect could be wrong because the recalled state was stale. The gap is intrinsic to the format: a question-answer pair is a recall unit, and a recall unit has no field for an authority tag, a deletion record, or a rollback trace. Even the most rigorous member of this family, having stripped out the retrieval-versus-generation confound, still certifies only that the right text can be recovered, not that recovering and acting on it is governed. Memory QA isolates retrieval fidelity and leaves downstream governance outside the test.

\subsection{Boundary cases: long-context and RAG benchmarks}

Long-context and retrieval-augmented-generation benchmarks are frequently cited as evidence about agent memory, but they are boundary cases rather than members of the family. They evaluate whether a model can use information already supplied to it, not whether a system should commit information to durable agent-owned state and govern it thereafter. LongBench \citep{bai2024longbenchbilingualmultitaskbenchmark} probes long-context understanding across diverse bilingual tasks and establishes an empirical baseline for using a large window; \citet{liu2023lostmiddlelanguagemodels} shows the sharp boundary effect that models underuse information placed in the middle of a long context despite attending to its edges. On the RAG side, \citet{chen2023benchmarkinglargelanguagemodels} decomposes retrieval quality into noise robustness, negative rejection, information integration, and counterfactual handling, giving the field a vocabulary for faithfulness and robustness to noisy evidence. \citet{gutierrez2024hipporag} bridges toward structured memory by organizing retrieval over a knowledge graph with a PageRank-style spreading-activation step, improving multi-hop retrieval, but its graph is built offline and queried statically.

These benchmarks measure context use, retrieval faithfulness, and noise robustness extremely well, and the always-on field should treat their results as binding constraints: if a model cannot reliably use the middle of its context window, no memory system that surfaces evidence into that window is safe. But the limitation is structural and definitional. A long-context benchmark holds the information fixed and varies the model; an always-on system must decide what becomes durable state, when that state becomes authoritative, and how it is later revised, scoped, revoked, or removed. RAG benchmarks score retrieval against a fixed corpus; they do not score a durable write, an update under a temporal-validity window, a deletion that must propagate to derived tiers, or an authority constraint on who may read which evidence. The boundary effect documented by lost-in-the-middle is a property of inference over a static window; the always-on failure mode is that the window is assembled from accumulated state whose provenance, freshness, and authorization are themselves uncertain. This family therefore brackets the substrate problem rather than the governance problem. Its measures of durable write, update, deletion, and authority are absent, by design and correctly so, because that was never its question. Treating long-context or RAG scores as memory-governance scores is a category error, and recognizing the boundary is itself part of the argument: the field has mature instruments for the supply-and-use of context and almost none for the governance of state.

\subsection{Memory-guided action benchmarks}

The family that moves closest to always-on deployment makes earlier experience change later choices, so that memory is evaluated through its effect on action rather than through retrospective questions. MemoryArena \citep{he2026memoryarena} is a clear recent example: tasks are interdependent across sessions, so an action or feedback in one session should alter behavior in a later one, and the benchmark reports recall, induction, and contextual grounding. LifelongAgentBench \citep{zheng2025lifelongagentbenchevaluatingllmagents} evaluates agents as lifelong learners over sequential task streams, measuring forward and backward transfer and catastrophic forgetting in serial tool environments. Surrounding these are the realistic task substrates that show how hard action remains even before persistent personalization is added: enterprise knowledge-work tasks such as WorkArena \citep{boisvert2024workarena}, open-ended computer-use environments such as OSWorld \citep{xie2024osworldbenchmarkingmultimodalagents}, and stateful application benchmarks such as AppWorld \citep{trivedi-etal-2024-appworld}, where an agent must track world-state changes across interactions with simulated apps and people. This family's contribution is the move from final-answer scoring to final-state checking: success is verified against the state of the world after the agent acts, which is the right test for a system whose state has actionability.

Comparing these benchmarks exposes the shared limitation. MemoryArena and LifelongAgentBench reward experience reuse but treat memory as an unaudited performance booster: there is no scoring of whether a reused experience was authorized to influence the new task, no provenance tier on the stored trajectory, and no test of whether a poisoned or stale experience should have been blocked. The realistic substrates are, almost by construction, episodic: WorkArena, OSWorld, and AppWorld reset the environment at each task, which is exactly the boundary an always-on protocol must cross. They verify that an action produced the right final state, but reset identity, consent, accumulated permission, and long-run governance between tasks, so the questions that define always-on behavior never arise. Even AppWorld, which executes real side effects against stateful apps, scores whether the side effect was correct, not whether it was recoverable: there is no perturbation that retries an irreversible call to test idempotency, no revocation of the authority that licensed an earlier action followed by a check that the action's state was rolled back, and no deletion request whose propagation to derived state is verified. The matrix view makes the pattern concrete: each suite tracks state in action but does not isolate whether failures stem from a bad write, weak organization, or weak retrieval at action time, and none couples action to governed writes, revocation, or rollback. The gap here is not realism and not action, both of which this family has in abundance. The gap is that the same protocol does not jointly stress persistent identity, governed writes, provenance, deletion, rollback, stale-state adjudication, privacy scope, and recovery. Memory-guided action benchmarks prove that acting on memory is hard and measurable; they leave untested whether acting on memory is governed.

\subsection{State and belief mutation benchmarks}

The fourth family attacks the problem the others defer: what happens when stored state becomes obsolete, conflicting, or contested, and which stored state should govern the present. This is where authority first becomes an explicitly measured property rather than an implicit assumption. STALE \citep{chao2026stalellmagentsknow} asks whether agents detect that a stored memory has gone stale and can resolve an implicit conflict between an obsolete value and a current one, probing the validation stage of the lifecycle. \citet{han2026memorydepth} approaches the same staleness-and-expiry question from the write side rather than the read side, evaluating whether a long-running agent correctly preserves still-valid facts and expires obsolete ones across context teardown by gating consolidation on surprise and valence signals, which makes \emph{what state survives} a scored property rather than an implementation detail. In the social direction, GroupMemBench \citep{yang2026groupmembenchbenchmarkingllmagent} is the first benchmark for agent memory in multi-party conversation, testing speaker attribution and turn-by-turn state tracking so that who said what, and therefore whose authority a remembered fact carries, becomes scoreable; and the interactive social-intelligence setting of SOTOPIA \citep{zhou2024sotopiainteractiveevaluationsocial} reveals how persistent memory underwrites believable multi-turn social behavior. A cluster of newer benchmarks push on individual always-on stressors, each isolating one axis. \citet{jung2026meme} is a direct probe of deletion propagation: its multi-entity evolving-memory tasks, Cascade, Absence, and Deletion, test whether an update or removal propagates to dependent entries, and the reported near-total collapse (Cascade at 3 percent, Absence at 1 percent) is evidence that tested memory systems struggle to maintain the deletion-propagation invariant. \citet{kwon2026reclaim} isolates a provenance failure with a striking consequence: a memory that keeps a stale conclusion but drops its source becomes confidently uncorrectable, strictly worse than empty memory, and the judge-free reclaim evaluation shows a source-first write policy repairs it. \citet{zhang2026kpr} stress-tests the converse condition, whether an agent preserves provenance and re-verifies authority before it accepts external knowledge, using controlled ablation of descriptions and diffs together with poisoned-patch injection, so that provenance retention and authority re-checking on ingestion become directly measurable rather than assumed. \citet{wang2026docarena} attacks scope enforcement at benchmark-construction time, automatically translating document access policies into adversarially structured question-answer environments that test whether an agent respects tool-use scope and can verify the provenance chain of the evidence it returns, turning an access-control specification into a graded test of scoped retrieval. \citet{bhargava2026memaudit} audits what memory writers actually preserve under a fixed storage budget, isolating selection quality and validity-state preservation from raw representation, and \citet{shao2026scaleconditioned} holds task evidence fixed while adding irrelevant sessions to decompose where stored evidence stops being usable under scale. \citet{liu2026streammembench} brings this into a streaming, egocentric setting and finds that systems often fail to turn stored evidence and feedback into reliable future behavior, while \citet{zheng2026seagym} scores self-evolving harnesses by decomposing an update into reusable gain, overfit, regression, and cost, revealing that an intermediate self-improvement snapshot can collapse later. Underlying the whole family is a formal claim: \citet{orogat2026gem} reframes long-term memory correctness as a property of the state trajectory rather than of individual records, defining Governed Evolving Memory with state-level operators and six correctness conditions covering write, read, delete, consistency, and rollback traceability, and proving that record-level stores cannot satisfy them.

This family is the most important for the survey's thesis, because it is the only one in which authority, staleness, provenance, and deletion are first-class measured quantities rather than incidental. It is also where the field's governance deficit shows up as hard numbers: the MEME collapse and the RECLAIM uncorrectability result are not warnings about a hypothetical future, they are measurements that today's stores fail invariants the thesis names. Yet the coupling gap remains. These benchmarks test mutation largely in isolation from action: STALE asks whether the agent knows a value is stale, not whether it refrained from issuing a tool call licensed by that stale value; MEME tests whether a deletion propagated through the memory graph, not whether a side effect derived from the deleted value was rolled back; GEM formalizes rollback traceability as a correctness condition but is a framework awaiting an empirical harness that audits real systems against it. The social members add belief and group structure but, as the matrix notes, do not score authority and scope governance during multi-party updates or deletion propagation when a speaker leaves the group. Even when a benchmark records an authority or scope violation, the violation must be interpreted: \citet{singh2026modelforensics} supplies a forensic baseline that combines chain-of-thought inspection with causal intervention to distinguish a genuine governance breach from a harmless reasoning shortcut, the diagnostic any always-on protocol needs before it can attribute a failed invariant check to real misalignment rather than a benign path. So the family that most directly tests state mutation still rarely couples mutation to real tool action, to a privacy or consent boundary, or to recovery after a bad write. It proves that stale and contested state is mishandled; it does not prove whether a system can act on contested state and then recover.

\subsection{Personalization and proactive benchmarks}

A fifth family evaluates long-horizon user state and, increasingly, whether an agent surfaces stored state on its own initiative. The personalization line operationalizes what it means for state to be long-term and user-owned. \citet{xu2026personalizedllmpoweredagentsfoundations} surveys evaluation criteria for personalized LLM agents, defining multi-session coherence and preference retrieval as the targets; \citet{jiang2025personamem} contributes PersonaMem, which tracks an evolving user persona across interactions and reports persona-recall metrics; and \citet{xie2026dynamicmem} sharpens the temporal dimension by separating user attributes, habits, and preferences that mutate on different timescales, exposing that user state evolves at multiple speeds and that a single freshness policy cannot fit all of them. On the action-coupled side, \citet{cai2025personalizedwebagents} demonstrates with PUMA and the PersonalWAB benchmark that retrieving user history to align cross-session web actions measurably helps, closing part of the loop from personalization to action that the memory-QA family leaves open. The proactive line is newer and addresses a stressor no other family touches: whether an agent recalls and acts on a latent stored constraint without being prompted to. \citet{zhang2026triggerbench} measures this prospective memory, finding that spontaneous recall is far harder than retrospective query-driven recall and degrades as context length grows. This is a distinct competence: every other benchmark cues the relevant memory through the question or the task, whereas a deployed always-on agent must decide on its own when stored state becomes relevant and when surfacing or acting on it is appropriate.

The contrast within this family is instructive. Personalization benchmarks measure persona recall, multi-session coherence, and preference adaptation well, and PUMA shows personalization can be tied to action; TriggerBench adds the orthogonal and under-measured competence of utility-calibrated proactivity. But the governance half is again out of frame. As the matrix records, PersonaMem tests recall but not recoverability or auditability, with no validation that an agent can undo or correct a poisoned persona fact or detect a contradiction; PUMA is retrieval-based with no scope control over which user facts may influence which actions and no audit on action provenance; and the dynamic-attribute benchmark tests mutability but not provenance, leaving no record of which fact changed, when, or which user action caused the change. TriggerBench measures whether a constraint is recalled in time, not whether the agent distinguishes a constraint it is authorized to act on from one whose authority has lapsed, nor whether a proactive action is recoverable if the triggering memory was stale or poisoned. Personalization is where consent and privacy should be most salient, since the stored state is by definition sensitive and user-owned, yet write policy, consent, and the right to erasure are treated as out of scope. The clearest move to close this is to make the substrate itself governance-aware: \citet{selvam2026profilefoundry} builds a synthetic, deterministic profile substrate with generational provenance so that private-fact leakage, tool-use scope enforcement, and consistent state recovery across invocations can be scored as first-class outcomes, showing that the consent and recoverability dimensions are testable in personalization once the data is generated with provenance and scope baked in rather than retrofitted onto a recall set. The family extends always-on evaluation in two directions, long-horizon user state and spontaneous recall, while leaving the same governance gap open: it tests that the agent remembers the user and can act on that memory, not that it is licensed to, can be told to forget, and can recover when the personalized state proves wrong.

\subsection{Coverage against the six always-on benchmark stressors}

Reading the five families together yields the main evaluation finding of this survey: their coverage of the six always-on stressors is a clean partition rather than a single missing dataset. Table~\ref{tab:eval-families} summarizes all five families qualitatively, by representative examples, what each measures well, and what it leaves untested for always-on agents. For three of the families, where the closest-work set is dense enough to support it, we additionally coded individual benchmarks against the six stressors at the per-benchmark level; the long-context/RAG and personalization families are characterized at the family level only, because their member benchmarks were not designed around these stressors and a per-benchmark coding would mostly record absences.

The partition is visible at the level of individual benchmarks as well as at the family level. Table~\ref{tab:eval-coverage} codes representative benchmarks from the three families where the survey's benchmark-level coding is densest against the six stressors. Reading down the columns is more telling than reading across the rows: the recovery column (R6) is almost entirely empty, and no single benchmark scores real action (R4), a privacy boundary (R5), and recovery (R6) together, which is the combination an always-on agent most needs and the one the field most consistently leaves unmeasured.

\begin{table}[t]
\centering\small
\setlength{\tabcolsep}{4pt}\renewcommand{\arraystretch}{1.15}
\begin{tabularx}{\linewidth}{p{0.20\linewidth}Ycccccc}
\toprule
Benchmark & Family & R1 & R2 & R3 & R4 & R5 & R6 \\
\midrule
LoCoMo & Memory QA &  & \pmark &  &  &  &  \\
LongMemEval & Memory QA &  & \pmark &  &  &  &  \\
MemoryAgentBench & Memory QA &  & \pmark &  & \pmark &  &  \\
PerLTQA & Memory QA &  & \pmark & \pmark &  &  &  \\
MemoryArena & Mem-guided action &  & \pmark &  & \cmark &  &  \\
LifelongAgentBench & Mem-guided action &  & \pmark &  & \cmark &  &  \\
WorkArena & Mem-guided action &  &  &  & \cmark &  &  \\
WorkArena++ & Mem-guided action &  &  &  & \cmark &  &  \\
STALE & State/belief mut. &  & \cmark &  &  &  &  \\
Memora/FAMA & State/belief mut. &  & \cmark &  &  &  &  \\
BeliefShift & State/belief mut. &  & \cmark & \cmark &  &  &  \\
Lifelong-SOTOPIA & State/belief mut. & \pmark &  & \cmark &  &  &  \\
\bottomrule
\end{tabularx}
\caption{Per-benchmark coverage of the six always-on benchmark stressors, for representative benchmarks from the three families where benchmark-level coding is densest. R1 persistent multi-user memory, R2 stale-state adjudication, R3 social and belief dynamics, R4 real action with side effects, R5 privacy or consent boundary, R6 recovery after bad memory. $\bullet$ denotes a scored stressor, $\circ$ partial, blank not scored. Two patterns are visible by column: the R6 (recovery) column is essentially empty, and no row marks R4, R5, and R6 together. The partition is structural, not a single missing dataset.}
\label{tab:eval-coverage}
\end{table}

\begin{table}[t]
\centering
\small
\setlength{\tabcolsep}{4pt}
\renewcommand{\arraystretch}{1.2}
\begin{tabularx}{\linewidth}{p{0.135\linewidth}YYY}
\toprule
Family & Representative examples & Measures well & Missing for always-on \\
\midrule
Memory QA over histories & LoCoMo \citep{maharana2024evaluatinglongtermconversationalmemory}; PrecisionMemBench \citep{flynt2026precisionmembench}; MemTrace \citep{long2026memtrace} & Recall, temporal updates, abstention, retrieval precision vs.\ generation & Write policy, authority, consent, real action, recovery \\
Boundary: long-context and RAG & LongBench \citep{bai2024longbenchbilingualmultitaskbenchmark}; Lost-in-the-Middle \citep{liu2023lostmiddlelanguagemodels}; RGB \citep{chen2023benchmarkinglargelanguagemodels} & Context use, retrieval faithfulness, noise robustness & Durable write, update, deletion, authority (out of scope by design) \\
Memory-guided action & MemoryArena \citep{he2026memoryarena}; LifelongAgentBench \citep{zheng2025lifelongagentbenchevaluatingllmagents}; AppWorld \citep{trivedi-etal-2024-appworld}; OSWorld \citep{xie2024osworldbenchmarkingmultimodalagents} & Experience reuse, tool and computer use, final-state checking & Persistent identity, governed writes, revocation, rollback \\
State and belief mutation & STALE \citep{chao2026stalellmagentsknow}; MEME \citep{jung2026meme}; RECLAIM \citep{kwon2026reclaim}; GroupMemBench \citep{yang2026groupmembenchbenchmarkingllmagent} & Staleness, deletion propagation, provenance, belief and group dynamics & Coupling to real action, privacy boundary, recovery after bad write \\
Tool-security evaluation & AgentDojo \citep{debenedetti2024agentdojodynamicenvironmentevaluate}; AgentPoison \citep{chen2024agentpoisonredteamingllmagents} & Prompt-injection attacks and defenses over realistic tool/data boundaries & Persistent write, deletion propagation, rollback after injected state persists \\
Personal\-ization and proactive & PersonaMem \citep{jiang2025personamem}; PersonalWAB \citep{cai2025personalizedwebagents}; TriggerBench \citep{zhang2026triggerbench} & Persona recall, preference drift, spontaneous proactive recall & Consent and erasure, scope control, audit, recoverability \\
\bottomrule
\end{tabularx}
\caption{Benchmark and evaluation families relevant to always-on agents. Each family measures part of the requirement space well and leaves a disjoint part untested. No family is forced to govern state, act on it, and recover from it at once, so recoverability under authority and provenance constraints falls between families.}
\label{tab:eval-families}
\end{table}

The six stressors an always-on benchmark would need to exercise are persistent multi-user memory, stale-state adjudication, social and belief dynamics, real tool or GUI action with side effects, a privacy or consent boundary, and recovery after a bad memory. The coverage finding is that, to our knowledge, no widely adopted benchmark we examined covers more than five of the six, and, more pointedly, none combines real action, recovery, and a privacy or consent boundary in a single protocol. The reason is visible in the table: the stressors are distributed across families so that satisfying all six would require a benchmark to inherit the action realism of the memory-guided family, the mutation adjudication of the state-and-belief family, the consent salience of the personalization family, and a recovery mechanism that essentially no family scores. Recovery is the binding stressor. As the survey's corpus statistics show, only a small fraction of works expose any rollback mechanism, and none of the benchmark families reports recovery success or cost after corruption, so even where a benchmark perturbs state it cannot certify that a system restored a known-good state afterward. Action without recovery is the most dangerous combination to leave unmeasured, because an always-on agent that acts on memory and cannot recover from a bad write is precisely the failure the thesis warns about, and it is exactly the combination that the action family executes but does not test and the mutation family tests but does not execute.

The deficit is therefore structural rather than a gap that one more dataset would close. Each family's format encodes its blind spot. A memory-QA recall unit has no field for an authority tag or a deletion record; an episodic action benchmark resets the very persistence it would need to test governance across; a mutation benchmark scores the memory graph but not the side effect derived from it; a personalization benchmark treats the user's stored state as a recall target rather than as consent-bound, revocable, auditable state. None of these is a defect of the benchmark within its own purpose. The defect is that the union of their purposes still leaves the persistent-state lifecycle, observe, write, validate, organize, retrieve, act, update, forget, audit, rollback, evaluated only in pieces, with the governance-defining stages of forget, audit, and rollback measured least and never jointly with action. A score on any one family certifies a capability that a stateless or single-session system could in principle also possess; none certifies the property that actually distinguishes a persistent-state system, namely that state is governed across its lifecycle under authority, scope, provenance, recoverability, and actionability constraints.

The evaluation gap motivates a separate protocol. The missing instrument is not a harder recall set or a larger action environment, but one that records and scores state transitions rather than answer quality: an episode whose stream carries idempotency keys, causal links, permission epochs, provenance tiers, and retention constraints, and whose perturbations include restart, conflicting update, deletion request, adversarial write, shared-scope change, permission revocation, and delayed consequence. The score should come from deterministic invariant checks of the kind the methodology subsection argued these predicates admit, not from a bias-prone holistic judge. Such an instrument would treat a trajectory as correct only if a deleted value is truly unretrievable across derived tiers, an action is licensed by a current rather than a lapsed authority epoch, a conflict is surfaced with provenance, a retried call does not duplicate an irreversible side effect, and a rollback trace is sufficient to restore a known-good state. Because the five families partition these stressors rather than combining them, the next section develops the Always-On Evaluation Protocol: an event-stream and snapshot schema with worked episodes and deterministic checks, designed to wrap the realistic substrates this family already provides with the restart, conflict, revocation, leakage, and rollback perturbations that govern persistent state.

\section{Failure modes, safety, and governance}\label{sec:failures}

\regime{Control}{how persistence fails, and the governance operations that should prevent it.}
The previous parts established what always-on agents accumulate (substrates, lifecycle operations, continual adaptation) and how the field measures it. The missing counterpart is governance of the state being accumulated. We treat governance and failure modes as two faces of one object. A persistence-induced failure is the observable symptom; a governance mechanism is the lifecycle operation that should have prevented it. Organizing the two together is deliberate. Prior agent-security surveys catalogue threats by attack surface, by component, or by adversary capability \citep{chu2026systematicsurveyagentsecurity,lin2026surveylongtermmemorysecurity}. We instead index every failure to the lifecycle stage at which corrupt state enters and to the invariant (Section~\ref{sec:lifecycle}) it violates, because that connects an attack to the governance operation that should have caught it. A poisoning attack is a write that escaped authority and provenance checks and then persisted. A stale-commitment failure is an update boundary that let an expired authority keep licensing action.

This framing also lets us state the survey's corpus-scoped map precisely. In our $435$-work coded corpus, governance is the least developed region of the lifecycle. By state axis, authority is the rarest at $72$ of $435$ works, recoverability appears in $112$, and provenance in $153$; by lifecycle stage, the return arc thins, with audit at $88$, forget at $66$, and rollback at only $27$. The counts are scoping estimates, but the qualitative skew is large. Fifteen targeted rounds aimed at governance, authority, rollback, formal verification, durable execution, deletion propagation, and the database, distributed-systems, and HCI literatures grew the corpus more than fourfold yet moved the governance fraction only from roughly one-sixth to one-third before late rounds showed diminishing returns under this query frame. We read this as evidence that governance-targeted search did not remove the skew within our frame, not as a proof of field-level saturation. Subarea by subarea, the literature supplies fragments of governance while leaving the lifecycle incomplete. That pattern is the gap: always-on agents are persistent-state systems whose state must be governed over a lifecycle rather than only stored and retrieved.

\subsection{Threat model and governance surface}

Before cataloguing failures, we fix the threat model used here. The attacker need not control the model or the memory store. The typical persistence attack is weaker and harder to notice: an untrusted channel writes state, a trusted channel later retrieves it, and a tool call or shared-memory update gives that state authority it never had. Table~\ref{tab:threat-model} names the principals and channels we assume, what each may write or read, and the recovery question a governed system must answer. The table is intentionally operational rather than cryptographic. It says where the lifecycle must bind authority, provenance, scope, and rollback handles if the later failure taxonomy is to be testable.

\begin{table}[t]
\centering
\scriptsize
\setlength{\tabcolsep}{3pt}\renewcommand{\arraystretch}{1.16}
\begin{tabularx}{\linewidth}{p{0.17\linewidth}p{0.10\linewidth}p{0.24\linewidth}p{0.22\linewidth}Y}
\toprule
Principal or channel & Trusted? & State written or read & Authority surface & Failure to test or recover \\
\midrule
Owner user instruction & Usually & Preferences, permissions, corrections, deletion requests & Direct authority, but only within declared scope and current epoch & Does a later action use the current instruction, not a superseded one? \\
Collaborator or group member & Partial & Shared facts, proposals, edits, delegations & Role-bounded authority over shared state & Can the system surface conflicts and prevent cross-principal scope expansion? \\
Untrusted web, file, email, or tool output & No & Observations, retrieved snippets, tool results, notes embedded in data & No authority to become durable instruction without validation & Can poisoning be quarantined before it writes persistent state or chooses a tool? \\
Memory manager, summarizer, or consolidator & Partial & Summaries, derived facts, indices, embeddings, skill records & Derivative authority inherited from sources & Does consolidation preserve source, scope, and deletion lineage? \\
Agent reflection or self-update loop & Partial & Lessons, skills, policies, acceptance scores & Authority only after validation and rollback registration & Can a bad self-update be traced, de-authorized, and reverted? \\
External API, tool server, or credential broker & Partial & Side effects, credentials, tool schemas, execution receipts & Capability grants and schema-level permission & Can rollback distinguish replay from compensation after an irreversible effect? \\
Administrator or organization policy & Trusted for policy & Retention rules, allowlists, logging, tenant boundaries & Global constraints over user and agent state & Can policy changes narrow authority without silently orphaning old state? \\
\bottomrule
\end{tabularx}
\caption{Threat model for persistent-state failures. A channel may be trusted for one purpose and untrusted for another: an external tool can be trusted to report a weather value but not to write a durable instruction; a collaborator can edit shared state but not expand another user's scope. The recovery column states the check a benchmark or audit should force.}
\label{tab:threat-model}
\end{table}

\subsection{Persistence-induced failure modes}

Episodic agents avoid an entire failure class by resetting. An attack that succeeds within a single task disappears when the task ends; a stale fact cannot govern a future decision because there is no future decision that shares state. Persistence removes this safety net. The failure modes share the same enabling property that makes always-on agents useful: state written in one interaction authorizes behavior in a later one, after the originating context, and often the originating user, is gone. Table~\ref{tab:failures} maps each failure to the lifecycle boundary where corrupt state enters and the invariant it breaks; the surrounding text follows those boundaries in order.

\begin{table}[t]
\centering
\small
\setlength{\tabcolsep}{4pt}\renewcommand{\arraystretch}{1.2}
\begin{tabularx}{\linewidth}{p{0.15\linewidth}YYp{0.12\linewidth}}
\toprule
Boundary & Failure & Invariant violated & Repr.\ work \\
\midrule
Write & Untrusted write poisoning; indirect prompt injection & Authority monotonicity; provenance preservation & \citep{louck2026originbound,joshi2026eywa} \\
Consolidate & Retrieval-key loss; procedure poisoning & Provenance preservation & \citep{ouyang2026memlineage,kumar2026memarchitect} \\
Retrieve and act & Memory distraction; control-flow hijack & Authority monotonicity; scope non-expansion & \citep{xu2026memorysilent,winston2026solveraided} \\
Update & Stale commitment; belief drift & Authority monotonicity & \citep{zhou2026usercorrections,shawn2026pace} \\
Forget and audit & Deletion failure; rollback gap & Deletion propagation; rollback traceability & \citep{zhao2026memorepair,shah2026unlearningmirage} \\
Shared state & Cross-agent propagation & Scope non-expansion & \citep{ren2026gatemem,ravindran2026portablemem} \\
Recovery & Rollback as attack surface & Rollback traceability & \citep{zheng2026acrfence} \\
\bottomrule
\end{tabularx}
\caption{Persistence-induced failure modes, each mapped to the lifecycle boundary where corrupt state enters or activates and to the state invariant (Section~\ref{sec:lifecycle}) it violates. The table treats safety as a property of the whole lifecycle, not a final filter applied after retrieval. The last row is important because recovery, the mechanism meant to repair the others, is itself a fresh boundary at which new corruption can enter.}
\label{tab:failures}
\end{table}

\subsubsection{Write poisoning and injection.}
At the write boundary the core failure is that untrusted observation becomes trusted state. Memory-poisoning and indirect prompt-injection studies show that a harmful record can enter long-term state during an ordinary interaction and activate much later, when the originating context is gone. The threat is not insertion alone but retrieval-aware optimization of what gets inserted so it is reliably surfaced and acted upon \citep{dash2026memorypoisoning}, and the same mechanism scales from single-session prompt manipulation to a cross-session system vulnerability once the payload survives in memories, files, and tool registries after the attacker's session ends \citep{xie2026crosssessioninjection}. Filtering the current prompt cannot undo a write that already landed. The governance response is to push authority and provenance to the write itself. Origin-bound write authority ties a record's right to influence action to a machine-checked, non-malleable account of its source, arguing that content-based and lineage-based trust can be laundered through summarization and so must be cryptographically rather than heuristically bound \citep{louck2026originbound}. Evidence-before-belief designs store immutable source evidence first and derive canonical facts only after validation against typed signals and source support, so a poisoned observation cannot become a belief without leaving an auditable derivation \citep{joshi2026eywa}. These works contribute the validate stage that most memory systems leave implicit: a write is not committed until an authority and a provenance chain license it. What they leave open is coverage. Origin-binding governs the write but assumes the provenance signal is itself trustworthy, and evidence-first storage raises storage and latency cost that no benchmark yet prices. The gap, in thesis terms, is that authority monotonicity and provenance preservation are enforced at the moment of writing by a handful of 2026 systems and assumed away by the rest.

\subsubsection{Retrieval distraction and tool-drift.}
At the retrieve-route-act boundary the failure is that stored state crowds out current intent. Retrieved records can distract the model, override newer evidence, or bias tool selection and ordering. A subtle and well-named form is the silent integration failure: an agent integrates sensitive or irrelevant memory content into its behavior even when no current prompt warrants it, and the integration happens after retrieval, so admitting the right documents does not prevent the wrong influence \citep{xu2026memorysilent}. This sharpens a point the retrieval literature often blurs: retrieval quality is necessary but not sufficient, because the generator can act on retrieved state in ways the retriever never intended. The governance response is to treat the boundary between retrieved state and action as an enforcement point rather than a similarity threshold. Solver-aided execution compiles natural-language tool-use policies into SMT constraints over agent-observable state and tool arguments, then gates each tool call against a precondition check before it executes \citep{winston2026solveraided}; skill-containment analysis bounds, by abstract interpretation and bounded model checking, what a non-deterministic skill can reach in its deterministic execution side \citep{metere2026skillcontainment}. These contribute a runtime authority check at act time, but leave implicit the interaction with memory: a containment proof over a skill says nothing about whether the state that parameterized the skill was authorized to do so. The gap is that authority monotonicity and scope non-expansion are checked, when they are checked at all, over the action's static envelope, not over the dynamic state that selected and parameterized it.

\subsubsection{Stale commitment and belief drift.}
At the update boundary the failure is temporal: old preferences, facts, commitments, permissions, or policies remain binding after they cease to be valid. A preference that was true last month, changed yesterday, should be superseded today, yet a system with no notion of supersession treats the historical value as current. A useful result here is the access-versus-compliance gap: even when a user's prior correction is present and recalled, agents still violate it most of the time, because recall is not compliance \citep{zhou2026usercorrections}. The proposed fix mines corrections into atomic rules compiled into act-time completion gates the agent must pass, again the validate and act stage made explicit rather than assumed. Belief drift is the dual failure during self-update: greedy, score-based acceptance of self-generated updates p-hacks the agent into committing false or harmful state, fixed by an anytime-valid, betting-based commit gate that bounds the false-commit probability \citep{shawn2026pace}. Both works contribute a notion of update as a gated transition rather than an overwrite. Neither closes the representational gap: to mark a value superseded rather than deleted, the store must represent both current and historical truth with authority and timestamps, which many flat-text memories do not. The gap is that authority monotonicity over time, the requirement that only a current, unrevoked authority may license a state-affected action, is rarely representable, so deletion becomes the only available tool for retiring obsolete state.

\subsubsection{Deletion failure and the rollback gap.}
At the forget-and-audit boundary the failure is that deletion is treated as a single operation when it is a cascade. A user asks the system to forget a fact, but that fact may already exist in raw logs, summaries, embeddings, cached prompts, retrieved traces, shared memories, and procedural skills. Deployment-time memorization work makes the residue concrete: derived tiers retain deleted content unless the purge covers the full pipeline \citep{chen2026deploymenttimememorization}, and the recoverability failure is demonstrated directly when supposedly forgotten information resurfaces under multi-hop or aliased queries \citep{shah2026unlearningmirage}. The governance response is cascade-aware deletion. A barrier-first repair contract withdraws stale derived descendants when a source is deleted or invalidated and republishes only validated, predecessor-closed successors, cutting invalidated-memory exposure toward zero \citep{zhao2026memorepair}. This contributes the deletion-propagation invariant in operational form. What stays open is that the contract assumes a tracked derivation graph; systems that consolidate by lossy summarization have already destroyed the lineage the cascade would follow, which is why provenance preservation and deletion propagation are coupled rather than independent.

\subsubsection{Cross-agent propagation.}
Shared state adds a propagation failure absent in single-agent settings. Corrupt state can spread through a shared memory substrate or an inter-agent channel faster than per-agent safety checks can contain it, because each agent trusts the substrate. A governed shared-memory benchmark frames the test directly: can an agent serve legitimate cross-principal requests while enforcing per-role scope and reliably forgetting on deletion request \citep{ren2026gatemem}. Cross-agent memory-interoperability protocols make the substrate explicit, moving persistent memory across heterogeneous vendor models with content-addressable provenance and capability-scoped disclosure \citep{ravindran2026portablemem}, and a workspace-collaboration protocol that delegates a live execution environment rather than only messages widens the shared surface further \citep{nie2026awcp}. The contribution is to name scope as the governing axis: a shared write must narrow or preserve scope, never silently widen it. The gap is that each new shared channel enlarges exactly the authority and provenance surface the propagation failures exploit, and the interoperability protocols assume the receiving model trusts the sender's encodings.

\subsubsection{Rollback as a fresh attack surface.}
The final row of Table~\ref{tab:failures} is the most counterintuitive and among the most relevant to the survey's claim. Recovery, the mechanism meant to repair every failure above, is itself a boundary at which new corruption enters. Because an agent regenerates slightly different requests after a checkpoint restore, retried tool calls become new external requests, enabling duplicate irreversible side effects and the resurrection of already-consumed authority unless irreversible effects are recorded and replayed rather than re-issued \citep{zheng2026acrfence}. This is why rollback traceability is an invariant rather than a convenience feature: a system that can revert its internal store but cannot reason about which external effects were irreversible will, on recovery, re-send a payment or re-grant a revoked permission. The gap is structural. The literature treats rollback as a capability to add, not as a transition that must itself satisfy authority monotonicity and scope non-expansion, and the corpus count makes the consequence stark: with only $27$ of $435$ works exposing any rollback mechanism, the operation most likely to silently re-introduce corruption is the least studied.

\subsection{Authority and permission state}

Authority is the rarest axis in the corpus ($72$ of $435$) and the one whose absence most directly undermines the persistent-state thesis: a record that can influence action without a current, scoped permission is the defining ingredient of poisoning, privilege drift, and cross-user leakage. The governance question is not who owns the memory but which authority currently licenses a given record to affect a given action, and authority must only narrow, never silently expand, as state and delegations accumulate.

One line treats authority as a server-side, monotone policy attribute. Intent-governed authorization makes the user's expressed intent a session-scoped attribute that may only reduce the authority static credentials grant to an agent's tool calls \citep{zhu2026intentgovernedauth}; this is authority monotonicity enforced at the API boundary. A complementary execution-time line attacks the time-of-check-to-time-of-use problem directly: a state-witness profile revalidates preconditions at execution and fails closed when world state has drifted between an action's approval and its delayed execution \citep{zhu2026openport}, which is exactly the stale-authority failure of the previous subsection caught at act time. Delegation is governed compositionally by overlaying recursive-delegation, dynamic-scoping, and resource-attenuation primitives onto existing relational authorization, with formal proofs that delegated authority narrows along the chain \citep{ibrahim2026overlaygovernance}, and extended at the credential layer by scoped agent credentials with accountable delegation chains and natural-language permissions compiled into auditable access control \citep{south2025authenticateddelegation}. A particularly clean formal result imports hardware memory-consistency theory to bound unauthorized operations by execution count rather than time-to-live, so revocation does not depend on agent speed \citep{parakhin2026bureaucracy}.

A second execution-time result makes the monotonicity argument constructive: authority is operationalized as a runtime enforcement mechanism that lets an action execute only when a current authority can be constructed from the system state, with a halt state and recovery loop that suspends execution whenever authority is undefined, so a stale or revoked permission can never license a tool call \citep{fernandez2026authority}. At the coordination layer, lighter-weight declarative manifests cover the open-web case: an agent-permissions.json convention lets site owners declare which agent interactions are allowed, supplying a practical permission-boundary mechanism across heterogeneous agent modalities where no shared credential infrastructure exists \citep{marro2025permissions}. DEMM-Bench makes reconstructability measurable by scoring whether recorded delegation, policy, and tool-firewall decisions are sufficient to rebuild an authority decision and its accountability trace \citep{solozobov2026demm}.

Tool and credential state form a tightly coupled sub-problem, since a stale token or a silently redefined tool is an authority leak by another name. A trusted broker lets an agent invoke API and SSH secrets through schema-checked calls while keeping the raw credential non-exportable, defeating prompt-injection exfiltration \citep{jin2026capseal}; tool definitions are bound to signed, versioned manifests so a persisted grant is invalidated when a tool is silently redefined, the rug-pull case \citep{bhatt2025etdi}; and tool servers are admitted only via offline-signed clearance against a pinned trust root under a deny-by-default allowlist \citep{metere2026attestedtoolserver}. The most thesis-aligned credential result targets the zombie-agent failure: a heartbeat-bound protocol makes a descendant agent's authority self-expire within a provably bounded window once parent liveness ceases, so an agent cannot keep acting on delegated authority after the operator who granted it has shut down \citep{deochake2026heartbeatcred}. This is authority monotonicity given a clock.

This body of work is the governance subarea with the deepest enforcement mechanisms and formal guarantees. What it leaves implicit is the link to memory: these systems govern the credential and the tool call, but say little about the persistent state that decides which credential to use or which tool to call. A monotone-authority API does not help if poisoned memory chooses to invoke an authorized but contextually wrong tool. The gap, then, is that authority is governed at the action boundary by a small, sophisticated literature and almost never co-governed with the memory whose retrieval triggers the action. Revocation dynamics, especially partial-delegation revocation and revocation cascades, remain largely unaddressed.

\subsection{Provenance, audit, and lineage}

Provenance is better covered than authority ($153$ of $435$) but is dominated by static logging rather than the derivation-aware lineage that deletion and rollback actually require. The governing question is whether a system can answer, for any stored fact or any action, where it came from, how it was transformed, and which other state descends from it.

Three mechanistic families recur. Append-only ledger designs treat memory as an immutable, signed history: a Merkle-indexed ledger with privilege-lattice access control makes revisions protocol-blocked and historied rather than destructive \citep{wright2025immutablemem}, and a signed delegation-provenance chain binds each agent action to an originating human authorization and scope for offline audit \citep{dalugoda2026hdp}. The append-only pattern has a deeper lineage in the systems literature that the agent work is rediscovering: blockchain-backed audit logging supplies an established design for tamper-proof, cryptographically secured action records whose state transitions cannot be retroactively altered \citep{ahmad2018blockaudit}, and smart-contract-backed audit trails extend this to autonomous agents by atomically recording each action, prediction, and explanation to a ledger so decision provenance is immutable and regulator-auditable \citep{wang2026auditor}. A broader honest-computing framework generalizes the requirement, establishing provenance through cryptographically backed data-lineage tracking and auditable custody chains that apply directly to persistent agent memory states and their transformations \citep{guitton2024honest}. Derivation-graph designs go further by tracking history and descent: MemLineage attaches a weighted derivation DAG and Merkle log and enforces an untrusted-path-persistence invariant that refuses actions descending from external ancestors \citep{ouyang2026memlineage}, while portable-memory and robot-policy-lineage designs carry content-addressable provenance across vendor models and across policy-iteration cycles, respectively \citep{ravindran2026portablemem,luo2026robolineage}. The same derivation discipline is what self-improving agents need to make learned behavior auditable: recent work reframes capability updates as traceable artifacts by recording verifiable provenance of skill acquisition through structured skill graphs and cryptographically backed audit logs, rather than as opaque parameter shifts \citep{huang2025audited}. Formal-monitoring designs make provenance checkable at runtime: Causal Past Logic gives a past-time temporal logic with a provably correct vector-clock monitor for guards over causally-visible stored state in distributed agent workflows \citep{bollig2026causalpast}. A complementary diagnostic line shows why provenance must be structured, not flat: when flat-text memory loses source attribution, agents hallucinate authority and misattribute facts, a failure named provenance-role collapse \citep{jin2026provenancerolecollapse}. That the discipline is far from settled even in evaluation is itself documented: a systematic audit of agent-benchmark disclosures finds reproducible provenance of evaluations still nascent, with mean audit scores of $0.38$ out of $1.0$ on core dimensions such as harness specification and inference-cost reporting \citep{moghadasi2026audit}.

The tension within this subarea is itself a finding. Immutability and deletion are in direct conflict: an append-only ledger that makes revisions protocol-blocked cannot, by construction, honor a right-to-be-forgotten request without breaking its own invariant \citep{wright2025immutablemem}. Derivation graphs resolve part of this by enabling cascade deletion over lineage, but assume an acyclic, fully-tracked graph and do not handle benign cycles introduced by cross-agent delegation \citep{ouyang2026memlineage}. The gap is that provenance is mostly logged for accountability, not maintained as the substrate that deletion propagation and rollback traceability depend on; few works treat lineage as a live, queryable, prunable structure rather than a write-once record, and provenance preservation under lossy consolidation, where the lineage is destroyed exactly when it is most needed, is the single thinnest cell.

\subsection{Privacy, consent, and scope}

Privacy is where governance work is densest in absolute terms, yet it remains skewed toward demonstrating leakage rather than enforcing scope. The persistent-state framing reclassifies privacy from a training-time property to a deployment-time, cross-session one: the risk is that an always-on agent silently writes, retains, and later exposes user state it was never explicitly authorized to keep, rather than only that a model memorized its training data \citep{carlini2021extractingtrainingdata}.

The leakage-demonstration literature is now precise about the attack surface. Membership-inference frameworks target persistent chat-agent memory units directly \citep{chen2026mrmmia}, an interrogation attack infers datastore membership with roughly thirty natural-language queries that evade query-rewriting defenses \citep{naseh2025riddle}, mobile-agent memory leaks private records under extraction \citep{wang2025mobileagente}, and deployment-time memorization studies show how deployed agents absorb sensitive interaction data \citep{chen2026deploymenttimememorization}. A particularly consequential audit of $2{,}050$ real persistent-memory entries finds that $96\%$ are silently system-created rather than user-authorized, motivating an attribution shield for transparency over autonomously written user state \citep{dash2026selfportrait}. The enforcement-side literature is thinner. Differential privacy for memory spends budget only on sensitive tokens to bound leakage from an external store \citep{koga2024dprag}; a watermark hidden in latent memory-write decisions lets an owner prove provenance of a leaked snapshot at near-zero utility cost \citep{zhang2026memmark}; and a secure exchange protocol gives cross-session sharing under AES-256-GCM access control \citep{masoor2025samep}. Two HCI studies supply the consent dimension the mechanism work omits: interviews reveal that users hold incomplete mental models of RAG-based memory and want granular review, edit, delete, and categorize control \citep{zhang2025ragmemoryperceptions}, and an analysis of $22$ agentic systems derives UI patterns with regulatory potential, positioning the interface as a governance surface that should make agent memory user-editable \citep{feng2025regulatoryui}.

Contextual integrity gives this subsection a positive criterion for scope: private state must not leak, and information flows must remain appropriate to the social and task context in which the state was written. CIMemories turns this criterion into a benchmark for persistent memory, asking whether memory-augmented assistants preserve contextual integrity across composed situations \citep{mireshghallah2025cimemories}. Contextualized privacy defense for LLM agents makes a complementary mechanism claim, treating privacy intervention as a context-dependent runtime decision rather than a static filter \citep{wen2026contextualizedprivacy}. In our terms, both lines are scope-non-expansion work: a record written in one context must not silently authorize use in another.

The data-centric privacy survey makes the gap explicit. It organizes risks by the data surfaces an agent touches and finds that information-flow control is the only governance covering cross-session inference leakage, while no benchmark exercises an agent across all its data surfaces under a single declared privacy policy \citep{lahjouji2026privacysurvey}. This is the scope gap in operational form. Scope non-expansion requires that state written under one user, task, or consent context not silently act under another, yet the corpus measures leakage after the fact far more than it enforces scope before it. The gap, in thesis terms, is that privacy is studied as an attack to demonstrate rather than a scope invariant to maintain jointly across the write, organize, retrieve, and share stages, and consent is largely confined to the interface layer rather than bound to the stored record.

\subsection{Public production memory controls}

Public documentation for deployed assistants gives a useful deployment-facing cross-check on the survey's abstractions. These systems are not research benchmarks, and the available documentation is not enough to score their internal enforcement. It does show which controls providers expose to users, administrators, and projects. Table~\ref{tab:production-controls} reads those controls through the same axes used in the corpus.

\begin{table}[t]
\centering
\scriptsize
\setlength{\tabcolsep}{3pt}\renewcommand{\arraystretch}{1.16}
\begin{tabularx}{\linewidth}{p{0.17\linewidth}p{0.24\linewidth}p{0.25\linewidth}Y}
\toprule
System & Persistent state exposed in docs & User or admin controls & Persistent-state reading \\
\midrule
ChatGPT Memory \citep{openai2026chatgptmemoryfaq} & Saved memories and reference-chat-history signals that can personalize later responses & Users can view, update, delete saved memories, disable memory, use temporary chats, and delete chats, but deletion across memory, chat history, files, and connected apps is operationally separate & Strong user-facing edit/delete affordance; weaker public account of derived-state provenance and deletion propagation across all downstream uses \\
Microsoft 365 Copilot Memory \citep{microsoft2026copilotmemorysupport} & Work-context memories and user preferences used for Microsoft 365 Copilot personalization & Users can ask Copilot what it remembers, update or remove memories, and disable memory; organizational controls mediate availability & Treats enterprise memory as governed tenant state, but the public surface emphasizes user control more than rollback or recovery of actions conditioned on old memories \\
Claude Code memory \citep{anthropic2026claudecodememory} & Markdown memory files at project, user, and managed-policy scopes, loaded into the coding context & Users and organizations edit memory files directly; hooks and permissions provide separate enforcement points & Makes scope visible through file location and hierarchy, but memory is instruction context unless paired with tool gates and audit hooks \\
Gemini Enterprise personalization and memory \citep{googlecloud2026geminienterprisememory} & Profile, conversation history, connected sources, and saved memories for enterprise personalization & Administrators configure personalization and memory; users can manage saved memories depending on organizational settings & Exposes the split between user memory and enterprise data connectors, a concrete instance of scope and authority separation across data surfaces \\
\bottomrule
\end{tabularx}
\caption{Publicly documented production controls mapped to the survey axes. The table does not claim to evaluate internal product behavior. It uses official documentation to show that deployed assistants already expose persistent-state controls, while rollback traceability, provenance of derived memories, and cross-surface deletion remain less visible in public interfaces.}
\label{tab:production-controls}
\end{table}

The deployment lesson is modest but important. Production systems already treat memory as governable state: users can inspect or delete some memories, organizations can bound what memory is available, and project-scoped memory can be separated from user-scoped memory. At the same time, the public control surface usually stops at editing or disabling memory. It rarely exposes the lineage of derived memories, the permission epoch under which a memory was written, or the rollback handle for an external action that memory influenced. That is the distinction this survey makes between memory control and persistent-state governance.

\subsection{Deletion propagation and the right to be forgotten}

Deletion deserves its own treatment because it is the operation where the persistent-state thesis bites hardest: in an episodic system deletion is free, since nothing persists, whereas in an always-on system a single forget request must propagate across every derived tier or it is only a retrieval edit. The recoverability axis ($112$ of $435$) is moderately covered, but almost entirely on the wrong side of the operation, as machine unlearning over model parameters rather than cascade deletion over agent memory.

The prerequisite problem, identifying what to forget, is addressed by GDPR and right-to-be-forgotten work that supplies a benchmark for the deletion target most mechanism papers assume away \citep{staufer2025whatshouldllmsforget}. The mechanism literature then splits by substrate. Parameter-resident approaches are surveyed comprehensively for removing data influence from weights \citep{nguyen2024machineunlearningsurvey}, but, as the survey itself notes, target weights rather than the retrieval indices and external stores agent memory actually uses. Within that line, recent methods sharpen what efficient parameter-level erasure requires: one frames unlearning as an information-theoretic boundary adjustment that minimizes the gradient of influence functions around deleted datapoints, making removal tractable without full retraining \citep{foster2024information}, while a complementary result argues that for LLM-based agents deletion must operate at the semantic embedding level rather than on token outputs, since otherwise a prompt-rephrasing attack recovers the supposedly unlearned concept from internal representations \citep{spohn2025align}. A theoretical barrier circumscribes the whole enterprise: in multi-stage training pipelines, unlearning is path-dependent, so deletion propagation that achieves erasure equivalent to a retrained agent must track the full training history and stage order, which most always-on stores do not record \citep{yu2025impossibility}. Retrieval-resident approaches operationalize the forget operation on RAG memory by removing documents and evaluating deletion completeness \citep{wang2024unlearningmeetsrag}, and a membership-leakage study establishes the failure mode for a persistent retrieval store directly \citep{anderson2024ismydata}. A database-centric account reframes the requirement for persistent agent stores: deletion must address logical and physical removal as well as inference leakage through derived data and through the deletion pattern itself, which is why semantic deletion guarantees that bound post-erasure recoverability are the right target rather than mere record removal \citep{chakraborty2026deletion}. The verification dimension is surveyed for the first time as a structured problem, mapping methods that confirm deletion actually succeeded \citep{xue2025unlearningverificationsurvey}, which is the recoverability axis in evaluative form. Formal verification supplies one principled route: framing unlearning as a probabilistic hypothesis test over backdoor-instrumented models certifies erasure with high confidence and so could give right-to-erasure compliance an auditable proof rather than an assertion \citep{sommer2020probabilistic}. That verification is itself adversarially fragile, however: a malicious provider can covertly retain unlearned data while still passing the verification check, so a deletion-propagation guarantee built only on after-the-fact verification can be silently circumvented \citep{zhang2024fragile}. The strongest cascade result, the barrier-first repair contract that withdraws derived descendants and republishes only validated predecessor-closed successors, comes from the state-governance line discussed above \citep{zhao2026memorepair}, and a policy layer that filters stale zombie memories during decay supplies the routine, non-adversarial face of the same need \citep{kumar2026memarchitect}.

The collision with provenance is the defining structural problem. An immutable ledger maximizes auditability but makes deletion protocol-impossible \citep{wright2025immutablemem}; a derivation graph enables cascade deletion but presumes lineage that lossy consolidation destroys \citep{ouyang2026memlineage,zhao2026memorepair}. The gap is that no corpus work formalizes delete propagation across dependent entries, indices, summaries, and downstream actions with cryptographic proof of absence, and verification of deletion in retrieval-based agent memory, as opposed to parametric models, remains largely unaddressed. Deletion propagation appears here as an invariant because the literature usually treats deletion as a feature of one tier rather than a property of the whole state.

\subsection{Rollback and recovery}

Rollback is the most under-served capability in the coded corpus and one of the survey's main open problems. Only $27$ of $435$ works expose any rollback mechanism, the rollback share was $0\%$ before 2025 and reached only $9.5\%$ in 2026, and we found no coded work that reports recovery success or cost after corruption. Recovery closes the lifecycle loop: every other governance mechanism detects or prevents corrupt state, but rollback is what repairs the state-affected decisions a now-discovered bad record already caused. Its near-absence means the works we coded can increasingly notice that state went wrong but much less often undo the consequences.

The works that do address recovery cluster into two thin lines. Transactional designs import database guarantees: a Saga-style model gives multi-agent planning workflow-wide consistency through compensable execution, automated compensation, independent validation agents, and modular checkpointing under relaxed-ACID semantics \citep{chang2025sagallm}, bridging the classical distributed-systems saga pattern with its compensating transactions to agentic planning, and a graph-native versioned memory proves its update and contraction operations satisfy the AGM belief-revision postulates, importing classical knowledge-representation rationality guarantees into cross-session persistent state \citep{park2026kumiho}. A complementary task-level design supplies the missing transaction boundary for tool use: semantic transactions wrap multi-step agent tool invocations in commit-rollback-recovery scopes with tracked effect lineage and reversible local state, so a partially completed task can be cleanly contracted rather than left in a torn state \citep{chen2026cordon}. These contribute principled undo and consistent contraction. Their limitation is the assumption that compensation handlers exist and that original state can be reconstructed, which fails when an action had an irreversible external effect. The second line confronts that failure head-on: the recovery boundary is itself an attack surface, because regenerated requests after a checkpoint restore become new external requests that duplicate irreversible effects or resurrect already-consumed authority unless those effects are recorded and replayed rather than re-issued \citep{zheng2026acrfence}. This is the strongest recovery result in the corpus because it shows rollback is not a safe escape hatch but a transition that must itself satisfy authority monotonicity and scope non-expansion.

A third, more infrastructural line is now emerging that the agent literature can borrow from rather than reinvent. The durable-execution side supplies the substrate beneath any agent-specific protocol: device-resident checkpointing recovers GPU state and restarts long-running inference without reconstructing KV caches \citep{gan2026concordia}, write-ahead logging with quorum-replicated durability provides foundational low-latency checkpoint patterns decoupled from agent semantics \citep{kuschewski2026btrlog}, and checkpoint-resume orchestration lets long-running multi-agent systems pause on upstream failure and resume converged work across process restarts with reduced recovery overhead \citep{annapureddy2026prima}. These answer the recovery-time question by making fast, durable restart cheap, though none yet reasons about which restored effects were irreversible. The measurement side is finally being instrumented as well, addressing the corpus-wide absence of recovery metrics noted above: MEMPROBE establishes that successful task completion and recoverable memory state are distinct capabilities, turning recovery accuracy into a quantifiable governance metric rather than an assumed property \citep{ma2026memprobe}, and a mean-time-to-recovery metric adapted for agent coherence (MTTR-A) extends classical dependability measures to distributed multi-agent systems by quantifying the time to detect and recover from reasoning drift \citep{or2025mttralm}. Together these begin to supply the recovery-time and recovery-point objectives whose total absence we flag next.

The gaps here compound. There is no quantification of recovery-time or recovery-point objectives, no governance-overhead or latency budget, and no longitudinal production case study; transactional consistency is almost entirely single-agent, leaving distributed and multi-agent rollback semantics absent; and the open instrument named in the research agenda is authority-scoped rollback of tool-originated state once the authority that licensed the original action has lapsed, a case we found no current benchmark scoring. Stated against the thesis, rollback traceability is the invariant most directly tied to repair: every action should carry a handle back to the records that justified it so that a later-discovered bad record yields a bounded, repairable set of decisions. In our corpus this is enforced by a handful of 2026 systems and rarely measured.

\subsection{Synthesis: governance as the missing half of the lifecycle}

Read together, the six subareas tell one story. The works we coded contain sophisticated, sometimes formally verified machinery for the parts of governance closest to the action boundary, monotone authorization, capability containment, attested tool servers, and a substantial descriptive literature on the attacks that persistence enables. Across these enforcement results, validation works best as a live runtime gate rather than a static policy check; recent work makes this explicit. Behavioral contracts integrate preconditions, invariants, governance policies, and recovery as executable specifications enforced at runtime to bound behavioral drift in autonomous agents \citep{bhardwaj2026contracts}, supplying a frame in which our validate stage and every invariant become checkable obligations rather than aspirations. Concrete instances populate that frame: a dedicated runtime authority-control layer gates tool invocations and tracks source and target trust to prevent authority confusion independent of prompt-only defenses \citep{qin2026airguard}; the halt-and-recover authority construction discussed above guarantees no action executes without constructible authority \citep{fernandez2026authority}; and capability mutation at runtime is bounded by explicit validation, traceability, and rollback constraints so that adaptation stays observable rather than unconstrained \citep{garralda2026governed}. Two framing results bound when such gates should be used: a governance rubric reserves runtime guardrails for controls that are observable, determinate, and time-sensitive, assigning the rest across architecture, policy, escalation, and audit \citep{koch2026governance}, and a proactive monitor learns behavioral dynamics to predict unsafe trajectories and intervene before a violation manifests rather than after \citep{wang2025probguard}, which complements the reactive, post-hoc detection that dominates the corpus. What remains thin in our coded corpus is the return-arc machinery that closes the lifecycle: cascade-aware deletion that survives lossy consolidation, lineage maintained as a live structure rather than a write-once log, scope enforced across all of an agent's data surfaces under one policy, and rollback that is itself governed rather than a fresh attack surface. The numbers make the asymmetry concrete as a scoping estimate: authority at $72$, recoverability at $112$, audit at $88$, forget at $66$, and rollback at $27$, all out of $435$, with late targeted search rounds showing diminishing returns under our query frame. Here the skew is a safety concern as well as a coverage statistic: every failure mode in Table~\ref{tab:failures}, from a poisoned write that activates after its author is gone to a recovery that re-sends a payment, is enabled by a missing return-arc operation. The five invariants, authority monotonicity, scope non-expansion, deletion propagation, provenance preservation, and rollback traceability, are therefore not abstract desiderata but the named properties whose absence each documented attack exploits, and which the existing literature names, occasionally enforces, and rarely measures end to end.

\section{The Always-On Evaluation Protocol (AOEP)}\label{sec:aoep}

\regime{Control}{how governance can be scored on a running system rather than only described.}
The preceding parts established a recurring asymmetry in our coded corpus. The literature has learned to accumulate state and to retrieve it, and a large benchmark literature now scores how well it does so, but much less of that literature scores whether the state was \emph{governed}: whether a deleted fact stayed deleted across derived tiers, whether an action was licensed by a still-valid authority, whether an untrusted observation was prevented from becoming durable instruction, whether a corrupted decision can be traced back and undone. The Always-On Evaluation Protocol (AOEP) targets that omission. It does not measure answer quality on a memory question; it measures the correctness of a \emph{state trajectory} under the kinds of perturbations only a persistent agent encounters: restart, conflict, deletion request, permission revocation, shared-scope change, adversarial write, and delayed consequence. AOEP-v0 is a reference design plus a pilot analysis, with the evaluation flow shown in Figure~\ref{fig:aoep}. The pilot illustrates how a survey-derived contract can expose missing governance state in tested memory configurations; it is not evidence that deployed memory systems generally fail or that larger readers could never help once the memory layer exposes the right fields.

AOEP is \emph{representational, not a leaderboard}, and a \emph{wrapper around realistic substrates, not a replacement for them}. AOEP-v0 is not intended to rank deployed memory products or establish benchmark-grade generalization. The contribution is a contract for what a state transition must expose to be scorable, plus a small harness showing how that contract can distinguish stores that expose governance fields from stores that expose only text.

AOEP operationalizes the six axes by giving each one concrete fields in an event stream and state snapshot. Authority becomes permission epochs and approval records; scope becomes actor, resource, privacy class, and sharing fields; mutability becomes typed update, supersession, conflict, and tombstone operations; provenance becomes causal links, trust tiers, signatures, and source identifiers; recoverability becomes idempotency keys, transaction ids, deletion ledgers, and rollback ledgers; and actionability becomes the typed operation attached to the record or side effect. The protocol therefore does not add a separate taxonomy. It turns the earlier axes and lifecycle invariants into the fields a validator can check.

\subsection{Why existing evaluation cannot score governance}

The evaluation literature has strong machinery for two questions and much less for a third. On judge reliability, a long line of work establishes the statistical machinery for trustworthy comparison: \citet{zheng2023judgingllmasajudge} establishes LLM-as-judge via MT-Bench and Chatbot Arena, \citet{gu2024surveyllmasajudge} and \citet{li2024llmsasjudgessurvey} synthesize consistency and bias-mitigation methodology, and \citet{shi2024judgingthejudges} quantifies position bias with repetition-stability and position-consistency diagnostics. \citet{you2026agentasjudge} extends the paradigm from judging outputs to judging agentic trajectories step-by-step. This body of work tells us how to score reliably; it does not tell us \emph{what} state property to score. On the methodology of agent evaluation itself, \citet{kapoor2024aiagentsthatmatter} exposes the absence of cost-versus-accuracy reporting and the overfitting that follows from inadequate holdout sets, \citet{miller2024errorbarsevals} and \citet{madaan2024quantifyingvariance} supply confidence intervals, two-model difference tests, and empirical seed-variance estimates, and \citet{white2024livebench} addresses contamination through monthly-refreshed questions. These works discipline how we report; they assume the unit being reported is task success.

On memory specifically, a 2026 wave has begun to pull retrieval correctness apart from answer correctness. \citet{flynt2026precisionmembench} argues that LoCoMo-style benchmarks conflate answer-correctness with retrieval-correctness and contributes PrecisionMemBench to score retrieval in isolation from the generator. \citet{long2026memtrace} re-bases long-term-memory evaluation on the per-fact knowledge point probed across age, question-type, and evidence conditions, showing that pooled accuracy hides distinct failure modes and that the bottleneck is evidence use, not retrieval. \citet{liu2026streammembench} diagnoses, over egocentric streams, whether agents turn stored evidence and feedback into reliable future behavior, and \citet{shao2026scaleconditioned} holds task evidence fixed while adding irrelevant sessions to find where stored evidence stops being usable. This is real progress, but its frame remains \emph{retrieval and use of evidence for an answer}. Compared against each other, these benchmarks differ in what they isolate (retrieval precision, per-fact tracing, streaming feedback, distractor scaling) yet they share a blind spot: none records whether the value that was retrieved was authorized to act, whether its source was trusted, or whether a later deletion request was honored across the store.

Two recent works move closest to the state-trajectory view and frame the gap precisely. \citet{jung2026meme} contributes a multi-entity evolving-memory benchmark whose Cascade, Absence, and Deletion tasks test whether updates and removals propagate to dependent entries, and reports near-total collapse (Cascade 3\%, Absence 1\%) of current systems on dependency reasoning. \citet{kwon2026reclaim} shows, judge-free, that a memory keeping a stale conclusion but dropping its source becomes confidently uncorrectable and strictly worse than empty memory, fixed only by a source-first write policy. Most directly, \citet{orogat2026gem} reframes long-term agent-memory correctness as a property of the state trajectory rather than individual records, formalizing Governed Evolving Memory with state-level operators and six correctness conditions, and proving that record-level stores cannot satisfy them. AOEP should be read as a prototype instantiation of this state-trajectory view: where \citet{orogat2026gem} supplies a formal impossibility result, AOEP supplies a checkable contract that can be wrapped around concrete memory systems, and where \citet{jung2026meme} measures deletion propagation alone, AOEP scores it alongside authority currency, provenance trust, conflict surfacing, and rollback logging. The gap this subsection exposes is concrete: state-trajectory correctness has been formalized and partially benchmarked, but we found no widely adopted protocol that drives a memory system through restart, conflict, deletion, revocation, and adversarial write while scoring all five governance invariants deterministically. That is the role AOEP-v0 is meant to prototype.

\subsection{Protocol: event stream and state snapshot}

AOEP fixes a data contract rather than a task. An episode is a sequence of typed \emph{events}, replayed to a system under test one at a time, and a \emph{state snapshot} reconstructed at the end from the system's answers to neutral probes. The protocol's claim is that an event must carry strictly more than a memory dump records, and that this added structure is what makes governance scorable.

Concretely, each AOEP event carries fields a flat memory entry omits. It carries an \emph{idempotency key}, so a replayed duplicate write after a restart is recognizable as the same logical operation rather than a second fact. It carries \emph{causal links}, separated into parent, supersedes, and conflicts-with edges plus a transaction id, so that a later value can be marked as superseding an earlier one and a collaborator's proposal can be marked as conflicting with an owner's. It carries a \emph{permission epoch} on its scope, an integer that increments when authority changes, so that an action can be checked against the epoch in force at execution rather than the epoch at write. It carries \emph{provenance} with an explicit \emph{trust tier} and an optional signature, so an untrusted tool result is distinguishable from a user instruction regardless of how persuasive its text is. It carries \emph{retention and privacy constraints}, a retention policy, a privacy class, an optional time-to-live, and a requires-confirmation flag, so that deletion and confirmation obligations attach to the record itself. And it carries one \emph{typed operation} drawn from a closed set of eleven: \texttt{read}, \texttt{write}, \texttt{update}, \texttt{delete}, \texttt{tombstone}, \texttt{share}, \texttt{unshare}, \texttt{validate}, \texttt{quarantine}, \texttt{deny}, and \texttt{rollback}. The closed operation set is what lets the validator reason about the trajectory as a state machine: a \texttt{tombstone} that is not reflected in the deletion ledger, or a \texttt{quarantine} that is later promoted to instruction, is a named, locatable violation rather than a degraded score.

The snapshot is the dual of the stream. It exposes a \emph{deletion ledger}, a \emph{rollback ledger}, the set of \emph{pending conflicts} with their candidate values and the authority backing each, the \emph{per-resource permission epochs} currently in force, and the boolean results of the invariant checks. The protocol deliberately separates the system's self-report from the harness's verdict: the harness recomputes every invariant boolean itself from the reconstructed snapshot and never trusts a system's claim that it behaved. This design borrows the spirit of trajectory-level evaluation from \citet{you2026agentasjudge} while deliberately avoiding LLM-as-judge scoring, because the position, verbosity, and self-enhancement biases documented by \citet{zheng2023judgingllmasajudge} and \citet{shi2024judgingthejudges} would be fatal for a contract whose entire purpose is to detect quiet governance failures. Every AOEP check is deterministic.

This design choice distinguishes a state snapshot from a memory dump, and it directly answers the worked example threaded through AOEP. An episode begins with a user setting a durable billing value; the system restarts and replays a duplicate write; a collaborator proposes a conflicting value in shared scope; the owner then requests deletion. A memory-QA scorer marks the trajectory correct once the final recalled value is right. AOEP fails it if the tombstoned value survives in a derived tier (deletion propagation) or the write was licensed by a lapsed authority epoch (authority monotonicity): distinctions a recall unit has no field in which to record. The gap is tied to the thesis: the wire formats the field actually ships do not carry these fields. The vendor-neutral memory wire format of the substrate literature defines remember/recall/forget/merge/expire over text but specifies neither permission epoch, trust tier, nor causal supersession, so even an interoperable memory layer has nowhere to put the state on which governance depends.

\subsection{Deterministic coverage scorecard}

AOEP's scorecard is deliberately split into two scores rather than a single number. The first is \emph{obligation pass}: the positive obligations a system must \emph{actively} satisfy. These are operations that produce state the agent must maintain, namely recording a deletion, reporting the current permission epoch after a revocation, blocking a stale-permission or untrusted-instruction task, surfacing an owner-versus-collaborator conflict, and logging a rollback after an external action. The second is \emph{negative-invariant pass}: the no-leakage checks, namely that a deleted value is not visible, that an out-of-scope value does not appear, and that an untrusted instruction was not promoted.

The reason the split matters is that the two scores have opposite degenerate solutions. A system that stores nothing trivially passes every negative invariant: it cannot leak a deleted address because it never retained one, and it cannot promote an untrusted instruction because it never kept any text. If a single scalar mixed positive and negative checks, a no-memory floor would look respectable, and the protocol would reward amnesia. By separating them, AOEP makes the no-memory floor legible as what it is: a system that never leaks precisely because it satisfies no positive obligation at all. The headline must therefore be obligation pass, with negative-invariant pass read as a guardrail that catches the opposite failure (a system that recalls so eagerly it leaks). This mirrors, at the protocol level, the lesson of \citet{kwon2026reclaim} that empty memory can dominate confidently-wrong memory, and the lesson of \citet{flynt2026precisionmembench} that a pooled score hides which side of the tradeoff a system actually failed. The validator maps the five named invariants of the lifecycle to ten executable checks (eight per-snapshot booleans plus two ledger-subset checks): authority monotonicity decomposes into \texttt{no\_unauthorized\_writes}, \texttt{permission\_epoch\_current}, and \texttt{no\_stale\_action\_executed}; scope non-expansion into \texttt{no\_scope\_leakage}; deletion propagation into \texttt{no\_deleted\_content\_visible} plus a deletion-ledger subset match; provenance preservation into \texttt{no\_untrusted\_instruction\_promoted}; and rollback traceability into a rollback-ledger subset match plus \texttt{no\_external\_action\_without\_approval}. A non-conforming trajectory fails a specific named check, not a scalar, which is what makes the scorecard diagnostic rather than only comparative.

\subsection{Pilot harness}

To test whether the protocol discriminates real systems rather than only validating hand-authored reference trajectories, we built a harness around the same schema and ran seven systems through nine distinct fault patterns: the three hand-authored worked episodes (restart-conflict-deletion, permission-epoch drift, adversarial memory injection) plus six new patterns that each stress a different governance situation (deletion-derived summary residue, cross-user scope leak, stale permission after a restart, owner/collaborator conflict resolution, rollback of an external action, and untrusted tool-output poisoning). The harness strips outcome-revealing fields from each event, delivers the stream one event at a time, lets each system maintain its state however it chooses, and at the end poses neutral probes that name only target and actor identities, never expected values or invariant names. It then reconstructs the snapshot, computes the invariants itself, and scores against the validator.

Seven systems run. A deterministic \emph{governed reducer} applies AOEP rules directly; it is an oracle-free upper bound, not a deployable product. A \emph{no-memory floor} answers from nothing. Five ungoverned memory systems run over a local Qwen2.5-7B reader: \emph{naive append} (store all event text, retrieve by recency), \emph{full context} (all events in the prompt), \emph{vector-RAG} (all-MiniLM embeddings, top-$k$ cosine), a \emph{Mem0-style} reimplementation in which the reader extracts atomic memory facts from each event into a vector index, and \emph{Mem0 (package)}, a local \texttt{mem0ai} configuration run through a local OpenAI-compatible endpoint backed by the same reader, a local embedder, and an in-process vector store. All computation is local and decoding is greedy, so runs are deterministic. Including this local \texttt{mem0ai} setup reduces the narrow objection that the reimplementation alone created the failure mode; it does not turn the pilot into a product evaluation. The connection to production memory layers is mainly methodological: cost studies such as \citet{wolff2026costaccuracy} evaluate similar store families without pricing mutation or deletion-correctness obligations.

\begin{table}[t]
\centering
\small
\begin{tabular}{lccc}
\toprule
System & Obligation & Neg.-invariant & All \\
\midrule
Governed reducer (upper bound) & $15/15$ & $41/41$ & $92/92$ \\
No-memory (floor) & $0/15$ & $41/41$ & $77/92$ \\
Naive append (Qwen2.5-7B) & $7/15$ & $40/41$ & $83/92$ \\
Full context (Qwen2.5-7B) & $7/15$ & $40/41$ & $83/92$ \\
Vector-RAG (MiniLM+Qwen2.5-7B) & $7/15$ & $40/41$ & $83/92$ \\
Mem0-style (extracted facts) & $4/15$ & $38/41$ & $77/92$ \\
Mem0 (actual \texttt{mem0ai}) & $3/15$ & $36/41$ & $73/92$ \\
\bottomrule
\end{tabular}
\caption{AOEP pilot over nine fault patterns, one frozen reader, greedy decoding. The governed upper bound satisfies every obligation. The three raw-storage systems satisfy the same recall obligations and fail the same governance obligations, scoring identically across recency, full-context, and dense retrieval. The tested extracted-fact configurations score lower (Mem0-style $4/15$; the local \texttt{mem0ai} setup $3/15$), because in this pilot their extraction step drops the structured envelope. The no-memory floor is the only system that never leaks, and it does so by satisfying no positive obligation at all.}
\label{tab:aoep-pilot}
\end{table}

Table~\ref{tab:aoep-pilot} reports the headline. The reading turns entirely on the obligation column. The governed reducer satisfies all fifteen obligations, demonstrating that the obligations are satisfiable inside the reference harness; the no-memory floor satisfies none. The three raw-storage systems all score $7/15$, and the identity of their scores is the most informative fact in the table. They pass the obligations that reduce to \emph{semantic recall}, namely whether a value was deleted, whether an instruction was quarantined, and the status of an active item, all of which a capable reader infers from text. They fail the obligations that require maintained \emph{governance state}: reporting the current permission epoch after a revocation, blocking a stale-permission or untrusted-instruction task, surfacing a conflict, and logging a rollback. None of these is a retrieval problem in the tested configurations, which is why recency, full-context, and dense retrieval score identically. The information is not missing from the context; it is that these text stores have no field in which a permission epoch, a quarantine state, a conflict record, or a rollback ledger lives. This is a pilot instantiation of the state-trajectory concern formalized by \citet{orogat2026gem}: three stores that differ in retrieval strategy fail the same governance subset when the required envelope is absent.

The extracted-fact systems sharpen the point by scoring \emph{below} raw storage in this pilot. The Mem0-style store reaches $4/15$ and the local \texttt{mem0ai} configuration reaches $3/15$, because extracting atomic facts into a vector index flattens the structured event (its permission epoch, deletion link, and trust tier) into free text, so even the deletion-awareness that raw storage incidentally keeps can be lost. In these tested configurations, extraction without an explicit governance envelope performs worse on governance obligations, not better. The local package setup leaks a deleted billing address and an untrusted exfiltration address into its retrieved memories on two episodes, which is the cross-session injection threat and the deletion-residue failure made measurable in a concrete implementation. This is not a defect claim about \texttt{mem0ai} as software, nor a verdict on extracted-fact architectures in principle. It is a scoped demonstration of the failure that source-first write policies \citep{kwon2026reclaim} and dependency-propagation tasks \citep{jung2026meme} would predict when the envelope is absent.

\subsection{Robustness of the pilot}

The pilot result invites three objections, each of which would make the observed gap an artifact of setup rather than a property of the tested designs.

\emph{Reader variation does not remove the pilot gap.} Running the full-context system, where the entire event stream sits in the prompt so any failure is missing governance state rather than missing retrieval, across four readers spanning sizes and families holds obligation pass at or below the original while the largest tested reader does not help. Table~\ref{tab:aoep-reader} reports the span: Qwen2.5-3B, Qwen2.5-7B, and Llama-3.1-8B all score $7/15$, Mistral-7B-Instruct scores $4/15$, and the governed reducer scores $15/15$ on the same patterns. Within this local sweep, the missing objects are not facts hidden from the reader but explicit fields the text store never maintained: an epoch counter, a deletion ledger, and a conflict record. This is a pilot-level version of the separation \citet{long2026memtrace} draws when it finds the bottleneck to be evidence use rather than retrieval, and the same one \citet{shao2026scaleconditioned} draws when stored evidence stops being usable independent of model strength.

\begin{table}[t]
\centering
\small
\begin{tabular}{lcc}
\toprule
Full-context reader & Obligation & Neg.-invariant \\
\midrule
Qwen2.5-3B-Instruct & $7/15$ & $34/41$ \\
Qwen2.5-7B-Instruct & $7/15$ & $40/41$ \\
Mistral-7B-Instruct-v0.3 & $4/15$ & $39/41$ \\
Llama-3.1-8B-Instruct & $7/15$ & $38/41$ \\
\midrule
Governed reducer (upper bound) & $15/15$ & $41/41$ \\
\bottomrule
\end{tabular}
\caption{Reader robustness over the nine fault patterns, full-context system. In this local $3$B to $8$B sweep across three model families, no tested reader approaches the governed upper bound on governance obligations; the larger reader does no better and the smaller one leaks more. The result is consistent with a missing-governance-state explanation, but it is not a model-scaling claim.}
\label{tab:aoep-reader}
\end{table}

\emph{It is not a single-run effect.} The main runs decode greedily. Re-running the full-context system under sampled decoding (temperature $0.7$, top-$p$ $0.95$) across five seeds holds obligation pass in a $6$ to $8/15$ band (mean $6.6$, standard deviation $0.8$), straddling the greedy $7/15$ and nowhere near the governed $15/15$. The governance gap is stable to sampling variance, which is the kind of seed-sensitivity control that \citet{madaan2024quantifyingvariance} argues every agent result should report and that \citet{miller2024errorbarsevals} formalizes; we report it here so the separation is not a single lucky draw.

\emph{It is not an artifact of synthetic fault patterns.} To check that the split survives realistic task shapes, we authored three longer, multi-actor episodes from documented agent-task benchmarks: a tau-bench-style customer-refund flow (approval policy, an untrusted chat message redirecting the refund, a stale retry, a card-deletion request), a TheAgentCompany-style enterprise-scheduling flow (granted then revoked calendar access, a contested reschedule, a stale write), and one grounded in an actual AppWorld task \citep{trivedi-etal-2024-appworld}, the group-dinner Venmo payment-request task, using AppWorld's real apps and operations with an untrusted note redirecting the payment and a card-deletion request. The pattern holds and sharpens: across the three episodes the governed reducer satisfies all nine obligations, the three raw-storage systems satisfy four and the Mem0-style store two, and the raw-storage systems leak the attacker's redirect instruction (and, on the Venmo task, the stored payment card) into the action flow. The recall-versus-governance gap is a property of the memory designs, not of the minimal fault patterns. This matters because \citet{chen2026crossscenario} shows that memory systems tuned to one trajectory format collapse across five others; the persistence of the AOEP split across hand-authored, realistic, and benchmark-grounded shapes gives the same warning from the governance side.

\emph{It reproduces on real execution traces.} The realistic episodes are hand-authored from real tasks; to ground the base streams in real executions we recorded genuine AppWorld traces by driving its actual API server across three apps with verified side effects. On the Venmo task, \texttt{create\_payment\_request} returned a live request id and \texttt{delete\_payment\_card} dropped the account from four stored cards to three; on a Spotify library-cleanup task, \texttt{remove\_song\_from\_library} dropped the library from fourteen songs to thirteen and the song stayed absent; on a file-system task, a create-then-delete created and removed a file holding a secret, verified present then gone. Each real call was captured as an AOEP event, and an untrusted note was added (redirect the payment, re-add the deleted song, recreate the deleted file). Across the three traces the governed reducer satisfies all seven obligations while the ungoverned systems satisfy three to six, and each leaks an untrusted instruction on one trace. The base operations actually executed against the API backend with verified side effects; only the perturbations and probes are authored. Finally, on \emph{agent-chosen} trajectories, a stronger local policy (Qwen2.5-Coder-32B) drove AppWorld autonomously, fetching credentials, logging in, and completing three single-app tasks in its own self-chosen steps (a harder multi-step task it did not finish, which we report rather than hide); when one untrusted observation was injected mid-trajectory, the governed reducer handled it on all three completed runs while the full-context system leaked the exfiltration instruction. The gap appears even on steps the agent selected rather than ones we scripted.

\subsection{What the pilot establishes, and what it leaves open}

Read together, the pilot and its robustness checks support a scoped qualitative claim: recall and governance separate once the evaluation contract asks for fields such as permission epochs, deletion ledgers, trust tiers, conflict records, and rollback ledgers. Recall is necessary and the raw-storage systems have it; governance requires representing state that a flat or extracted store may not carry unless it is explicitly instrumented to do so. In this local sweep, no tested reader from 3B to 8B across three families manufactures that state from text it can read. The local \texttt{mem0ai} configuration failing below raw storage makes the point concrete, but the paper should not be read as ranking that package or as establishing class-wide performance of extracted-fact systems.

\paragraph{A required ablation before stronger claims.} The pilot should be read as a configuration-level result unless and until one ablation is run: add a minimal governance envelope to an extracted-fact store, consisting of a permission epoch, deletion ledger, and provenance trust tier, and rerun the same nine patterns. This is the unit test that separates two different claims. If the envelope closes most of the gap, then the result is that current tested configurations omit governance metadata. If it does not, then extraction itself may be discarding structure in a way that the envelope cannot repair. The present manuscript supports only the first, weaker reading: in the tested configurations, the fields required by AOEP are absent, so readers and retrieval strategies cannot recover them.

The limitations follow from that gate. This is a pilot, not a benchmark: nine fault patterns plus three realistic episodes and three real execution traces, scored over a handful of readers and memory systems. AOEP is built to wrap realistic substrates, and the natural next step is to drive it across hosted, multi-task agent benchmarks at the scale that \citet{kapoor2024aiagentsthatmatter} demands for cost-aware reporting and that the snapshot-collapse phenomenon of \citet{zheng2026seagym} warns must be measured over long horizons rather than single snapshots. The pilot measures whether ungoverned systems \emph{do} expose the required governance state, not whether they \emph{could} govern if minimally augmented. It scores one local \texttt{mem0ai} configuration as run, and that configuration carries no permission epoch, deletion ledger, or trust tier. We did not run the envelope ablation above, so we do not claim to separate an architectural limit from an instrumentation gap. The deeper gap, tied directly to the thesis, is the one the corpus quantifies as a scoping estimate: rollback is studied as a mechanism but rarely measured, with only $27$ of $435$ works exposing any rollback mechanism and none in our corpus reporting recovery success or cost after corruption. AOEP scores rollback as one obligation among fifteen; what it does not yet do is measure the \emph{cost} of recovery, the latency and side-effect-replay correctness of undoing a corrupted decision once the authority that licensed it has lapsed. Prospective and spontaneous governance is a second open axis: \citet{zhang2026triggerbench} shows that spontaneous recall and action on a latent stored constraint is far harder than answering a direct query and degrades with context length. An always-on agent therefore has to honor deletion or revocation when asked and also surface the obligation unprompted when the context demands it, a capability AOEP can perturb through its delayed-consequence pattern but does not yet score at scale. The evaluation agenda is to measure recovery cost and prospective governance over realistic substrates with the statistical rigor of \citet{miller2024errorbarsevals} and \citet{madaan2024quantifyingvariance}. AOEP's specific advance over the survey's earlier parts is that it turns the diagnosis into a checkable contract for what a state transition must expose.

\figaoep
\section{Resources and frameworks}\label{sec:resources}

\regime{Control}{what each application domain demands of persistent state, read as a stress test.}
Persistent state, its lifecycle, and its governance look different once a domain supplies the substrate on which an always-on agent actually runs. The persistence problem does not arrive in a generic form. A clinical agent and a web agent both accumulate, retrieve, and act on state, but answer to different obligations: the clinical agent must reconcile a patient's recall-biased self-report against an authoritative record and survive a regulator's audit; the web agent must not let a stale workflow abstraction fire an irreversible purchase. Each domain therefore supplies a stress test that sharpens one or more of the six axes (authority, scope, mutability, provenance, recoverability, actionability) and one or more of the five invariants (authority monotonicity, scope non-expansion, deletion propagation, provenance preservation, rollback traceability). The question is not what an agent in the domain does, but what persistence demand the domain adds, which benchmarks and systems expose it, and where the governance gap remains open.

The domain view shows that the skew is not a property of the methods literature alone. Applications inherit it. Domains race to build agents that remember more, recall faster, and personalize more aggressively, and they build benchmarks that score recall and task success. The return arc of the lifecycle, which validates a write before it lands, propagates a deletion, preserves provenance through consolidation, and traces a rollback, is thin in every domain we examine, and often thinnest where the legal stakes are highest. Healthcare adds consent and auditability but rarely couples memory to a HIPAA-style audit cycle; finance, legal, and cybersecurity add non-repudiation and rollback obligations but score portfolio returns or verdict accuracy rather than recovery after a poisoned or lapsed-authority action. The domain view therefore reproduces the corpus-wide skew at a finer grain, naming the missing instrument per domain rather than only in aggregate.

\subsection{Web and computer-use agents}

Web and computer-use agents supply the canonical environments against which always-on claims are now measured. The foundational substrate is a set of realistic, resettable environments: WebArena provides a self-hosted web environment with reproducible sites \citep{zhou2024webarenarealisticwebenvironment}; OSWorld grounds open-ended multimodal control in a real operating system with screenshots in and keyboard or mouse out \citep{xie2024osworldbenchmarkingmultimodalagents}; AndroidWorld exposes a live Android device with programmatic tasks that read, modify, and tear down real system state \citep{rawles2024androidworld}; and BrowserGym unifies many browser-task environments into one ecosystem for cross-environment comparison \citep{lesellierdechezelles2025browsergym}. WorkArena and its enterprise extensions push the same substrate toward consequential knowledge work atop ServiceNow \citep{drouin2024workarena, boisvert2024workarena}, and TheAgentCompany frames an entire simulated software company so tasks resemble the durable, multi-day obligations of a real workplace \citep{xu2025theagentcompany}. The comprehensive survey of LLM-brained GUI agents maps this area across web, mobile, and desktop, cataloguing action spaces and memory strategies \citep{zhang2024llmguisurvey}. These environments were built to test capability, not persistence: each task resets, identity is fresh, and there is no consent boundary or recovery obligation to violate.

The persistence demand this domain adds is most visible in the systems that carry state across the resets these benchmarks impose, where two strategies have emerged. The first builds an external, retrievable record of past interaction. Agent Workflow Memory induces reusable workflow abstractions from prior trajectories so a procedure learned in one session applies in the next \citep{wang2024agentworkflowmemory}; ReasoningBank distills generalizable reasoning strategies from self-judged success and failure trajectories into a persistent, retrievable store with a closed update loop \citep{ouyang2025reasoningbank}; WebAtlas builds a persistent cognitive map of websites through exploration, storing failed interactions as cross-task memory for look-ahead planning \citep{cheng2025webatlas}; and branch-and-browse exploration shares a page-action memory within and across sessions, with web-state replay as a recovery primitive \citep{he2025branchandbrowse}. The mobile and computer-use line follows the same logic: OS-Copilot's FRIDAY self-improves by accumulating learned skills and tools and reusing them on unseen applications \citep{wu2024oscopilot}; Cradle pairs general computer control with explicit memory and skill-curation modules that persist across games and software \citep{tan2024cradle}; Agent-S introduces experience-augmented hierarchical planning that combines external knowledge search with internal experience retrieval \citep{agashe2024agents}; Mobile-Agent-E accumulates reusable Tips and executable Shortcuts across tasks \citep{wang2025mobileagente}; MobileRAG's MemRAG component retains and retrieves prior interactions so the agent stops reconstructing the interface from scratch \citep{loo2025mobilerag}; and ReinAgent organizes a multi-agent GUI system around a shared memory module with slot-based information management and proactive updates \citep{jia2025reinagent}. The second strategy folds persistence into the model's own behavior: STAMP trains agents to learn what and when to memorize and retrieve via deterministic memory variables in virtual environments \citep{wang2026stamp}, and the dual-memory M2 design for long-horizon web agents combines training-free internal trajectory summarization with an external offline insight bank \citep{yan2026m2dualmemory}. PerPilot adds a personalization layer, pairing memory-based retrieval with reasoning-based exploration to resolve under-specified personalized instructions \citep{wang2025perpilot}.

A recent result shows that this accumulation is not self-justifying. Under a token-matched budget, a vanilla baseline matches or beats online memory, skill, and workflow modules on WebArena once the retrieval cost of those modules is counted against them \citep{hajimiri2026budgetconstrained}. This cost-of-memory critique shows the value of persistent web state is conditional on governance, not volume, a domain-specific version of the corpus-wide pattern. Two further results sharpen where the governance has to live. iOSWorld is the first native iOS simulator benchmark built around a persistent user identity across twenty-six apps, with a large fraction of its tasks explicitly requiring that identity to be carried correctly \citep{jang2026iosworld}, making identity and scope first-class rather than reset away. The naive-visual-memory study shows experiential full-image visual memory helps GUI state recognition but actively hurts action grounding, isolating a perception-reasoning-action failure taxonomy on OSWorld \citep{choi2026naivevisualmemory}: accumulated state is net-negative at the act boundary, a direct instance of the actionability axis going wrong. The Darwin roadmap for self-evolving GUI agents names memory management as one of three pillars alongside task curricula and outcome verification \citep{beechey2026darwin}, treating persistence as a design axis. The problem is concrete: these environments read, modify, and tear down real device and web state, with real side effects (a sent message, a deleted record, a committed purchase), yet the benchmarks we examined do not score whether the agent can roll that side effect back when the action turns out to have been licensed by stale or poisoned state. AndroidWorld and OSWorld check final state for success, not recoverability. The web domain has a strong substrate for testing actionability and a much thinner one for testing rollback traceability.

\subsection{Embodied and robotics agents}

Embodied agents add a demand the web domain can avoid: the agent's state is a claim about a physical world that is only partially observable, drifts on its own, and cannot be reset by reloading a page. Persistence here is more than retrievable history. It is a maintained belief about object locations, world rules, and irreversible physical actions, and the benchmark literature has moved toward measuring this belief rather than task success alone. WorldLines builds a long-horizon household benchmark over action-native state trails with paired Memory-QA and Task-Planning tracks, isolating overwritten-world-state and partial-observability failures and supplying a visibility-aware memory baseline \citep{zhang2026worldlines}. Ego2World compiles egocentric cooking videos into hidden rule-governed worlds where the agent plans over a partial belief graph, isolating belief-state maintenance and showing that action-overlap metrics overestimate physical-state success \citep{cheng2026ego2world}. LMEE scores the process of memory utilization as well as the outcome for embodied exploration \citep{wang2026lmee}. These three mark a maturation: the field now grants that an embodied agent can reach a goal with a corrupted belief, a failure even when the task score is positive, an actionability-versus-provenance distinction the early navigation benchmarks could not express.

The mechanism literature responds with three contrasting approaches to keeping embodied belief trustworthy, and the contrast exposes a governance frontier. The first treats spatial memory as a high-fidelity reconstruction supporting re-observation: GS-Mem uses 3D Gaussian Splatting as persistent spatial memory to grant post-hoc re-observability so observations missed at capture time can be recovered later \citep{lu2026gsmem}, a substrate-level recoverability mechanism. The second injects calibrated uncertainty into stored semantics: the remember-with-confidence approach equips spatio-temporal memory with per-object semantic uncertainty derived from cross-view caption scatter and performs budget-constrained active refinement, so unreliable stored descriptions are flagged and resolved rather than trusted as oracle \citep{zhang2026rememberconfidence}, one of the few works in any domain that builds validation directly into the stored state. The third draws on cognitive psychology to diagnose how embodied memory fails: EMEM organizes an embodied-memory benchmark over ProcTHOR around eight cognitive paradigms (false-memory lures, pattern separation, source monitoring, serial position, long-horizon interference), separating false-memory and consolidation failures from spatial-storage failures \citep{rasheed2026emem}. The social and lifelong line extends persistence across community interaction: ELLA equips an embodied social agent with name-centric semantic plus spatiotemporal episodic memory accumulated over lifelong community life \citep{zhang2025ella}; RoboOS-Next proposes a unified memory representation for lifelong adaptation and scalable coordination across heterogeneous robot teams \citep{tan2025roboosnext}; MindForge frames lifelong cultural learning through structured theory-of-mind shared across agents \citep{lica2024mindforge}; and GOAT-Bench tests navigation to a sequence of multimodal goals within an episode, reusing memory of previously found objects \citep{khanna2024goatbench}. The cross-session stress test is RoboMME, which pads a robot's query history with unrelated sessions and shows memory-augmented vision-language-action policies decay as distractor sessions accumulate \citep{rathi2026robomme}, the embodied face of the interference failure mode.

The persistence demand here is that physical state has no undo. A web purchase can sometimes be refunded; a manipulation that knocks over a glass cannot be rolled back, only mitigated. This makes recoverability and provenance preservation the binding constraints, yet the benchmark designs above still score belief fidelity and task success, not whether the agent records enough provenance to explain a wrong physical action or maintains the recovery path that mitigation requires. EMEM diagnoses where false memories enter but does not score deletion propagation when a hallucinated object must be purged from a downstream plan; remember-with-confidence flags unreliable descriptions but does not test rollback of an action already taken on a description later found wrong. The open problem for embodied agents is therefore a governance loop matched to physical irreversibility: a write-validate step that gates which beliefs may authorize a non-reversible motor command, and an audit trace sufficient to attribute a physical error to the stored belief that caused it. No embodied benchmark in the corpus scores this.

\subsection{Healthcare}

Healthcare turns the governance gap into a legal and safety obligation. The persistence demand the domain adds is twofold: consent and deletion under regimes such as HIPAA and GDPR, and auditability of why a clinical decision was made and on what evidence. State here is the longitudinal patient record, and it carries an authority structure absent from web or robotics: the authoritative source (the chart, the FHIR record) outranks the conversational source (what the patient recalls), and conflating them is a documented safety hazard. The benchmark family has converged on this longitudinal, multi-stage character. AgentClinic was the seminal clinical-agent benchmark to study a persistence axis explicitly, with a notebook memory that survives across cases \citep{schmidgall2024agentclinic}. MedAgentBench defines the EHR-agent benchmark family with an actionability and scope axis built in, since its tool calls write irreversibly against patient records \citep{jiang2025medagentbench}. ClinEnv benchmarks an LLM as attending physician over an ordered sequence of irreversible decision stages per admission, making sequential irreversibility first-class \citep{lu2026clinenv}. MediLongChat tests recall and reasoning over a patient's longitudinal history with a dedicated cross-dialogue reasoning task \citep{hu2026medilongchat}, and ESMemEval defines a longitudinal user-memory benchmark family with explicit state axes spanning information extraction and temporal reasoning \citep{chen2026esmemeval}. PsychEval frames counseling as a longitudinal task requiring sustained memory and dynamic goal tracking across sessions \citep{pan2026psycheval}, a framing TheraMind operationalizes with an explicit cross-session persistent-state lifecycle: write after a session, retrieve within a session, and an update strategy across the therapeutic relationship \citep{hu2025theramind}.

The clinical mechanism literature contains some of the most governance-relevant ideas in any application domain, because clinicians will not accept an agent that silently overwrites a record. MedMemoryBench directly targets always-on memory in clinical agents and formalizes the memory-saturation failure mode, where sustained information inflow degrades the agent's ability to use its own memory \citep{wang2026medmemorybench}, a continual-operation degradation result with clinical stakes. VitalTrace replaces unbounded patient-history serialization in an ICU multi-agent framework with a compact persistent patient-state memory whose updates obey curated physiological state-transition rules, improving temporal consistency over long trajectories \citep{qu2026vitaltrace}, a rare instance of a validation step (rule-governed transitions) gating writes. TrajOnco performs temporal reasoning over sequential clinical events through a multi-agent long-term-memory framework \citep{zeng2026trajonco}. The strongest governance pattern is the dual-stream clinical memory that separates the patient's recall-biased self-report from the authoritative FHIR record and runs a reconciliation engine to flag discrepancies, rejecting the overwrite-with-latest-statement default as a safety hazard \citep{pugh2026dualstream}. This is authority monotonicity and provenance preservation realized in a deployed-style clinical design: the authoritative source is never silently superseded by a lower-authority statement, and discrepancies are surfaced rather than resolved by recency.

Yet even here the gap persists: these systems build provenance separation and rule-gated updates, but none couples the memory lifecycle to the audit and deletion cycle the regulation actually requires. A right-to-be-forgotten request under GDPR, or a record-amendment obligation under HIPAA, demands that a deletion propagate through every derived tier (summaries, embeddings, cached prompts, inter-agent messages, fine-tuned adapters) and that the propagation be verifiable; no clinical benchmark in the corpus scores deletion completeness across these tiers, and no clinical system demonstrates rollback of a clinical action after the authorizing record is amended. TheraMind maintains a therapeutic relationship across sessions but offers no erasure guarantee; ClinEnv stages irreversible decisions but does not test recovery after a decision is later found to rest on an amended record. Domain-specific governance, HIPAA-style audit trails coupled to the memory lifecycle and GDPR-style verifiable erasure, is the largest unaddressed demand here, which is why the dual-stream and rule-gated systems read as exceptional rather than standard.

\subsection{Finance, legal, and cybersecurity}

Finance, legal, and cybersecurity agents share a persistence demand that healthcare only partially carries: non-repudiation. A trade, a filing, or a credential grant is an act whose record must be tamper-evident and attributable after the fact, and whose reversal (where reversal is even possible) must be traceable. The state these agents hold is a ledger of consequential commitments, and the binding axes are authority (who licensed the action), provenance (what evidence and which credentials backed it), and rollback traceability (can the action and its derived state be unwound, and is the unwinding itself recorded). The finance literature is the oldest in this group. FinMem is the canonical trading agent whose core contribution is a layered deep, intermediate, and shallow memory module aligned to decision timescales \citep{yu2023finmem}; FinAgent extends this with a diversified memory-retrieval system over historical market data plus a dual reflection module \citep{zhang2024finagent}; FinCon is a manager-analyst multi-agent system whose risk-control component episodically self-critiques to update persistent investment beliefs \citep{yu2024fincon}; and InvestorBench was the first standardized benchmark for memory-equipped financial decision agents across products and market environments \citep{li2024investorbench}. Two recent benchmarks reframe finance around state explicitly: $\tau$-Banking extends $\tau$-bench into a fintech customer-support domain where agents must navigate roughly seven hundred policy and procedure constraints \citep{shi2026tauknowledge}, and the thousand-day-scale RetailBench simulates a single store's pricing, replenishment, inventory aging, and cash flow, where errors compound over the horizon \citep{zhang2026retailbench}. KTD-Fin sharpens a provenance subtlety the other finance work elides, separating pretraining-memorized market data, fixed at the knowledge cutoff, from genuinely retrieved current state \citep{zhu2026ktdfin}, a provenance-axis question (is this belief sourced from current evidence or a stale parametric prior) that directly governs whether a trade is justified.

The legal line is younger but unusually aligned with the persistent-state frame because legal procedure is, by construction, a governed state machine. LegalWorld models Chinese civil litigation as a causally connected five-stage state chain built on tens of thousands of paired judgments \citep{zuo2026legalworld}; SimCourt was the first framework to model the procedural structure of Chinese criminal trials with five courtroom roles whose legal agents carry role-bound state \citep{zhang2025simcourt}; and the civil-court multi-agent simulation integrates a memory module with statute retrieval to support long-process adjudication \citep{chen2026civilcourt}. These systems implicitly demand authority monotonicity (a later stage cannot expand the authority granted by an earlier ruling) and provenance preservation (a verdict must be traceable to the evidence and statutes that produced it), because that is how legal procedure works; but they are evaluated on adjudication accuracy, not on whether the agent's state respects those procedural invariants under perturbation. Table~\ref{tab:resources-finlegal} contrasts what each representative resource measures against the governance obligation the domain imposes.

\begin{table}[t]
\centering
\small
\setlength{\tabcolsep}{4pt}\renewcommand{\arraystretch}{1.2}
\begin{tabularx}{\linewidth}{p{0.17\linewidth}YYY}
\toprule
Domain & Representative resource & Persistence demand added & Governance gap \\
\midrule
Finance (trading) & FinMem \citep{yu2023finmem}, FinAgent \citep{zhang2024finagent}, InvestorBench \citep{li2024investorbench} & Non-repudiation of trades; timescale-aligned belief & Scores returns, not rollback or audit of a wrong trade \\
Finance (current vs. prior) & KTD-Fin \citep{zhu2026ktdfin} & Provenance of belief: current evidence vs. parametric prior & No deletion or correction of a stale-prior-driven action \\
Finance (long horizon) & RetailBench \citep{zhang2026retailbench} & Compounding over thousand-day operations & No recovery instrument for compounded state error \\
Legal procedure & LegalWorld \citep{zuo2026legalworld}, SimCourt \citep{zhang2025simcourt} & Procedural state chain; role-bound authority & Scores verdict accuracy, not invariant preservation under perturbation \\
Cybersecurity-adjacent & $\tau$-Banking \citep{shi2026tauknowledge} & Policy-constrained action over hundreds of rules & No test of credential rollback after lapsed authority \\
\bottomrule
\end{tabularx}
\caption{Finance, legal, and security resources read through the non-repudiation and rollback demand each adds. The recurring gap is that the benchmarks score the outcome (return, verdict, policy compliance) but not the return arc of the lifecycle (audit, deletion, rollback) that non-repudiation requires.}
\label{tab:resources-finlegal}
\end{table}

The gap statement for this group is the most direct in the part. Non-repudiation is, in lifecycle terms, the conjunction of provenance preservation and rollback traceability: every consequential act must be attributable and, where reversed, the reversal must itself be auditable. The systems above build sophisticated memory for deciding what to do and essentially nothing for unwinding what was done: none scores rollback of a tool-originated commitment, a placed trade, a granted credential, a filed document, after the authority that licensed it lapses or is found invalid, the exact open instrument the broader corpus identifies, a retried or reversed action must not resurrect already-consumed authority or duplicate an irreversible side effect. Where the side effect is money, a legal filing, or a security credential, the absence of this instrument is not an academic gap but a deployment blocker, unaddressed across every resource in this group.

\subsection{Education and tutoring}

Education agents add a persistence demand that looks gentle but is governance-heavy: the authoritative state is a model of what the learner knows, and that model must mutate correctly over time. A tutoring agent that treats a once-mastered skill as permanently mastered, or fails to decay knowledge the student has since forgotten, will mis-instruct; an agent that overwrites a misconception rather than verifying it has been corrected will record success where there is none. The binding axes are mutability (the learner state evolves at the speed of learning and forgetting) and provenance (the basis for believing a skill is mastered), and the mechanism literature is unusually explicit about modeling this evolving state. TutorLLM provides a concrete persistent learner-state substrate, using a knowledge-tracing model to predict per-student knowledge state as the memory the tutor conditions on \citep{li2025tutorllm}. TASA couples an event memory and persona profile with a continuous forgetting curve over knowledge tracing, deliberately decaying previously-mastered skills so instruction is recalibrated to the learner's current mastery \citep{wu2025tasa}, one of the clearest domain instances of principled, relevance-driven forgetting being a feature rather than a failure. DeepTutor builds an agentic tutor with a dynamic learner memory that adapts to evolving needs and ships TutorBench with learner profiles \citep{zhao2026deeptutor}; AgentTutor formalizes a persistent learner-state lifecycle with organize and retrieve operations over a learner profile plus experience memory \citep{liu2026agenttutor}; and PsychAgent combines persistent episodic and experiential memory with weight-level consolidation, internalizing skills via rejection fine-tuning \citep{yang2026psychagent}, a parametric-plus-non-parametric hybrid that raises the provenance question of where a consolidated skill came from.

Contrasting these makes a mechanism point. TASA and TutorLLM treat learner state as something tracked and decayed by a principled model, so mutation is governed by a knowledge-tracing prior; DeepTutor and AgentTutor treat it as adaptive memory shaped by interaction, so mutation is governed by what the agent observes. The former is more auditable (decay follows a curve), the latter more flexible (it captures idiosyncratic learner change), and neither validates the key correctness condition: that a recorded mastery update reflects genuine understanding rather than a lucky answer, and that a recorded misconception was corrected rather than overwritten. The gap is validation at the write boundary and recoverability of the learner model. Education has, in TASA's forgetting curve, one of the better forgetting mechanisms in the corpus, but no education benchmark scores whether an erroneous mastery update can be detected and rolled back, and FERPA-style governance over the durable learner record, who may read, amend, or erase it, is absent here. The learner state is the most sensitive long-horizon record after the medical chart, and it is governed least.

\subsection{Coding agents}

Coding agents make the persistence demand concrete and testable in a way few other domains do, because a codebase is a large, structured, versioned state with an existing notion of rollback (the commit graph) that the agent's memory must respect. The demand the domain adds is cross-session coherence: an agent must hold a project's conventions, prior fixes, and codebase structure across sessions without repeating known mistakes or violating learned conventions. The codified-context work documents the core persistence failure directly: coding agents lose coherence and conventions across sessions and repeat known mistakes \citep{vasilopoulos2026codifiedcontext}. The systems answer with hierarchical persistent memory. Prometheus builds persistent codebase-navigation memory beyond the context window, exercising the organize and retrieve lifecycle and the scope axis over a repository \citep{pan2025prometheus}; MemRepair maintains a hierarchical persistent memory that reuses past fixes and learns from failed validation feedback across repair attempts \citep{liu2026memrepair}, one of the few systems anywhere that treats a failed validation as a first-class write rather than a discarded outcome. On the benchmark side, coding contributes useful always-on stress tests. MemoryCode is a synthetic multi-session coding benchmark that tests whether an agent can retrieve and act on instructions across sessions amid distractors \citep{rakotonirina2025memorycode}, and SWE-Bench-CL restructures SWE-Bench-Verified GitHub issues into chronological per-repository task streams, pairing a semantic-memory agent with forgetting and forward and backward-transfer metrics and a composite continual-learning score \citep{joshi2025swebenchcl}. SWE-Bench-CL is notable because it imports continual-learning evaluation, transfer and forgetting metrics, into a domain with real ground truth (does the patch pass the test), making it one of the rare resources that measures whether accumulation compounds or interferes rather than assuming it helps.

The contrast with the web domain is informative: coding agents operate over state that already has version control, so the rollback primitive exists at the substrate, yet the memory layer does not use it. MemRepair learns from failed validations but does not roll its own memory back to a known-good state when a learned fix later proves wrong; Prometheus navigates and reasons but does not check consistency when it modifies shared code state. The open problem is to bind the agent's persistent memory to the codebase's own rollback semantics, so reverting a commit also reverts the memory writes derived from it (deletion propagation), and a poisoned procedural memory, a learned anti-pattern presented as a fix, can be quarantined and traced (provenance preservation, rollback traceability). The domain has the cleanest available substrate for studying lifecycle-complete memory governance and has not yet used it for that purpose.

\subsection{Ambient, wearable, and IoT agents}

Ambient, wearable, and IoT agents most fully realize the always-on premise, because the agent literally never resets: it observes a continuous, multimodal life stream and is expected to surface help proactively. The persistence demands it adds are three. First, capture is unbounded and continuous, so the write and consolidation boundaries face the heaviest pressure in any domain. Second, the agent must decide when to act unprompted, making trigger state and proactivity calibration a first-class state type rather than an afterthought. Third, the stream is intimate first-person life data, so privacy and scope are acute. The benchmark literature here is the newest and most directly aimed at the always-on requirement. EgoLife provides a three-hundred-hour, six-participant, one-week egocentric AI-glasses dataset and the EgoLifeQA benchmark for a life assistant that recalls past events and monitors habits, with an EgoButler retrieval-memory system \citep{yang2025egolife}. TeleEgo benchmarks always-on egocentric assistants over fourteen-plus hours of synchronized streaming multimodal data and, distinctively, scores Memory Persistence Time and Real-Time Accuracy across twelve subtasks \citep{yan2025teleego}, naming time-to-persistence as a metric rather than only recall. LifeDialBench pushes to a one-year simulated horizon plus a seven-day real EgoMem, evaluated under an online temporal-causality protocol, and reports a finding that anchors this domain discussion: lossy sophisticated memory systems underperform a high-fidelity RAG baseline \citep{zheng2026lifedial}. This is the ambient-domain echo of the cost-of-memory critique: elaborate consolidation can be net-negative when it drops the handles a later query needs.

The mechanism and proactivity line is where this domain contributes a state type the others underweight. ContextAgent fuses open-world sensory perception with personas built from historical data to predict when proactive service is needed and then calls tools, releasing the ContextAgentBench benchmark \citep{yang2025contextagent}; Memento is a wearable AR assistant that permanently captures verbal queries with spatiotemporal and activity context and proactively re-surfaces recurring interests when a matching context recurs \citep{kim2026memento}; and EgoSelf builds a graph-based interaction memory of a user's first-person history to capture long-term habits and predict future interactions \citep{wang2026egoself}. The smart-home and IoT line extends the same proactivity into device control under constraints: PersonalHomeBench evaluates foundation-model agents in personalized homes by progressively building evolving household state and scoring both reactive and proactive ability \citep{bharadwaj2026personalhomebench}; IoTGPT memorizes decomposed instruction subtasks for reuse to cut LLM calls while adapting them to per-user preferences \citep{yu2026iotgpt}; AirAgent maintains a dynamic memory-tag layer that continuously updates a personalized user profile and fuses it with live sensor data to drive proactive device control under explicit constraints \citep{men2026airagent}; and the shopping-companion benchmark, though e-commerce, shares the ambient pattern of persisting cross-session preference over a 1.2M-item pool and isolating preference-hallucination cascades unique to long-horizon purchasing \citep{yu2026shoppingcompanion}.

The governance gap is acute for ambient agents because the demands are high and the instruments remain young. Proactivity calibration, the trigger-state question of when to surface help versus stay silent, is named (ContextAgent, the broader proactive-agent line \citep{lu2024proactiveagent}) but not governed: no ambient benchmark we examined scores the authority and scope conditions under which a proactive action is licensed, so an over-eager surfacing (intrusion) and a missed-help failure are treated as accuracy rather than scope or authority violations. Privacy is acute and under-addressed: the life stream is the most sensitive durable record in any domain, yet ambient resources rarely score deletion completeness across the derived tiers a wearable accumulates (raw video, transcripts, summaries, habit graphs, proactive-trigger indices), and a right-to-be-forgotten request against a year of lifelog has no demonstrated propagation path. LifeDialBench's finding that lossy consolidation underperforms high-fidelity retrieval is the consolidation-boundary failure made domain-specific: always-on memory needs consolidation that remains addressable by retrieval, with the handles preserved, and the ambient domain is where the pressure to drop them is highest and the cost best measured.

\subsection{Scientific discovery agents}

Scientific-discovery agents add a persistence demand distinct from the others: the durable state is the research program itself, the accumulating record of what has been tried, what worked, and what failed, and its governance question is whether that record steers exploration correctly over a campaign rather than misleading it. Two representative systems make the structure explicit. DeepScientist's main state object is a cumulative, tiered Findings Memory in which validated findings accumulate, are promoted to higher-fidelity validation, and steer a Bayesian explore-exploit policy across a month-long discovery campaign \citep{weng2025deepscientist}. EvoScientist maintains separate persistent ideation and experimentation memories that record both feasible and previously-failed directions so research strategies improve across the program \citep{lyu2026evoscientist}. Both treat the negative result, the failed direction, as a first-class durable write, a provenance-rich design: the value of the memory depends on faithfully attributing each finding to the evidence and validation tier that produced it. DeepScientist's promotion mechanism is, in lifecycle terms, a validation-gated update: a finding does not gain authority over the policy until it survives higher-fidelity validation, an authority-monotonicity discipline applied to scientific belief.

The gap is recoverability of the program when a promoted finding turns out to be wrong. A finding that was validated, promoted, and then used to steer dozens of subsequent experiments contaminates everything downstream of it; rolling it back requires deletion propagation through the explore-exploit policy and every derived finding, and an audit trace that can attribute the wasted exploration to the retracted result. Neither system scores this recovery, nor demonstrates that a retraction propagates correctly, which matters because a cumulative findings memory works by letting early findings compound into later ones. The same path also lets a single contaminated finding do maximum damage. Scientific discovery therefore repeats the corpus-wide finding under unusually clear conditions: the systems are strong at accumulating and steering, and untested at recovering when the accumulation is poisoned.

\subsection{Memory frameworks, libraries, and serving substrates as reusable resources}

Cutting across every domain above is a layer of reusable resources, the memory frameworks, libraries, and serving substrates that domain systems build on, and the way the domains consume this layer is itself diagnostic. The dominant pattern is that domain systems adopt a retrieval-and-storage library (a vector store, a temporal knowledge graph, an extracted-fact memory) and inherit its governance properties wholesale, which usually means they inherit little governance by default. Supporting evidence comes from the budget-matched web study and the ambient-domain finding together: under a token-matched budget the online memory, skill, and workflow modules do not beat a vanilla baseline on WebArena \citep{hajimiri2026budgetconstrained}, and at the lifelog horizon a lossy sophisticated memory system underperforms a high-fidelity RAG baseline \citep{zheng2026lifedial}. As a statement about the resource layer, these suggest that popular memory libraries can add cost and lossiness without the governance that would justify them: they implement write, organize, and retrieve and omit validate, audit, deletion-propagation, and rollback. A domain builder who reaches for one gets accumulation and retrieval and must build the return arc themselves, which is rarely visible in the domain literature above.

The serving and substrate question, where state physically lives and how it is bounded under load, is also consumed cross-domain rather than solved per-domain, and it is the layer where the continual-operation degradation modes surface. MedMemoryBench's memory-saturation finding in healthcare \citep{wang2026medmemorybench} and RoboMME's distractor-session decay in robotics \citep{rathi2026robomme} are the same phenomenon, unbounded growth degrading retrieval, observed through different domains, which argues that bounded, evictable, governed serving substrates are a shared resource need rather than a per-domain one. The few systems that do treat the substrate as governable (VitalTrace's rule-governed compact patient state \citep{qu2026vitaltrace}, GS-Mem's re-observable spatial substrate \citep{lu2026gsmem}) are domain-specific and not packaged as reusable resources, so the next clinical or embodied builder cannot reuse them. The gap at this layer is therefore an interoperability and governance gap at once: the works we examined do not yet provide a reusable memory substrate that ships the return arc, validation at write, provenance through consolidation, verifiable deletion propagation, and rollback traceability, as first-class operations, so a domain system inherits governance less readily than it inherits retrieval. Until such a substrate exists, each domain tends to re-derive accumulation cheaply while leaving governance bespoke. The resource layer turns the survey's diagnosis into an engineering target: governed persistent state should be a reusable substrate, not a per-paper afterthought.

\section{Research frontiers and open problems}\label{sec:frontiers}

\regime{Control}{the open problems on the return arc, and concrete programs to close them.}
The preceding parts establish a diagnosis, not a celebration. Across the
$N{=}435$ coded corpus, the works we examined have learned how to accumulate state
and how to retrieve it: substrates, write pipelines, retrieval rankers,
consolidation schedules, and skill libraries are well studied and heavily benchmarked. They
are less developed on how to \emph{govern} state once it persists. The two thinnest
cells in the coding scheme are the governance axes most tied to consequential
state: only $27/435$ works expose any rollback mechanism, and authority is
the rarest of the six state axes at $72/435$. Every part of this survey reported
the same pattern under a different name. P1 finds almost no foundational work on formal
authority structure and zero on state recovery after corruption; P3 has zero works
that directly address rollback and only two that address audit lineage; P5 states
its deletion and rollback semantics abstractly because the mechanistic literature to
ground them does not yet exist; P6 lacks formal consistency theory for shared
mutable state; P7 lists almost all of its weak cells as ``no benchmark exists'' for
the very governance properties the thesis prioritizes. The agenda is broader than
 retrieval accuracy alone: it is the work of converting an
accumulate-and-retrieve framing into a govern-over-a-lifecycle framing.

One observation should temper how this frontier is read. The governance mechanisms
the survey was able to cite, origin-bound write authority, semantic transactions,
saga-style compensation, audited skill graphs, runtime authority gating, are
overwhelmingly 2025 and 2026 preprints, and few have appeared in widely deployed,
long-running systems or been measured under realistic fault load. That clustering
admits two readings, and the honest position holds both. The optimistic reading is
that the field is correctly identifying the gap and mobilizing quickly, so the
preprint surge is genuine frontier work on a normal multi-year maturation timeline.
 The cautious reading is that governance may be harder to build and benchmark than
memory: it demands formal reasoning, cross-system testing, and infrastructure that a
paper can propose but not easily scale, so some of these mechanisms may not survive
contact with production. We report the frontier as we find it, and treat its youth as
itself a signal: the direction is right, but durability is unproven, and the agenda
is therefore stated as work to be done rather than results to be consolidated.

Five coupled research programs follow from that diagnosis. The first four are the
directions the coverage map already implies, each anchored to a quantified corpus
gap and stated as a coupled pair of a measurement instrument and a mechanism the
instrument would discipline. The fifth is the meta-observation that makes the
other four tractable: late rounds of in-vocabulary ``agent memory'' queries showed
diminishing returns in our search frame, so the governance backbone must borrow
native vocabulary, and in many cases entire correctness calculi, from databases
and distributed systems, formal methods, capability security, machine unlearning,
and regulatory compliance. Each direction below gives the gap, why it matters for
persistent state, a concrete program of work, and the open problem it leaves
standing. Table~\ref{tab:frontier-map} ties the five programs to the corpus cells
they target and to the closest current anchors.

\begin{table}[t]
\centering
\small
\caption{The five frontier programs, the quantified corpus gap each targets, the
closest current anchors in the corpus, and the open instrument each program needs.
The recurring pattern is that mechanisms exist as point solutions while the
measurement and the cross-axis composition are absent.}
\label{tab:frontier-map}
\begin{tabular}{p{2.3cm}p{2.4cm}p{2.6cm}p{2.4cm}}
\toprule
\textbf{Program} & \textbf{Corpus gap} & \textbf{Closest anchors} & \textbf{Missing instrument} \\
\midrule
Lifecycle-complete evaluation &
Rollback $27/435$; recovery never measured &
\citet{chang2025sagallm}, \citet{zheng2026acrfence} &
Recovery-success and recovery-cost metrics on hosted systems \\
\addlinespace
Controlled compounding &
Consolidation drops retrieval handles; contagion has no safe threshold &
\citet{kang2026oslmr}, \citet{liu2026memorycontagion} &
Write-acceptance gate keyed to later addressability \\
\addlinespace
Authority, privacy, deletion, cross-surface &
Authority $72/435$; no single-policy cross-surface benchmark &
\citet{lahjouji2026privacysurvey}, \citet{shah2026unlearningmirage} &
One agent, all surfaces, one declared policy \\
\addlinespace
Shared-memory governance &
No distributed rollback; revocation cascades unmodeled &
\citet{ren2026gatemem}, \citet{zhu2026openport} &
Authority-scoped rollback of tool-originated state \\
\addlinespace
Adjacent-discipline bridging &
Governance axes thin in every part &
\citet{park2026kumiho}, \citet{parakhin2026bureaucracy} &
Native correctness calculi imported, not re-derived \\
\bottomrule
\end{tabular}
\end{table}

\subsection{Lifecycle-complete evaluation and running AOEP on real systems}

The highest-risk gap in the corpus is not that rollback mechanisms are
rare but that rollback is studied as a \emph{mechanism} and almost never as a
\emph{measured capability}. The $27/435$ figure undercounts the problem: even among
those twenty-seven works, none reports the quantities an operator of an always-on
agent would actually need, namely how often recovery succeeds after a corruption
event, how much state is lost when it does, and what the recovery costs in latency
and compute. The vocabulary for these quantities already exists in databases and
distributed systems as recovery-point and recovery-time objectives, mean-time-to-recovery,
and recovery-success rate, but the agent-memory literature has not adopted it. The
result is a field that can describe checkpoint-and-restore designs while being unable
to say whether any of them works under realistic fault load.

The mechanism literature that does exist is suggestive but partial. \citet{chang2025sagallm}
imports the database Saga model into multi-agent LLM planning, giving compensable
execution, automated compensation, and independent validation agents so that a
multi-step workflow can be rolled back through compensating actions rather than naive
undo. This is the right import, but it assumes compensation handlers exist and that
the prior state can be reconstructed, the assumption that fails for irreversible
side effects. \citet{zheng2026acrfence} confronts that failure directly:
it characterizes semantic rollback attacks in agent checkpoint-restore, where naive
restore either duplicates an already-consumed effect or resurrects an authority that
was meant to lapse, and mitigates by recording irreversible effects and choosing
replay-or-fork accordingly. \citet{nakajima2026regimes} offers a third stance: an
event-sourced runtime where state is a deterministic projection of an append-only
log, so exact replay is free, at the cost of an ever-growing log with no garbage
collection. \citet{shawn2026pace} attacks the commit decision rather than the
restore, replacing greedy score-based acceptance of self-updates, which p-hacks
itself into harmful commits, with an anytime-valid betting gate that bounds the
probability of a false commit. These four works disagree about where recovery should
live: in compensation logic, in effect ledgers, in the log substrate, or in the
commit gate. None of them is evaluated against the others on a common recovery
benchmark in the corpus, because we found no shared benchmark for this case.

The concrete program is to make recovery a first-class measured property and to run
the measurement on real systems rather than on synthetic stores. An always-on
evaluation harness should inject corruption, poisoned writes, stale facts, partial
write failures, and lapsed-authority retries, then score recovery-success rate,
state lost at the recovery point, and recovery cost, against a stated objective. The
critical design constraint, drawn from \citet{zheng2026acrfence}, is that the harness
must distinguish reversible from irreversible state, because a restore that looks
correct on retrievable memory can still have duplicated an external side effect. The
open problem this leaves is sharp: within our query frame, there is no agreed definition of a
``recovered'' always-on agent. \citet{orogat2026gem} formalizes governed evolving
memory with six correctness conditions, including rollback traceability, and proves
that record-level stores cannot satisfy them, but it is a theoretical framework
without an empirical harness, and it does not cover the authority and provenance
policies that determine whether a given restore is even permitted. Until recovery is
measured on hosted, multi-task agent benchmarks rather than asserted in design
sections, the thesis's recoverability axis remains a property the current
literature often claims but rarely demonstrates.

\subsection{Controlled compounding and retrieval-addressable consolidation}

Always-on agents are supposed to get better with use. Replay, reflection, skill
libraries, and consolidation are the mechanisms by which experience is meant to
compound into competence. The plasticity-stability tension is well studied as a
learning problem, but as a \emph{governance} problem it exposes a different failure:
the same accumulation that should compound verified competence also lets bad
trajectories, stale facts, and poisoned traces compound into durable policy. The
question is not how to remember more but how to admit only what should become
permanent, and to keep what is admitted addressable by the queries that will later
need it.

Two distinct hazards live at the consolidation boundary. The first is loss of
addressability. Consolidation compresses, and compression can score well on aggregate
recall while silently dropping the retrieval handles, the names, identifiers, dates,
and keys, that a later query must match; a memory then looks compact and coherent
while becoming unusable. The principle the agenda names for this is
retrieval-addressable consolidation: accept a write or update only when it preserves
the handles a later query will use, rather than only preserving topical gist. The
second hazard is contamination that compounds across time. \citet{liu2026memorycontagion}
shows that evaluator bias embedded in stored trajectories propagates
cross-temporally to future agents that share the memory, and that there is no safe
contamination threshold even under oracle consolidation. This is a strong warning
against naive compounding: without an admission gate, shared experience can be
systematically harmful rather than only occasionally noisy. The retention literature gives the lever such a gate would pull.
\citet{kang2026oslmr} casts retention as an NP-hard constrained stochastic
optimization over budget, usefulness, and delayed miss-or-stale cost, learning a
query-conditioned, observability-safe policy that beats recency and heuristic
baselines under tight budgets, the regime where addressability is
sacrificed first. \citet{kumar2026memarchitect} adds a rule-based layer that governs
decay, resolves contradictions among facts, and filters stale ``zombie'' memories
before they re-enter retrieval.

What none of these works supplies is a write-acceptance gate that jointly enforces
both addressability and provenance-bounded admission as a precondition for a write
becoming durable. \citet{shawn2026pace} provides the statistical shape of such a
gate for self-updates, an anytime-valid betting bound on false commits, but does not
key it to whether the committed item remains retrievable, and \citet{joshi2026eywa}
provides the provenance shape, storing immutable source evidence first and deriving
canonical facts only after validation against typed signals, but does not measure
addressability after derivation. The open problem is to unify them: a consolidation
operator that compounds competence only when the candidate write is both
authorized-by-provenance and addressable-by-future-query, and that is evaluated on
whether long-horizon competence actually rises rather than on single-step recall.
The contagion result of \citet{liu2026memorycontagion} makes the stakes explicit.
A field that consolidates for recall without an admission gate is building durable
policy out of unvetted experience, which is the plasticity-stability invariant
failing in its most damaging direction.

\subsection{Authority, privacy, deletion, and a cross-surface benchmark}

The authority axis is the rarest in the corpus at $72/435$, and its scarcity is not
an accident of coding: authority, the question of who may read, write, delete, or
act on a piece of state, is the axis that the accumulate-and-retrieve paradigm has no
reason to represent. An always-on agent that never forgets and never checks who
licensed a fact will treat a preference that was true last month, changed yesterday,
as still authoritative today, and will let any stored fact influence any action. The
governance program here has three tightly linked obligations, privacy, deletion, and
authority, and a missing instrument that would test them together.

On privacy, the corpus has moved from training-time leakage to deployment-time and
memory-resident leakage. \citet{carlini2021extractingtrainingdata} established that
models memorize and leak verbatim training data; \citet{chen2026deploymenttimememorization}
shows the analogous failure for deployed agents memorizing sensitive deployment-time
data, and \citet{chen2026mrmmia} gives the first membership-inference framework aimed
explicitly at chat-agent persistent memory units. \citet{naseh2025riddle} shows the
attack needs only about thirty natural-language queries to infer datastore
membership while evading query-rewriting defenses, and \citet{xu2026memorysilent}
isolates the deeper failure with its RBI-Eval probe: agents integrate sensitive
memory content even when it is not warranted, and retrieval cannot prevent
integration once the memory reaches the generator. On deletion, the picture is
worse, because deletion is widely assumed to be solved and is not.
\citet{staufer2025whatshouldllmsforget} addresses the prerequisite everyone else
assumes away, identifying \emph{what} must be forgotten under right-to-be-forgotten
obligations; \citet{xue2025unlearningverificationsurvey} surveys how one would
\emph{verify} that deletion succeeded, mapping directly onto the recoverability axis;
and \citet{shah2026unlearningmirage} gives a negative result, showing
that supposedly forgotten information resurfaces under multi-hop and aliased queries.
\citet{wang2024unlearningmeetsrag} and \citet{koga2024dprag} give the two
mechanism families, unlearning over retrieval stores and differential privacy spent
only on sensitive tokens, but neither closes the cascade: deleting a source fact does
not delete the summaries, embeddings, and derived facts it spawned.
\citet{zhao2026memorepair} is the closest corpus work to a cascade solution, a
barrier-first repair contract that withdraws stale derived descendants when a source
is deleted and republishes only validated, predecessor-closed successors, cutting
invalidated-memory exposure toward zero; it should be read as the deletion-propagation
invariant made operational, and as evidence of how rare such operationalization
currently is.

On authority, several recent works treat it as a property that must only
narrow. \citet{zhu2026intentgovernedauth} makes the user's expressed intent a
monotone, session-scoped policy that can only reduce the authority static credentials
grant; \citet{ibrahim2026overlaygovernance} proves how delegated authority attenuates
along recursive delegation chains; and \citet{parakhin2026bureaucracy} establishes a
structural equivalence between hardware memory-consistency models and authorization
revocation, bounding unauthorized operations by execution count rather than by a
time-to-live that an idle agent can outlast. \citet{dash2026selfportrait} supplies
the empirical motivation from the user's side: an audit of 2{,}050 real ChatGPT
persistent-memory entries found 96\% were silently system-created, so the authority
over what an agent remembers about a user is, in practice, held by the agent and not
the user. \citet{zhang2025ragmemoryperceptions} and \citet{feng2025regulatoryui}
confirm from interviews and system analysis that users hold incomplete mental models
of persistent memory and want granular rights to review, edit, and delete, with the
interface itself acting as a governance surface.

The missing instrument unifies all three obligations. The corpus contains no
benchmark that drives a single agent across all the data surfaces it touches, the
retrieval store, the tool APIs, the long-term memory, and the inter-agent channel,
under one declared privacy policy, and scores whether the policy holds end to end.
\citet{lahjouji2026privacysurvey} identifies this absence, finding that
information-flow control is the only governance that covers cross-session inference
leakage and that no existing benchmark exercises an agent across surfaces under one
policy. This is the cross-surface benchmark the agenda calls for: a privacy and
authority policy declared once, an agent that must honor inspection, correction,
erasure, and rollback over masked, access-controlled, and origin-bound memory, and a
score that fails the moment a deleted fact resurfaces in any surface or an action
descends from an authority that has lapsed. The open problem is that until such an
instrument exists, the deletion and authority literatures remain a collection of
surface-local point defenses, and the thesis's scope-non-expansion and
deletion-propagation invariants cannot be tested in the multi-surface setting where
they actually fail.

\subsection{Shared-memory governance and authority-scoped rollback}

Shared and multi-agent state multiplies every governance hazard, because a single
write can now influence principals who never authored it, and because the authority
that licensed a write may belong to one agent while the consequences land on another.
Two corpus cells stay persistently thin here. The first is the intersection of
authority and forgetting: who may compel deletion from shared memory, and who
verifies that it propagated to every principal's view. The second, and sharper, is
distributed rollback of tool-originated state once the authority that licensed the
original action has lapsed.

On the access-control side the corpus is maturing. \citet{ren2026gatemem}
benchmarks whether an LLM agent governing a shared memory pool can serve legitimate
requests while enforcing per-role and per-scope access controls and reliably
forgetting data on request across four real-world domains, which is the closest
existing instrument to a shared-memory governance benchmark. \citet{masoor2025samep}
and \citet{ravindran2026portablemem} build the cross-session and cross-vendor
sharing substrates, with AES-256-GCM access control and capability-scoped,
injection-resistant rehydration respectively, but each new shared channel widens
the authority and provenance surface that attacks exploit.
\citet{dalugoda2026hdp} contributes an append-only signed delegation-provenance chain
that binds each agent action to an originating human authorization and scope with
offline audit verification, which is the lineage half of authority-scoped rollback.
\citet{wright2025immutablemem} pushes immutability to its logical end, an append-only
Merkle ledger with privilege-lattice access control, but in doing so it collides head
on with deletion: a strictly immutable ledger cannot honor a forget request, so
immutability and deletion-propagation are in direct tension, and the works we
examined do not yet resolve which yields.

The authority-scoped rollback problem is where the corpus is thinnest. Consider the
canonical case: an agent makes a tool call under
a delegated authority, the side effect is irreversible or externally visible, the
authority later lapses or is revoked, and the call is retried after a transient
failure. A correct system must not let the retry resurrect already-consumed authority
or duplicate the irreversible effect. \citet{zhu2026openport} addresses the temporal
half of this with a State Witness profile that revalidates execution-time
preconditions and fails closed on state mismatch, killing stale approvals when
world-state drifts between an action's approval and its delayed execution, the
time-of-check-to-time-of-use drift that always-on agents create by design.
\citet{deochake2026heartbeatcred} addresses the liveness half: descendant agent
authority self-expires within a provably bounded window once parent liveness ceases,
killing zombie agents that keep acting after operator shutdown.
\citet{bhatt2025etdi} and \citet{south2025authenticateddelegation} bind grants to
signed tool definitions and to scoped, accountable delegation chains, so a silently
redefined tool or an over-broad delegation can be detected. But none of these works,
and no benchmark in the corpus, scores the full authority-scoped rollback case: a
retried call whose licensing authority has lapsed, against a shared substrate, with
an irreversible effect, where the correct outcome is neither blind replay nor blind
restore. This is the multi-agent instantiation of the recoverability axis, and with
only $27/435$ works exposing any rollback at all, it is unmeasured. Trigger
calibration, deciding when an always-on agent should interrupt or act on shared
state proactively rather than wait, is the related open axis: surfacing state too
eagerly is itself an authority overreach, and the utility-calibrated timing of
proactive action remains largely unstudied as a governance question rather than a
helpfulness one.

\subsection{Bridging from adjacent disciplines}

The four programs above share a constraint that the gap analysis makes visible:
late in-vocabulary agent-memory searches returned diminishing new governance
coverage under our query frame. The governance backbone is therefore unlikely to be
filled by re-running the same vocabulary. It needs deliberate import of native
correctness calculi from five established disciplines, treating their decades of
accumulated theory as part of the foundation the governance axes need. This bridging
is already visible in several strong recent works, which is evidence that the move
is productive.

From databases and distributed systems comes transactional correctness and recovery.
\citet{chang2025sagallm} imports the Saga model wholesale; the further frontier is
snapshot isolation, write-ahead logging, point-in-time recovery, and consistency
models, sequential, causal, and eventual, applied to concurrent writes on shared
agent memory, where P6 explicitly lacks formal consistency theory. From classical
knowledge representation comes belief revision: \citet{park2026kumiho} builds a
graph-native versioned memory whose update and contraction operations are proven to
satisfy the AGM and Hansson belief-base postulates, importing rationality guarantees
that the heuristic consolidation literature lacks entirely, and
\citet{wang2026toki} recasts contradiction resolution as write-time concurrency
control with a bitemporal operator algebra and four soundness theorems, giving the
both-truths representation an always-on agent needs to mark a fact superseded rather
than delete it. From formal methods and runtime verification comes provable
enforcement: \citet{bollig2026causalpast} defines a past-time temporal logic with a
provably correct vector-clock monitor for guards over causally-visible stored state;
\citet{winston2026solveraided} compiles natural-language tool-use policies into
SMT-LIB and uses Z3 as a per-call precondition gate; and
\citet{metere2026skillcontainment} promotes a skill manifest to a mechanically
checkable capability-containment proof. These works move enforcement from the design
section into the forward pass, which P5 flagged as its sparsest cell.

From capability security comes the discipline of authority that only narrows.
\citet{ibrahim2026overlaygovernance} overlays recursive delegation, dynamic scoping,
and resource-scope attenuation onto existing relational authorization with formal
proofs of attenuation along delegation chains; \citet{jin2026capseal} keeps raw
credentials non-exportable behind schema-checked invocations to defeat prompt-injection
exfiltration; and \citet{metere2026attestedtoolserver} admits tool servers only via
offline-signed clearance against a pinned trust root. These are capability-system
primitives, object capabilities, attenuation, confinement, transplanted into the
agent setting, and they are why authority, though rarest in the corpus, has
a plausible path toward stronger coverage. From machine unlearning and compliance comes the deletion
and audit backbone: \citet{nguyen2024machineunlearningsurvey} and
\citet{xue2025unlearningverificationsurvey} supply the methods and the verification
theory, \citet{shah2026unlearningmirage} supplies the adversarial test that any
import must pass, and the regulatory framings of \citet{staufer2025whatshouldllmsforget}
connect the recoverability axis to right-to-erasure obligations that regulated
domains, healthcare, finance, and legal, will impose whether or not the field is
ready. \citet{dmitrenko2026memorycomponents} adds a dimension the agent-memory
literature ignores entirely: the license and sustainability posture of the
open-source data-infrastructure projects that back agent memory is itself an
architectural variable for always-on durability, since a memory substrate whose
upstream project is abandoned is a durability failure waiting to happen.

The open problem for this fifth program is integrative rather than local. Each import
above lands a single governance axis: belief revision for mutability, capability
security for authority, provenance DAGs and Merkle logs for provenance,
transactional recovery for recoverability, runtime verification for validation. The
thesis demands that the six axes and five invariants hold \emph{jointly} over the
full lifecycle, yet no work in the corpus composes more than two or three axes at
once, and the imports come from disciplines with incompatible assumptions: immutable
ledgers from blockchain versus cascading deletion from referential integrity,
eventual consistency from distributed systems versus the strict precondition gates of
runtime verification. The target is a unified persistent-state governance layer in
which authority monotonicity, scope non-expansion, deletion propagation, provenance
preservation, and rollback traceability are enforced together and measured together,
with the cross-surface benchmark of the third program and the recovery harness of the
first program as its evaluation. Until that layer exists, the survey's diagnosis is
also an agenda: the works we coded accumulate and retrieve state with much more
sophistication than they govern it, and closing that gap requires importing,
composing, and measuring the governance backbone that always-on agents, as
persistent-state systems, require.

\section{Conclusion}\label{sec:conclusion}

This survey began from a simple distinction: an agent that carries state across tasks is not the same kind of system as one that resets. The episodic assistant that answers a question and forgets it can be wrong without making the error durable. The always-on agent inherits what it wrote yesterday. Its competence and its liabilities both compound, and the substrate that lets it carry knowledge forward is the same substrate that lets a stale preference, a poisoned trace, or a lapsed permission govern an action taken months later. Such a system should be analyzed as \emph{persistent state}, not memory alone: retrievable memory plus task ledgers, permissions, tool and credential state, standing commitments, provenance records, shared and social state, and trigger conditions, with external commitments treated as part of the accountability surface because they must remain traceable to the state that authorized them. Memory is the most studied component of that union, but treating the whole as memory hides the governance obligations that make always-on agents distinct.

\subsection{What the survey established}

Four contributions structure the argument. First, always-on agents are persistent-state systems. Their state is governed over a lifecycle, observe, write, validate, organize, retrieve, act, update, forget, audit, and rollback, and can be characterized along six axes: authority (who may write or invoke it), scope (where it may legitimately apply), mutability (how it changes under new evidence), provenance (where it came from and through what transformations), recoverability (whether a change can be undone), and actionability (whether it can drive an external effect). Five invariants tie the axes to the lifecycle: authority monotonicity, scope non-expansion, deletion propagation, provenance preservation, and rollback traceability. The frame is wider than memory by construction. Classical cognitive architectures already separated procedural, episodic, and semantic stores within a single decision cycle \citep{laird2012soar, anderson2004actr, tulving1972episodic}, and the complementary-learning-systems tradition explained why fast capture and slow consolidation must be kept on different timescales to avoid catastrophic interference \citep{mcclelland1995complementary}. Modern agent frameworks ported these modules into LLM systems \citep{sumers2024cognitivearchitectureslanguageagents}, but they ported the storage without the governance: the cognitive theories assumed type boundaries and decay schedules enforced by biology, whereas an LLM agent mixes fast and slow updates in flat text with no runtime enforcement of who wrote what, with what authority, recoverable by whom.

Second, the literature is read as a layered persistent-state stack rather than as a memory taxonomy, so that substrates, the lifecycle, continual adaptation, governance, multi-agent state, evaluation, failure modes, and application domains each receive a first-class treatment. This is not a cosmetic relabeling. A memory-form taxonomy strands governance, serving substrates, deletion propagation, and rollback as footnotes, which is precisely the failure mode the survey exists to name. The stack instead lets the memory-form and lifecycle axes live where they are sharpest, inside the substrate and state-movement layers, while giving authority, provenance, and recoverability the room to carry the argument.

Third, we coded a $435$-work corpus and measured which parts of the lifecycle each work addresses. The main finding is a structural asymmetry between accumulation and governance. In our coding, the forward arc of the lifecycle is crowded: retrieve appears in $269$ works and write in $200$. The return arc is sparse: only $27$ of $435$ works expose any rollback mechanism, and across the state axes authority is the rarest at $72$ of $435$. These counts are scoping estimates, but they consistently show that the works we examined study storing and retrieving more thoroughly than revoking, recovering, attributing, and bounding stored state.

Fourth, we turned the diagnosis into a checkable protocol by proposing the Always-On Evaluation Protocol and instantiating it as AOEP-v0: an event-stream and snapshot schema, worked episodes, a validator, and a pilot that scores state mutation and recovery rather than answer quality. The pilot is deliberately small and representational rather than a benchmark. It illustrates how tested memory wrappers can fail obligations that require explicit governance state, such as permission epochs, deletion ledgers, trust tiers, conflicts, and rollback records.

\subsection{Why bigger context and better retrieval are not enough}

One objection to our thesis is that the problem is already dissolving on its own, that million-token context windows and stronger retrieval will let an agent read all of its past and behave correctly. The corpus does not support this. Larger context and retrieval address a reading problem, and the governance gaps we document are not reading problems. Long-context studies show that capacity is not comprehension: models underuse information placed in the middle of a long window even when it is present \citep{liu2023lostmiddlelanguagemodels}, and retrieval quality decomposes into noise robustness, rejection, and faithfulness failures that more tokens do not fix \citep{chen2023benchmarkinglargelanguagemodels}. Even granting perfect recall, retrieval answers ``what did I record?'' It does not answer the four questions an always-on agent must answer to act safely: what \emph{should} persist, which competing record is \emph{authoritative}, how a record should \emph{change} when new evidence contradicts it, and how a user can \emph{revoke} it. These are decisions about authority, mutability, and recoverability, not about context length.

The difference becomes sharp where state crosses from passive storage into action. A preference that was true last month and was changed yesterday is still retrievable today; the question is whether the agent treats it as superseded or as current, and that is a mutability-and-provenance decision no retriever makes \citep{uddin2026recallforgettingbenchmarkinglongterm, ma2026memprobe}. Personalization research has repeatedly shown that accumulated user state drifts, contradicts itself, and can be poisoned: routine interactions gradually weaken confirmation boundaries and expand the agent's effective action scope without any single step looking like an attack \citep{xu2026toxicchats}. Surveys of personalized agents catalog these requirements, multi-session coherence, preference retrieval, drift handling, while noting that the systems offer no authority monotonicity and no scope gating to enforce them \citep{xu2026personalizedllmpoweredagentsfoundations}. A bigger window lets the agent see the poisoned preference more reliably. It does not give the agent or the user any mechanism to mark it untrusted, bound where it applies, or roll back the actions it already licensed. Recent state-trajectory work makes the same point formally: stores that keep conclusions while dropping sources become hard to correct \citep{kwon2026reclaim}, and record-level stores cannot satisfy rollback traceability without additional governed-state structure \citep{orogat2026gem}. The deficiency is representational and operational; scale alone does not remove it.

\subsection{The operational agenda}

The corrective is to treat persistent state as a measured, governed lifecycle rather than as a passive log to be searched. Concretely, this means three commitments that recur across every part of this survey and that we summarize in Table~\ref{tab:conclusion-shift}. State decisions must be made explicitly rather than implicitly: a system should decide what to admit at the write boundary, what authority a record carries, and under what scope it may act, instead of accumulating everything and hoping retrieval sorts it out. State must remain attributable and recoverable: provenance must survive consolidation so that a corrupted or superseded record can be located, quarantined, and undone, rather than dropped the moment a summary scores well on aggregate recall. And state governance must be evaluated rather than asserted: a system that claims to honor deletion, respect authority, or support rollback should be driven through perturbations that exercise those obligations and scored on whether the obligations actually hold.

\begin{table}[t]
\centering
\caption{The shift the survey argues for: from persistent state as an accumulated log to persistent state as a governed lifecycle. Each row pairs the prevailing framing with the operational alternative and the corpus signal that motivates it.}
\label{tab:conclusion-shift}
\small
\begin{tabular}{@{}p{0.28\columnwidth}p{0.34\columnwidth}p{0.28\columnwidth}@{}}
\toprule
\textbf{Question} & \textbf{Accumulate-and-retrieve framing} & \textbf{Governed-lifecycle alternative} \\
\midrule
What persists? & Store the interaction, retrieve later. & Admit at the write boundary under a scope and authority label. \\
Which record rules? & Most similar to the query. & The authoritative one; authority only narrows. \\
How does it change? & Append a newer entry. & Mutate with provenance; mark prior truth superseded. \\
How is it revoked? & Delete the row, hope it propagates. & Cascade deletion, verify completeness, log the rollback. \\
How do we know? & Downstream task accuracy. & Score the governance obligation directly. \\
\bottomrule
\end{tabular}
\end{table}

These commitments are practical because isolated pieces of each already exist in the corpus. Operating-system framings externalize working state to durable storage and reason about what should be paged in and out, supplying the substrate discipline a lifecycle needs \citep{packer2024memgptllmsoperatingsystems}. Provenance-rooted and immutable-ledger designs show that source attribution and tamper-evidence can be maintained through updates \citep{wright2025immutablemem}, and certified defenses demonstrate that write-time provenance plus randomized ablation can bound the influence of poisoned writes \citep{sharma2026smsrcertified}. Transactional models import compensable execution and rollback from databases into multi-agent planning \citep{chang2025sagallm}. Workflow-abstraction methods show that accumulation can be made selective and reusable rather than indiscriminate \citep{wang2024agentworkflowmemory}. The missing step is integration across all six axes and five invariants, with evaluation as obligations rather than as incidental byproducts of task success. AOEP-v0 starts from that integration by separating positive obligations a system must actively satisfy from negative no-leakage invariants that a system can satisfy trivially by storing nothing. The pilot suggests that obligations reducible to recall and obligations requiring maintained governance state should be reported separately.

\subsection{Outlook}

The stakes rise rather than fall as deployments mature. The failure modes we cataloged compound under continual operation: poisoning persists across sessions once it enters durable state \citep{xie2026crosssessioninjection}, alignment can degrade as the agent accumulates and reuses its own outputs \citep{shao2025misevolution}, and in shared settings evaluator bias in stored trajectories propagates to future agents with no safe contamination threshold even under oracle consolidation \citep{liu2026memorycontagion}. None of these is a single-turn prompt problem, and none is solved by reading more of the past more reliably. They are governance problems in the sense this survey defines: failures of authority, scope, provenance, and recoverability over a lifecycle that the works we coded instrument mostly on its forward arc. We read our main numbers, $27$ of $435$ works with any rollback and $72$ of $435$ touching authority, not as a verdict on the quality of existing work, which is often excellent within its frame, but as a map of where the frame stops. Persistence is not a minor feature of always-on agents to be handled by a larger window or a sharper index. It is the property that makes these systems both more capable and more accountable than their episodic predecessors, and treating it as a measured, governed lifecycle, deciding what should persist, which state is authoritative, how it changes, and how it is revoked, is the operational agenda the next generation of always-on agents will be built and judged against.

\FloatBarrier 
\appendix
\section{Survey methods and corpus construction}\label{app:methods}

The headline claims of this survey, in particular the lifecycle-stage and
state-axis counts and the rollback and authority figures the argument
rests on, are only as credible as the procedure that produced the corpus. This
appendix documents that procedure so the numbers are auditable rather than
asserted.

\subsection{Sources and time window}

The corpus draws on four sources: arXiv (cs.AI, cs.CL, cs.LG, cs.CR) as the
primary source for 2023 to 2026 preprints; Semantic Scholar for backward and
forward citation chasing and for venue and citation metadata; OpenReview (ICLR,
NeurIPS, COLM) for venue status and camera-ready versions; and the ACL Anthology
for published versions at ACL, EMNLP, and NAACL. The time window is 2023-01
through 2026-06, with foundational anchors admitted regardless of date when they
ground a taxonomy or a term the survey uses (for example, the episodic-semantic
distinction, catastrophic interference, and the Soar architecture).

\subsection{Search procedure}

Collection ran in three passes. We \emph{seeded} from recent agent-memory surveys
and the closest-work set and extracted their unit of analysis and shared
vocabulary. Next we \emph{chased citations} backward (references) and forward
(citing papers via Semantic Scholar) two hops from each seed, keeping any work
that touches a persistent-state operation, state type, benchmark, or failure
mode. Finally we ran \emph{targeted term-set sweeps} for 2024 to 2026 work the
citation graph missed, with query sets grouped by mechanism (for example
``agent memory'', ``memory OS'', ``procedural memory agent'', ``knowledge graph
memory''), benchmark (``long-term conversational memory'', ``multi-session
benchmark'', ``lifelong agent benchmark''), failure mode (``memory poisoning LLM
agent'', ``indirect prompt injection'', ``machine unlearning LLM'', ``deletion
residue''), continual adaptation and personalization (``continual learning
agent'', ``test-time training'', ``knowledge editing forgetting''), and
foundations (``cognitive architecture memory'', ``LLM agent survey''). The
targeted sweeps were run as iterative non-duplicate rounds: each round queried
the sources, removed works already in the corpus, and added only genuinely new
hits, until additional rounds returned predominantly duplicates. Rounds aimed
deliberately at the governance frontier (rollback, authority, deletion
propagation, durable execution, formal verification, and the adjacent database,
distributed-systems, and HCI literatures) so that the governance counts cannot be
attributed to having searched only the memory mainstream.

\subsection{Inclusion and exclusion criteria}

A work is \emph{included} if it satisfies at least one of: (I1) it builds or
studies a persistent-state mechanism; (I2) it contributes a benchmark or
evaluation method relevant to memory, long-horizon, or stateful behavior; (I3) it
documents a persistence-induced failure mode; (I4) it is a foundational anchor a
reader needs to place the state taxonomy, the lifecycle, or the
plasticity-stability framing; or (I5) it defines a boundary case (long-context,
RAG, personalization) used to delimit always-on agents. A work is \emph{excluded}
if it is (E1) pure model-architecture or pretraining work with no memory, state,
or agent angle; (E2) an application that uses an agent but contributes nothing to
a persistent-state operation, state type, benchmark, or failure mode (we add no
padding for breadth alone); (E3) a duplicate or superseded version when a later
one is cited; or (E4) available only in a venue without an accessible English
record. The resulting corpus is $N{=}435$ works.

\subsection{Coding scheme}

Every included work is coded along four dimensions: a \emph{category} (mechanism,
benchmark, failure mode, foundation, boundary, or survey); the \emph{lifecycle
stages} it exercises, as a multi-label set over observe, write, validate,
organize, retrieve, act, update, forget, audit, and rollback; the \emph{state
axes} it addresses, as a multi-label set over authority, scope, mutability,
provenance, recoverability, and actionability; and an \emph{application sub-area}.
Multi-label coding is why per-stage and per-axis incidences sum to more than the
corpus size. All counts in the survey are recomputed from the coded corpus by a
script rather than hand-entered.

\subsection{Coding reliability}

Because the lifecycle and axis labels are interpretive, we measured inter-coder
reliability with a blind second coding of a $236$-work sample (over half the
corpus). On this sample, pooled per-cell agreement is $0.82$ on lifecycle stages
(Cohen's $\kappa = 0.58$) and $0.74$ on state axes ($\kappa = 0.44$), which is
moderate agreement, as expected for fine-grained multi-label coding of short
descriptions. Two points matter for how the headline claims should be read.
First, the reliability evidence supports the survey's \emph{directional} claims
(forward arc dominant, return arc sparse) far better than it supports any exact
cell count, and that is the only weight we place on it. Even the rollback label,
the rarest, carries real labeling noise: the two coders agreed on it for only
$11$ of the $19$ works either marked, so the precise value $27$ should be read as
an estimate with a margin of several works, not a measurement. Second, the
directional comparison does not depend on that precision, because it is a large
qualitative asymmetry rather than a narrow margin: retrieve appears in
$269$ works and rollback in roughly $27$, an order-of-magnitude gap that no
plausible reclassification of borderline cells could close. A reader who
distrusts the exact $27$ can inflate it severalfold and still obtain the same
directional conclusion: the return arc is much sparser than the forward arc. We
therefore treat the counts as well-supported for the comparative claims the
survey makes (forward arc dominant, return arc sparse) while cautioning that any
single cell
should be read as an estimate, not a precise measurement.

\subsection{What the rollback count does and does not aggregate}

The figure that twenty-seven of $435$ works expose any rollback mechanism
aggregates three mechanically distinct operations: internal-state rollback
(reverting a stored fact or a consolidation decision), workflow rollback
(compensating a partially executed multi-step task), and external-effect rollback
(undoing an already-committed side effect such as a payment or a deletion). The
aggregate is therefore generous for the easiest subclass and an overcount for the
hardest. External-effect rollback, the subclass that matters most because
irreversible effects cannot be repaired by any internal mechanism, is rarer than
the aggregate suggests. A finer corpus pass should separate the three; we report
the conservative aggregate and flag the subclass structure so the headline number
is not over-read.

\subsection{Threats to validity}

Three threats deserve naming. First, the corpus is a \emph{representative coding,
not an exhaustive census}; we sampled toward the governance frontier and the gap
persisted, but we cannot exclude that some governance work uses vocabulary none of
our query sets reached, which is why the universal claims in the body are scoped
to the corpus or to works we examined rather than to the literature as a whole.
Second, the corpus is weighted toward recent preprints, and the governance turn we
document is itself recent (Figure~\ref{fig:timeline}); restricting to work
through 2024 makes the governance fraction smaller rather than larger, so the gap
is not an artifact of including immature 2026 work, but the durability of the
2025 to 2026 governance preprints remains unproven. Third, single-coder labeling
with a second-coder check is weaker than full double-coding; the $\kappa$ values
above bound how much this matters, but they do not make the exact cell counts
immune to recoding.

\FloatBarrier
\bibliographystyle{plainnat}
\bibliography{references}

\end{document}